\documentclass[12pt]{article}
\usepackage{mathrsfs}
\usepackage{cite}
\usepackage{hyperref}
\hypersetup{colorlinks=true,backref=true,linkcolor=black,anchorcolor=black,citecolor=black,filecolor   =black,menucolor=black,pagecolor=black,urlcolor=black}
\usepackage[english]{babel}
\newlength{\abstractwidth}
\usepackage{epsf}
\usepackage{color}
\usepackage{graphicx}
\usepackage{amssymb}
\usepackage{latexsym}

\usepackage{color}
\definecolor{dgreen}{rgb}{0,0.70,0.30}
\definecolor{gold}{rgb}{0.85,.66,0}
\definecolor{purple}{rgb}{1.0,0.3,0.6}
\definecolor{red}{rgb}{1.0,0.0,0.0}

\usepackage{tikz}
\usetikzlibrary{calc,decorations.pathreplacing}

\newcommand{\nn}{\nonumber}

\newcommand{\eps}{\epsilon}

\def\ba{\begin{align}}
\def\ea{\end{align}}
\def\bse{\begin{subequations}}
\def\ese{\end{subequations}}
\def\Im{\,{\rm Im}\,}

\def\cA{{\cal A}}
\def\cB{{\cal B}}
\def\cC{{\cal C}}
\def\cD{{\cal D}}
\def\cE{{\cal E}}
\def\cF{{\cal F}}
\def\cG{{\cal G}}

\def\cJ{{\cal J}}

\def\cL{{\cal L}}
\def\cM{{\cal M}}

\def\cO{{\cal O}}

\def\cR{{\cal R}}

\def\cT{{\cal T}}

\def\cV{{\cal V}}
\def\cW{{\cal W}}

\def\ba{{\bf a}}

\def\bp{{\bf p}}

\def\mC{\mathfrak{C}}
\def\mD{\mathfrak{D}}
\def\mE{\mathfrak{E}}
\def\mF{\mathfrak{F}}

\def\mL{\mathfrak{L}}

\def\mS{\mathfrak{S}}

\def\mm{\mathfrak{m}}

\def\mp{\mathfrak{p}}
\def\mq{\mathfrak{q}}

\def\bp{{\bf p}}
\def\p{\partial}
\def\half{{1\over 2}}

\def\f{\varphi}
\def\no{\nonumber}
\def\sm{\smallskip}

\def\CC{{\mathbb C}}
\def\NN{{\mathbb N}} 
\def\RR{{\mathbb R}}

\def\ZZ{{\mathbb Z}}

\def\M{\mathcal M}
\def\tE{E}
\def\RE{non-holomorphic  Eisenstein series}

\def\tet{\vartheta}
\def\ep{\varepsilon}

\def\half{ {1\over 2}}

\renewcommand{\thefootnote}{\fnsymbol{footnote}}
\renewcommand{\thanks}[1]{\footnote{#1}}
\newcommand{\starttext}{
\setcounter{footnote}{0}
\renewcommand{\thefootnote}{\arabic{footnote}}}

\newcommand{\bea}{\begin{eqnarray}}
\newcommand{\eea}{\end{eqnarray}}
\newcommand{\be}{\begin{eqnarray}}
\newcommand{\ee}{\end{eqnarray}}

\newcommand{\e}{\epsilon}

\begin{document}
\starttext
\setcounter{footnote}{0}

\begin{flushright}
\today\\
DAMTP-08-02-2015 \\
IPHT-T15/012 \\
IHES/P/15/04

\end{flushright}

\bigskip

\begin{center}

{\Large \bf On the modular structure of the genus-one  Type II superstring low energy expansion}

\vskip 0.7cm

{\large \bf Eric D'Hoker$^{(a)}$, Michael B. Green$^{(b)}$ and Pierre Vanhove$^{(c)}$}
\vskip .2in

{ \sl (a) Department of Physics and Astronomy }\\
{\sl University of California, Los Angeles, CA 90095, USA} 

\vskip 0.08in

{ \sl (b) Department of Applied Mathematics and Theoretical Physics }\\
{\sl Wilberforce Road, Cambridge CB3 0WA, UK}

\vskip 0.08in

{\sl (c) Institut des Hautes \'Etudes Scientifiques\\
 Le Bois-Marie, 35 route de Chartres,
 F-91440 Bures-sur-Yvette, France, \\
 and \\
Institut de physique th\'eorique,\\
 Universit\'e Paris Saclay, CEA, CNRS, F-91191 Gif-sur-Yvette}

\vskip 0.2in

{\tt \small dhoker@physics.ucla.edu; M.B.Green@damtp.cam.ac.uk; pierre.vanhove@cea.fr}

\end{center}

\vskip -1 in

\begin{abstract}
The analytic contribution to the low energy expansion of  Type II string amplitudes at genus-one is a power series in space-time derivatives with coefficients that are  determined by integrals of modular functions over the complex structure modulus of the world-sheet torus.  These modular functions are associated with world-sheet vacuum Feynman diagrams and given by multiple sums over the discrete  momenta on the torus.  In this paper we exhibit  exact differential and algebraic relations for a certain infinite class of such modular functions by showing that they satisfy Laplace eigenvalue equations with inhomogeneous terms that are polynomial in non-holomorphic Eisenstein series.   Furthermore, we argue that the set of modular functions that contribute to the coefficients of interactions
 up to order $D^{10} \cR^4$  are linear sums of functions in this class and quadratic polynomials in  Eisenstein series and odd Riemann zeta values.  Integration over the complex structure results in coefficients of the low energy expansion  that are rational numbers multiplying  monomials in odd Riemann zeta values.  
 \end{abstract}

\newpage
\setcounter{tocdepth}{2}
 \tableofcontents

\newpage

\section{Introduction}
\setcounter{equation}{0}
\label{intro}

The low energy expansion of the Type II superstring effective action generates an infinite set of higher derivative interactions that generalize the Einstein--Hilbert action of classical general relativity.  The dependence of these interactions on the vacuum expectation values of the scalar fields  is highly constrained by space-time supersymmetry and  duality. The detailed study of these higher derivative interactions is motivated by the desire to obtain a better understanding of the full non-perturbative structure of string theory, and has  revealed some striking systematics of the modular structure with respect to the dualities of string theory. The simplest nontrivial example is provided by uncompactified Type IIB theory, which has maximal supersymmetry and for which the S-duality group is $SL(2,\ZZ)$, and the higher derivative interactions depend upon the complex axion-dilaton   through non-holomorphic modular forms.   In this case the low order terms in the low energy expansion preserve a fraction of the supersymmetry.  These BPS interactions are expected to be ``protected'' in some sense, and are known to possess a small number of perturbative contributions, However, little is known about the structure of the modular forms associated with higher order terms.  Important information concerning  these higher derivative interactions  would be provided by determining their contribution order by order in string perturbation theory.

\sm

The low energy expansion of string scattering amplitudes, from which the higher derivative interactions are obtained,  has a rich mathematical structure already in uncompactified string theory, at any order in perturbative theory.  For example, the higher derivative interactions that arise at tree-level in Type II string amplitudes have coefficients that are rational multiples of odd multi-zeta values, and these in turn are related to multiple poly-logarithms.    Contributions from higher loop amplitudes are more difficult to analyze  due, in part, to the presence of non-analytic behavior associated with massless intermediate states.  However, the procedure for isolating those non-analyticities and extracting local effective higher derivative interactions is by now well-established, at least at low orders in the low energy expansion. 

\sm

Perturbative calculations in Type II string theory in ten-dimensional Minkowski space-time have been carried out to various string loop orders and various orders in $\alpha'$, and have been compared against the predictions of space-time supersymmetry and duality when available~\cite{Green:1997tv,Green:1997as,Green:1998by,Green:1999pv,Green:1999pu,Sinha:2002zr,D'Hoker:2005jc,D'Hoker:2005ht,Green:2005ba,Green:2014yxa,Gomez:2013sla,D'Hoker:2014gfa}. Specifically, the strength of the $\cR^4$ interaction was evaluated at one-loop order in \cite{Green:1997tv}, and the cancellation at two loop order, predicted from space-time supersymmetry and duality,  was verified in \cite{D'Hoker:2005jc}. The strength of the tree-level and two-loop orders of the $D^4 \cR^4$ interaction were predicted on the basis of supersymmmetry and duality in \cite{Sinha:2002zr,Green:1999pu}, and its one-loop contribution was predicted to vanish, which was verified in \cite{Green:1999pv}.   The non-zero predicted value  at two-loop order was verified in \cite{D'Hoker:2005ht}. The predictions of supersymmetry and duality made  for the $D^6\cR^4$ interactions in \cite{Green:2005ba}, were verified at two-loop order in \cite{D'Hoker:2014gfa}, and at three-loop order in  \cite{Gomez:2013sla} (although in this last instance a suspicious factor of 3 still needs to be elucidated). These comparisons, made originally for the case of ten-dimensional  Minkowski space-time, have been extended to toroidal compactifications, which are associated with higher rank duality groups,   in a number of papers, including 
\cite{Green:1997di,Kiritsis:1997em,Obers:1998fb,Pioline:1998mn,Obers:1999um,Basu:2007ru,Basu:2007ck,Green:2010wi,Pioline:2010kb,Green:2010kv,Pioline:2015yea}.

\sm

One theme that enters in many of these works is the consideration of homogeneous and inhomogeneous Laplace-eigenvalue equations in which the Laplace-Beltrami operator acts either on the scalar field vacuum expectation moduli or on the moduli of the worldsheet Riemann surfaces of string perturbation theory. 
In particular, in the matching of the $D^6\cR^4$ interaction at two-loop order, the relation between the integrand of its coefficient with the Zhang-Kawazumi invariant \cite{D'Hoker:2013eea} could be used fruitfully only once it was realized that this invariant satisfies a Laplace-eigenvalue equation on the moduli space of genus-two Riemann surfaces \cite{D'Hoker:2014gfa}.   Therefore, there appears to be strong evidence that a rich mathematical structure emerges not only for the low energy expansion of string scattering amplitudes, but also for their corresponding integrands on the moduli space of compact Riemann surfaces.

\sm 

The study of the genus-one expansion at low orders was initiated in
\cite{Green:1999pv, Green:2008uj}.  The results to be presented here
will give a more complete and unified understanding for certain
classes of the expansion coefficients,\footnote{They will also correct
  a number of numerical errors in earlier papers.}  and they hint at
an underlying elegant mathematical structure that generalizes the
multiple zeta values that arise at tree
level~\cite{Brown:2013gia,Stieberger:2009rr,Schlotterer:2012ny,Brown1,Stieberger:2013wea,Drummond:2013vz}. The expansion of the integrand  of the genus-one superstring four-graviton amplitude naturally  leads to generalizations of the classical  Kronecker-Eisenstein series~\cite{Weil:1977} closely  related to  series that have been considered in~\cite{Levin:1997,Goncharov:2008} for example.  We note that there has also been some recent progress in
understanding the mathematical structure of  $N$-particle genus-one open string amplitudes~\cite{Broedel:2014vla} based on elliptic multiple-polylogarithms constructed in~\cite{Brown:2011}.

\subsection{Overview of results}

In the present paper, we shall concentrate on the case of genus-one and consider a world-sheet torus $\Sigma$ with complex structure modulus $\tau$.  The modular functions of interest are then generated by vacuum Feynman diagrams for a free massless scalar field on $\Sigma$, and may be represented by multiple sums over the discrete momenta on the torus, subject to overall momentum conservation.  First, we shall consider the infinite class of contributions arising from $SL(2, \ZZ)$-invariant modular functions defined by the following multiple sums, 
\bea
\label{1a1}
C_{a_1, \dots, a_\rho} (\tau) = \sum _{ (m_r,  n_r ) \not= (0,0)} \! \!
\delta _{m,0} \, \delta _{n,0} \, \prod _{r=1} ^\rho \left (  { \tau_2  \over \pi  |m_r \tau + n_r|^2} \right )^{a_r}
\eea
where $\tau= \tau_1+ i \tau_2$ with $\tau_1, \tau_2$ real and $\tau_2 >0$. For $r=1,2, \dots,\rho$,  the integers $m_r,n_r \in \ZZ$ represent the world-sheet momenta $\bp_r = m_r \tau + n_r$ on the torus. Total momentum conservation is enforced by the Kronecker $\delta$ factors which require the vanishing of $m= m_1 + \dots + m_\rho$ and $n=n_1 + \dots + n_\rho$.  The functions $C_{a_1, \dots, a_\rho}(\tau)$ may be defined for all complex values of $a_r$ by analytic continuation, though the cases of interest here will be limited to integer values of $a_r$ such that the infinite momentum sums are convergent.  The function $C_{a_1, \dots, a_\rho}$ evaluates a  $(\rho-1)$-loop Feynman diagram on a world-sheet torus of modulus $\tau$. The set of functions  $C_{a_1, \dots, a_\rho}$ defines sub-classes of Feynman diagrams contributing to string amplitudes with arbitrary numbers of external massless states. 

\sm

The {\sl weight} $w=a_1+\dots + a_\rho$ of the diagram corresponds to the number of scalar Green functions in the diagram, and labels  the order in the low energy expansion  at which the modular function contributes, namely schematically $D^{2w} \cR^4$.  Such a diagram can contribute to the low energy expansion of the $N$-graviton amplitude with $N \ge w-1$.  It is important to stress that the notion of weight $w$ introduced here is different from the notion of {\sl modular weight} under $SL(2,\ZZ)$ transformations; the modular weight of every $C$-function defined above is zero.  In the present paper we shall determine all higher derivative effective interactions up to weight $w=5$ corresponding schematically to $D^{10}\cR^4$.

\sm

There are many other contributions to the coefficients of general higher derivative interactions that are defined by more general Feynman diagrams.  However, we will find relations between such contributions and the question of how many diagrams of weight $w$ are genuinely distinct is a fascinating one.  
  In particular, we shall examine one weight 5 contribution to the four-graviton amplitude which is not of the type of the $C$-functions above, but does contribute to the effective interaction $D^{10} \cR^4$. 
  
  \sm
  
We shall now provide a more detailed description of the results for the modular functions 
$C_{a_1, \cdots, a_\rho}(\tau)$ obtained   in this paper for increasing values of $\rho$.

\sm

$\bullet$ For $\rho=1$,  the function $C$ vanishes identically.

\sm

$\bullet$ For $\rho =2$ we have the relation, 
\bea
C_{a_1,a_2} (\tau) =\tE_{a_1+a_2} (\tau) 
\label{ca1a2}
\eea
where,
\bea
\label{1a2}
\tE_s(\tau)  = \sum _{(m,n) \not= (0,0)} { \tau_2 ^s \over \pi ^s |m \tau + n|^{2s}}
\eea
is the  $SL(2, \ZZ)$-invariant \RE.\footnote{The normalization factor of $\pi^{-s}$ in (\ref{1a2}) is slightly unconventional but has been included for later convenience. Henceforth, the dependence on $\tau$ will generally not be exhibited explicitly.} 
The Eisenstein series satisfies the following Laplace-eigenvalue equation, 
\bea
\label{1a3}
\Delta \tE_s = s(s-1) \tE_s
\eea 
where the Laplace-Beltrami operator on the upper half plane is defined by $\Delta = 4 \tau_2^2 \p_{\tau} \p_{\bar \tau}$. The Eisenstein series is known to be the only non-holomorphic modular function to satisfy (\ref{1a2}) and to have polynomial growth as $ \tau_2 \to \infty$.    The Eisenstein series $E_s$ must be  assigned a weight $w=s$ in line with the definition of the weight of $C_{a_1, \dots, a_\rho}$ given  earlier (and again,  $w$ should not be confused with the modular weight, which is zero for $E_s$).

\sm

$\bullet$ For $\rho \geq 3$, the functions $C$ are special cases of Kronecker-Eisenstein series \cite{Weil:1977}.   The simplest infinite class has $\rho=3$ for which we shall derive the most complete results in this paper. In particular, we shall show that each modular function $C_{a,b,c}$ is a linear combination, with rational coefficients,  of a basis of modular functions $\mC_{w; s; \mp}$, satisfying the inhomogeneous Laplace-eigenvalue equation,
\bea
\label{1a4}
\left ( \Delta - s(s-1) \right ) \mC_{w;s;\mp} = \mF _{w;s;\mp}( \tE_{s'} )
\eea
Here,  $w=a+b+c$ is the weight;  $s$ is given by $s=w-2\mm$ where $\mm$ may take one of the following values $\mm = 1, 2, \dots, [(w-1)/2]$; and the multiplicity index $\mp$ takes values $\mp =  0,  \dots , [(s-1)/3]  $. The function $\mF _{w;s;\mp}( \tE_{s'})$ is a polynomial of degree 2 and weight $w$  in the Eisenstein series $ \tE_{s'}$ with the integer $s'$ in the range $2 \leq s' \leq w$.

\sm

The cases of low weight $w$ may be written down simply in terms of the original functions $C_{a,b,c}$, and we will find,  for example,\footnote{The structure of these equations, and the equations for higher weight functions, is analogous to that of equations that arise in considerations of S-duality in \cite{Green:2005ba,Green:2008bf},  Such equations have solutions  that were discussed in \cite{Green:2014yxa}. } 
\bea
\label{1a5}
\Delta C_{1,1,1} & = & 6 \,  \tE_3
\no \\
(\Delta -2) C_{2,1,1} & = & 9  \tE_4 -  \tE_2^2
\no \\
\Delta C_{2,2,1} & = & 8 \,  \tE_5
\eea
For these low eigenvalues, no degeneracy occurs and the multiplicities are 1. The equations for higher weights are more involved and will be spelled out in section \ref{sec34}.

\sm

The structure of the spectrum of the Laplace-eigenvalue equations, summarized above, will be proven with the help of a generating function for the modular functions $C_{a,b,c}$. This function is closely related to the sunset diagram of a three scalar fields with three different mass parameters, and exhibits remarkable properties with respect to the non-compact group $SO(2,1;\RR)$. Its structure, which is governed by the representation theory of $SO(2,1;\RR)$ mod three, will provide a complete explanation of the spectrum announced in (\ref{1a4}).

\sm

The Laplace-eigenvalue equation for eigenvalue $s(s-1)=0$ at each odd weight $w$ is special in that it  exhibits a single inhomogeneous term proportional to $ \tE_w$. Therefore, it may  be integrated with the help of (\ref{1a3}) and, for each odd weight $w$,  will lead to one algebraic relation between the functions $C_{a,b,c}$ and Eisenstein series. The Laplace equation determines the algebraic equation up to an additive constant of integration which may be obtained from the asymptotic expansions of the equation. In the cases cited above, we find the relations, 
\bea
\label{1a6}
C_{1,1,1} & = &  \tE_3 + \zeta (3) 
\no \\
C_{2,2,1} & = & { 2 \over 5 }  \tE_5 + { \zeta (5)  \over 30} 
\eea
The result on the first line had  been obtained previously by Don Zagier \cite{Zagier:2014} using an artful direct evaluation of the multiple sums. For all higher odd weights, an explicit form may be obtained for the algebraic equation which results from  integrating the zero eigenvalue equation. It will be presented in equation (\ref{3g5}) in section \ref{sec3}.

\sm

When $s \geq 2$ the inhomogeneous term in (\ref{1a4}) is nonlinear in $ \tE_w$.  In such cases the inhomogeneous term in the equation for $\mC_{w;s;\mp}$  contains quadratic expressions in the Eisenstein series, as in the second equation of (\ref{1a5}), and the solutions are no longer Eisenstein series.  Whether the solutions to such equations can be expressed in terms of Poincar\'e series by applying the methods used in \cite{Green:2014yxa} remains to be established.  

\sm

$\bullet$ For $\rho \geq 4$, the situation is much less well-understood, and we shall not attempt to present a systematic treatment here. Instead, we shall limit attention to those modular functions that enter into the evaluation of higher derivative interactions resulting from the four-graviton amplitude up to order $D^{10} \cR^4$. In this case, a general diagram has 4 vertices and 6 edges labelled by their contribution to the weight $\ell_1 , \dots, \ell_6$. We shall use the notation introduced in \cite{Green:2008bf} in which such a diagram is denoted by the symbol $D_{\ell_1,\ell_2,\ell_3,\ell_4;\ell_5,\ell_6}(\tau)$
with the understanding that we drop any entry corresponding to $\ell_r=0$.   In \cite{Green:2008bf} various properties of these $D$ functions were studied, including their asymptotic behavior near the cusp.  Many, but not all, of the $D$-functions are of the form (\ref{1a1})\footnote{However,  contrary to the assertion in  \cite{Green:2008bf}, in order to evaluate the integral of these functions over the complex structure we need to determine the full non-perturbative structure of the modular functions.}. 

\sm

To order $D^8 \cR^4$, which corresponds to weight 4, the non-trivial contributions are from $D_{2,1,1}=C_{2,1,1}$ and $D_4=C_{1,1,1,1}$. The first has already been discussed. The second will be conjectured to satisfy an inhomogeneous Laplace-eigenvalue equation with eigenvalue 2, 
\bea
\label{1a7}
(\Delta -2) \left ( 5C_{1,1,1,1} - 15  \tE_2^2 -  18  \tE_4 \right ) = - 120 \,  \tE_2^2
\eea
Evidence for the conjecture is obtained by verifying the equation in the asymptotic expansion near the cusp where $\tau_2 \to \infty$, and we have done so for all pure power behaved terms, as well as for the leading exponential corrections
of order $e^{-2 \pi \tau_2}$. It would be valuable to obtain a full proof of the conjecture, but this is beyond the scope of this paper.
Equation (\ref{1a7}) by itself does allow for a simple integration, and therefore does not imply any simple algebraic relation of the type that we had derived for eigenvalue 0. However, the inhomogeneous $ \tE_2^2$ term that occurs in (\ref{1a7}) is precisely the one that also occurs in the middle equation in (\ref{1a5}), and upon elimination of this term, we obtain a Laplace-eigenvalue equation of the form, 
\bea
\label{1a8}
(\Delta -2) \left ( C_{1,1,1,1} -24 C_{2,1,1} - 3  \tE_2^2 + 18  \tE_4 \right ) = 0
\eea
The kernel of $(\Delta -2)$ on modular functions with polynomial growth is spanned by $ \tE_2$. Verification of the power asymptotic behavior near the cusp shows that such addition is absent, and we obtain, 
\bea
\label{1a9}
C_{1,1,1,1} = 24 \, C_{2,1,1} + 3 \, \tE_2^2 - 18 \, \tE_4
\eea
In appendix \ref{secD} we shall provide further evidence in support of this conjectured relation from direct numerical evaluation of the multiple sums.
From this relationship we conclude  that there is  only a single independent nontrivial weight 4 function in addition to products of Eisenstein series and odd zeta values.  

\sm

To order $D^{10} \cR^4$, which corresponds to weight 5, the  contributions which do not completely reduce to linear combinations of products of Eisenstein series and $\zeta$-values correspond to $D_{2,1,1,1}=C_{3,1,1}$, $D_{3,1,1}=C_{2,1,1,1}$, $D_5=C_{1,1,1,1,1}$, and $D_{2,2,1}$. Thus, all contributions at weight 5 except for $D_{2,2,1}$ are expressible as $C$-functions. By the same type of arguments as were used in the case of weight 4, we will be led to conjecture the following algebraic relations at weight 5,
\bea
\label{1a10}
40 C_{2,1,1,1} & = &  300 \, C_{3,1,1} +120 \,  \tE_2 \tE_3 - 276 \, \tE_5 + 7 \, \zeta (5) 
\no \\
C_{1,1,1,1,1} & = &   60 C_{3,1,1} +10 \, \tE_2 \tE_3 - 48 \, \tE_5+ +10 \, \zeta (3) \tE_2+ 16 \, \zeta (5) 
\no \\
10 D_{2,2,1} & = &   20 \, C_{3,1,1} - 4 \,  \tE_5 + 3\,  \zeta (5) 
\eea
Remarkably, these three relationships show that the four non-trivial weight 5 modular functions may be expressed as a sum of  a single non-trivial function in the $C_{a,b,c}$ series and products of  Eisenstein series $ \tE_s$  and $\zeta$-values of odd argument. 

\sm

This raises the interesting mathematical question of how many independent $C$- and $D$-functions there are at any given weight, and what the algebraic and differential structure is of the ring of such modular functions.  This question is the modular generalization of the problem of identifying a basis of independent generators of the polynomial ring of multi-zeta values that arise as coefficients in tree-level amplitudes.  However,    in order to resolve this issue it will be necessary to study the expansion of the $N$-graviton genus-one amplitude for all $N\ge 4$ (the fact that  extra diagrams appear at higher weight is clear from the $N=5$ and $N=6$ cases discussed in  \cite{Green:2013bza}). This will be discussed further in the final section. 

\sm

As a consequence of this analysis, we will show that the integrands on moduli space for the low energy expansion of the four-graviton amplitude in the Type II string to order $(\alpha ')^w $ with $w \leq 5$, may all be recast in the following general form,
\bea
\Delta (f_w) + \mF_w( E_{s'}, \zeta (s')) 
\eea
where $\Delta$ is the Laplace-Beltrami operator in $\tau$, $f_w$ is a modular function of weight $w$, and $\mF_w$ is a polynomial of degree 2 and weight $w$ in the integer weight Eisenstein series $E_{s'}$ and zeta-values $\zeta (s')$. This form of the integrand allows for an analytical evaluation of the integration over $\tau$, and thus of the contributions to the effective higher derivative interactions.

\subsection{Organization}

The remainder of this paper is organized as follows.
The structure and general features of the low energy expansion of the genus-one amplitude are reviewed in section  \ref{sec2}. The coefficients of the higher derivative terms  are described in terms of world-sheet Feynman diagrams which reduce to multiple sums over the discrete word-sheet momenta for an arbitrary  fixed value of~$\tau$.  In section \ref{sec3}, we present a detailed discussion and  derivation of the differential and algebraic relations obeyed by the modular functions $C_{a,b,c}$ of arbitrary weight, and we exhibit these relations explicitly  up to weight 11. In section \ref{sec9}, we introduce the generating function $\cW$ for the modular functions $C_{a,b,c}$, derive a differential relation for the Laplace operator on $\cW$, express this equation in terms of the $SO(2,1)$-invariant Laplacian, and use these ingredients to prove the structure of the inhomogeneous Laplace-eigenvalue equations conjectured in section \ref{sec3}. In sections \ref{sec4} and \ref{sec5} we discuss the conjectured relations for contributions at weights 4 and 5 respectively, and we present evidence for these conjectures based on the asymptotic expansions near the cusp. In section \ref{sec7} we review  the  one-loop results up to weight $w=3$ (order $D^6 \cR^4$) that were obtained in \cite{Green:2005ba,Green:2008bf}, and then make use of the results for weights  4 and 5 to derive the full higher derivative interactions up to order $D^{10} \cR^4$ by performing the $\tau$-integral of the corresponding modular functions.  In section \ref{sec8} we end  with a  discussion and comments on higher point functions, higher genus expansions, and toroidal compactification.   In appendices \ref{secB} and \ref{secC} we review and further develop the calculations of asymptotic expansions of non-holomorphic modular functions  near the cusp.   Finally, in appendix \ref{secD} we have included a description of a preliminary attempt  at a numerical check of our conjectured relations between the modular functions.

\section{Structure of the  low energy expansion}
\setcounter{equation}{0}
\label{sec2}

We will be particularly interested in the low energy expansion of terms in the perturbative expansion of the four-graviton amplitude in either of the ten-dimensional  Type II theories in flat space-time.  In this section, we shall  review and clarify both expansions, and give their formulation in terms of world-sheet Feynman diagrams. 

\subsection{Perturbative expansion of the four-graviton amplitude}

In either Type II theory, the perturbative expansion is in terms of the string coupling $g_s$. In Type IIA $g_s$ is the only (real) modulus, while in Type IIB $g_s$  is the inverse of the imaginary part of the single complex modulus that is sometimes known as the ``axion-dilaton''.   Type IIB  is invariant under the action of the non-perturbative duality group $SL(2,\ZZ)$ on this complex modulus.   In either Type II theory, the perturbative expansion is given as follows, 
\bea
\label{pertexpand}
\cA^{(4)}(\epsilon_i,k_i,g_s ) =\sum_{h=0}^\infty g_s^{2h-2}\,\cA_h ^{(4)} (\e_i,k_i) + \cO(e^{-1/g_s}) 
\eea 
The exponentially suppressed terms are generated by non-perturbative effects. The $h$-loop amplitude $\cA_h ^{(4)} (\e_i,k_i)$ results from the conformal field theory of matter and ghosts on genus-$h$ world-sheet (super) Riemann surfaces, and depends on the space-time momenta $k_i $ and  polarization tensors $\eps_i$ which characterize the four external graviton states labelled by $i=1,\dots,4$. Under the Lorentz group $SO(1,9)$, the $k_i$ transform in the vector representation, while the $\eps_i$ is a rank two tensor. Total momentum is conserved, the momenta $k_i$ are massless, and the polarization tensors are transverse.

\sm

As a consequence of space-time supersymmetry of Type II superstrings, the dependence of the four-graviton amplitude $\cA^{(4)}(\epsilon_i,k_i,g_s )$ on the polarization tensors $\eps _i$ is through a Lorentz scalar multiplicative factor, schematically denoted $\cR^4$. Here, $\cR_{\mu \rho \nu \sigma} \sim k_\mu k_\nu \e_{\rho \sigma}$ stands for the linearized Riemann tensor, and $\cR^4$ is the unique scalar contraction of four powers of the Riemann tensor compatible with maximal supersymmetry.\footnote{The explicit form of $\cR^4$ and the expressions for the tree-level and one-loop amplitudes to be given below may be found, for example, in  \cite{GSWII}.} 
The remaining dependence of the amplitude is through the dimensionless and Lorentz invariant  variables $s_{ij}=-\alpha ' k_i\cdot k_j/2$ for $1\leq i < j \leq 4$. In view of the zero mass condition for the graviton and total momentum conservation, the variables $s_{ij}$ obey a number of relations which are conventionally parametrized as follows, $s=s_{12}=s_{34}$, $t=s_{13}=s_{24}$, and $u=s_{14}=s_{23}$ subject to the condition $s+t+u=0$.

\sm

In terms of $\cR^4$, the tree-level amplitude is given by, 
\bea
\label{2a3}
\cA_0 ^{(4)} (\e_i,k_i)
 =  { \kappa _{10}^2  \, \cR^4 \over stu} 
{ \Gamma (1 - s )\Gamma ( 1-  t )\Gamma (1  -  u)
\over \Gamma (1+ s)\Gamma (1+t)\Gamma (1+ u)} 
\eea
where  $\kappa_{10}^2$ is Newton's constant in $10$ dimensions. 

\sm

The genus-one amplitude reduces to an integral  over the moduli space
$\cM_1$ of genus-one Riemann surfaces of a Lorentz scalar function
$\cB_1$ of the Lorentz-invariant variables $s,t,u$, 
\bea
\label{2a4}
\cA_1 ^{(4)} (\e_i, k_i)
= 
2\pi \,\kappa _{10}^2 \, \cR^4 
\int _{\M_1} d\mu_1\,  \cB_1 (s,t,u| \tau)
\eea
As usual, we identify $\cM_1$  with a fundamental domain for the
modular group $SL(2,\ZZ)$  in the complex upper half plane,
parametrized by the modulus $\tau=\tau_1+ i \tau_2$ with
$\tau_1,\tau_2 \in \RR$, $0<\tau_2$, $|\tau_1| \leq \half$, $1 \leq
|\tau|$, and $d\mu_1 = d\tau_1 \wedge d\tau_2/\tau_2^2$. With this normalization the   
volume of the fundamental domain is  $\int_{\M_1} d\mu_1={\pi\over3}$.  Although the 
variables $s,t,u$ are not independent, we shall keep all three as arguments in $\cB_1$ 
since Bose symmetry requires $\cB_1$ to be a symmetric function of $s,t,u$. 
The partial amplitude $\cB_1$ is given by the following integral representation,\footnote{This expression may be generalized to the case of a toroidal compactification of 10-dimensional space-time of the form $\RR^{10-d} \times T^d$, where $T^d$ is a $d$-dimensional torus specified by a point in the symmetric space $SO(d,d)/(SO(d)\otimes SO(d))$.} 
\bea
\label{2a5}
\cB_1 (s,t,u | \tau)
 = 
\frac{1}{\tau_2^4}\int _{\Sigma^4} \prod _{i=1}^4 d^2z_i 
\exp \left \{  \sum _{1 \leq i<j \leq 4} s_{ij}  \, G(z_i-z_j | \tau ) \right \}
\eea
Here, $\Sigma$ denotes the torus with modulus $\tau$;  the graviton
vertex insertion points $z_i$ are integrated over $\Sigma$ with volume form 
$d^2z = {i \over 2} dz \wedge d\bar z$ so that the area of $\Sigma$ equals
$\int_\Sigma d^2z=\tau_2$; and $G(z|\tau)$ is the scalar Green function on $\Sigma$ given by,
\bea
\label{2a6}
G(z | \tau ) = - \ln \left | { \tet _1 (z|\tau) \over \tet _1 ' (0|\tau)} \right |^2 
- { \pi \over 2 \tau_2} (z-\bar z)^2
\eea
in terms of the Jacobi $\theta$-function $\theta_1(z|\tau)$. The integral representation for ${\cal B}_1$ in (\ref{2a5}) at fixed $\tau$ is absolutely convergent only for limited ranges of $s,t,u$, but may be analytically continued throughout $\CC^2$, as was shown in  \cite{D'Hoker:1994yr}.    The function ${\cal B}_1$ can thereby be shown to possess poles at negative integer values of  $s$, $t$ and $u$  that arise from singularities in the product of the vertex operators for the gravitons. The amplitude ${\cal A}_1^{(4)}$ that results from the integral over $\tau$ in (\ref{2a4}) not only has poles in $s$, $t$ and $u$ corresponding to massive closed string intermediate states, but it also has the branch cuts expected by unitarity that arise from the propagation of two-particle (massless and massive) intermediate states.  
 
\sm

There are predictions based on supersymmetry and duality for the exact moduli-dependent coefficients of the three  lowest order terms in the low energy expansion of the four-graviton amplitude, namely the $\cR^4$, $D^4\cR^4$, and $D^6\cR^4$ interactions.  The explicit expansion of the genus-one amplitude up to this order  \cite{Green:1999pv,Green:2008uj} confirms the predicted values of the one-loop contributions to each of these coefficients.  

\sm

The genus-two amplitude $\cA_2 ^{(4)} (\e_i,k_i)$ is also known in terms of an integral representation over the moduli space of genus-two Riemann surfaces, along with an integration over the four graviton vertex operator insertion points. The  coefficient of the first term in its low energy expansion was evaluated in \cite{D'Hoker:2005ht}  and shown to match the two loop prediction based on space-time supersymmetry and duality for the  coefficients of the $D^4\cR^4$ interaction  (an analysis using the pure spinor formalism was subsequently given in \cite{Gomez:2010ad}).  Likewise, the coefficient of the $D^6\cR^4$ interaction was  evaluated in \cite{D'Hoker:2014gfa} and agrees with its predicted value. There is a candidate expression for the low energy limit of the genus-three four-graviton amplitude \cite{Gomez:2013sla}, which produces a value for the three-loop coefficient of the $D^6 \, \cR^4$ interaction with an apparent discrepancy of a factor of 3 with the predicted value.

\subsection{Low energy expansion of the four-graviton amplitude}

In brief, the low energy expansion of a string amplitude, such as the four-graviton amplitude of $\cA^{(4)}(\eps_i, k_i,g_s)$, is an expansion in the regime where $| s_{ij} | \ll 1$ for all $i,j$, namely when all momenta are small compared to the string scale, set by  $\alpha '$.

\subsubsection{Tree-level}

For tree-level, the expansion may be obtained directly from the explicit formula for the amplitude of (\ref{2a3}), and the expansion of the $\Gamma$-function, and is given by,
\bea
\label{2b1}
\cA_0 ^{(4)} (\e_i,k_i) = \kappa _{10}^2 \, \cR^4 \frac{1}{stu}\, 
\exp \left \{ \sum_{n=1}^\infty \frac{2 \zeta(2n+1)}{2n+1}\, \left ( s^{2n+1} + t^{2n+1} + u^{2n+1} \right ) \right \} 
\eea
The low energy expansion of the four-graviton amplitude is seen to be
quite special since all expansion coefficients will be polynomials  in
Riemann $\zeta$-values with rational
coefficients.\footnote{The above all-orders low energy expansion of the four-graviton amplitude has been generalized to amplitudes with $N$ scattering strings for arbitrary values of $N$ and the resulting coefficients for  $N>4$ are generally rational multiples of multi-zeta values of odd weight~\cite{Stieberger:2013wea,Schlotterer:2012ny}. For weights lower than or equal to 10, all multi-zeta values may be reduced to polynomials in ordinary $\zeta$-values, but above weight 11 irreducible multi-zeta values also appear. In the open-string case multi-zeta values enter for weights $\ge 8$. This raises the interesting question of whether the coefficients in the expansion of the one-loop $N=4$-string amplitude are again rational multiples of monomials of Riemann zeta values or whether they involve multi-zeta values, or more exotic objects.}  Note that only $\zeta$-values with odd argument occur  in view of the fact that the logarithm of the ratio of $\Gamma$-functions in (\ref{2a4}) is odd under reversing the signs of $s,t,u$.
These amplitudes produce higher derivative interactions  that are symbolically denoted by $D^{2w} \cR^4$, where  $D^{2w}$ stands for the scalar contraction of $2w$  covariant derivatives produced by the corresponding symmetric and homogeneous polynomial in $s,t,u$ of degree $w$.  A slight rearrangement of the expansion is obtained by expressing the symmetric polynomials in $s,t,u$,  in terms of its two generators $\sigma _2$ and $\sigma _3$, where,
\bea
\label{2b2}
\sigma _n = s^n + t^n + u^n
\eea
subject to the constraint $\sigma _1 = s+t+u=0$.  The contribution of weight $w$ then arises as a linear combination of monomials  $\sigma _2 ^p \sigma _3^q$ whose exponents $p,q$ are related to the weight by,
\bea
\label{2b3}
w=2p+3q
\eea
 The first few terms in this rearranged expansion are as follows,
{\small \bea
\label{2b4}
\cA_0 ^{(4)} (\e_i,k_i) = \kappa _{10}^2 \, \cR^4\left( \frac{3}{\sigma_3} + 2\zeta(3) + \zeta(5) \, \sigma_2 + \frac{2\zeta(3)^2}{3} \, \sigma_3 + \frac{\zeta(7)}{2}\, \sigma_2^2 + { 2 \over 3} \zeta(5) \zeta (3) \sigma _2 \sigma _3 + \dots \right)\quad
\eea}
Because $\sigma _{2n+1}$ is divisible by $\sigma _3=stu$ for all $n \geq 1$, only the leading term in the low energy expansion exhibits a pole for small $s,t,u$. This pole is associated with the exchange of massless states, and may be consistently isolated. The remaining terms contribute respectively to the higher derivative interactions of the form, $\cR^4$, $D^4 \cR^4$, $D^6 \cR^4$, $D^8 \cR^4$, and $D^{10} \cR^4$.

\subsubsection{One-loop}

For the genus-one amplitude $\cA_1 ^{(4)} (\e_i, k_i)$ of (\ref{2a4}), no poles occur at  $s$, $t$, or $u=0$ thanks to 
space-time supersymmetry, but the amplitude does have branch cuts starting at $0$ due to the fact that massless strings are being exchanged in closed loops. These non-analytic  contributions arise from the integration over the genus-one moduli space
in (\ref{2a4}). They may be isolated and subtracted to leave analytical terms which provide the  higher derivative interactions. 
Thus, a safe point of departure to construct the low energy expansion at one-loop  is the partial amplitude $\cB_1$ at fixed modulus $\tau$ of (\ref{2a5}).

\sm

The structure of the singularities in $s,t,u$ of the genus-one partial amplitude $\cB_1$ at fixed~$\tau$ is governed by the operator product expansion of the four graviton vertex operators. A singularity is produced  when two insertion points coincide, namely when $z_i-z_j \to 0$ for $i \not= j$. The  singularities result from local behavior on the surface only, and thus are  the same for genus-one as they were for tree-level, and consist of simple poles in $s,t,u$ at negative integers. Therefore, the integral representation 
for $\cB_1$ in (\ref{2a5}) admits an expansion in powers of $s,t,u$ which is absolutely convergent in a region of unit radius in $s,t,u$. The low energy expansion is then simply given by the Taylor expansion of the exponential in (\ref{2a5}), 
\bea
\label{2b5}
\cB_1 (s,t,u | \tau)
& = & \sum _{w=0} ^{\infty} { 1 \over w!} \, 
\frac{1}{\tau_2^4} \int _{\Sigma^4} \prod _{i=1}^4 d^2z_i \, \cL (s,t,u ; z_i | \tau ) ^w
\no \\
\cL (s,t,u ; z_i | \tau ) & = & 
\sum _{1 \leq i<j \leq 4} s_{ij}  \, G(z_i-z_j | \tau ) 
\eea
The term of order $w$ contributes to the effective interaction  schematically represented by $D^{2w} \cR^4$. For $w\leq 5$, the weight $w$ labels the interaction $D^{2w} \cR^4$ uniquely, but for weights $w \geq 6$, degeneracies appear as different Lorentz contractions of the momenta exist, such as  $\sigma _2 ^3 $ and $\sigma _3^2$ for $w=6$. These degeneracies are systematically resolved by using a more refined notation $\sigma _2^p \sigma _3^3 \cR^4$ for the interaction instead of $D^{2w} \cR^4$ with $w=2p+3q$. The corresponding expansion in powers of $\sigma _2$ and $\sigma _3$ is given as follows, 
\bea
\label{2b6}
\cB_1(s,t,u |\tau)=\sum_{p,q=0}^\infty j^{(p,q)}(\tau) \, \sigma_2^p\, \sigma_3^q 
\eea
 The coefficient functions  $j^{(p,q)}(\tau)$ are modular invariant and have weight $w=2p+3q$, following the same notation used for tree-level in (\ref{2b3}). The basis functions $j^{(p,q)}(\tau)$ in this expansion follow the conventions in \cite{Green:2005ba,Green:2008uj} and are suited to the four-graviton amplitude\footnote{The more general basis used in  \cite{Green:2013bza} generalizes to $N$-graviton amplitudes with $N>4$ and facilitates comparison of tree-level and loop amplitudes, and hence is adapted to comparison of the genus-one coefficients with the multi-zeta values that arise in the tree-level expansion.}.

\sm

The coefficient of the interaction  $\sigma_2^p\, \sigma_3^q\, \cR^4$
in the amplitude   (\ref{2a4}) is obtained  by integrating the modular
function  $j^{(p,q)}(\tau)$ over $\cM_1$.
 The resulting expansion of the amplitude therefore has the form,
  \bea
\cA_1 ^{(4) } (\e_i, k_i) &=&2\pi\kappa _{10}^2 \, \cR^4 \,\Bigl(\Xi^{(0,0)} + 
 \Xi^{(1,0)} \, \sigma_2 +\Xi^{(0,1)}  \, \sigma_3
+\Xi^{(2,0)} \, \sigma_2^2
+\Xi^{(1,1)} \, \sigma_2\, \sigma_3
\nn \\ && \hskip 0.7in 
+ \Xi^{(3,0)} \,\sigma_2^3 
+\Xi^{(0,2)} \, \sigma_3^2 +  \cdots +\  {\rm non-analytic\ terms}  \, \Bigr )  \quad
 \label{fourgravcons}
\eea
The analytic part of this expression resembles that of the tree-level
amplitude, with the monomials in zeta values in  (\ref{2a4}) replaced
by the  coefficients $\Xi^{(p,q)}$.

 As indicated earlier, such integrations will in general require regularization and subtraction of the non-analytic contributions in $s,t,u$ which arise from loops of massless strings.  The structure and evaluation of such integrals will be discussed in section~\ref{sec7}.  For the remainder of this section we will  be concerned with the structure of the modular coefficients $j^{(p,q)}(\tau)$, and their Feynman diagram representation. Our aim is to determine various  properties of the functions $j^{(p,q)}(\tau)$.

\subsection{World-sheet Feynman diagrams} 
\label{sec23}

In this subsection we will review the tools needed to study the low energy expansion of the  genus-one amplitude defined by (\ref{2a4}) and (\ref{2a5}).  Each $j^{(p,q)}(\tau)$ function  is given by   integration of the sum of products of $w=2p+3q$ Green functions over the toroidal world-sheet with complex structure $\tau$.   This defines a sum of  Feynman diagrams of weight $w$, each of which is a modular function of $\tau$.   We will now describe the systematics of this calculation, which covers and extends material in \cite{Green:1999pv, Green:2008uj,Green:2013bza}, and is  a prerequisite for the evaluation of higher order interactions in subsequent sections.   

\sm

The modular functions $j^{(p,q}) (\tau)$  in the expansion of (\ref{2b6}) are given by (\ref{2b5}) in terms of integrals over four copies of the torus $\Sigma$ of products of $w$ scalar Green functions. To evaluate these integrals, it will be convenient to use an alternative representation of the Green function $G$ which renders its double periodicity fully transparent. We parametrize a point $z$ on the torus $\Sigma$ of modulus $\tau$ by real coordinates $\alpha, \beta  \in \RR/\ZZ$, 
\bea
\label{2c1}
 z =\alpha  +\tau \beta 
\eea 
The Green function $G$ is then given by the Fourier series over integers $m,n \in \ZZ$,
\bea
\label{2c2}
G(z |\tau ) = G_0 (\tau) +  \cG (z|\tau) 
\eea
with
\bea
\label{2c3}
\cG (z|\tau) =  \sum_{(m,n) \neq(0,0)}  \cG (m,n | \tau ) \, e^{2\pi  i (m \alpha   - n \beta)}
\eea
which may also be viewed as  the definition of a Kronecker-Eisenstein series.
The Fourier modes $\cG(m,n| \tau )$ are given by,
\bea
\cG (m,n |\tau ) = {\tau_2\over \pi  |m\tau+n|^2} 
\eea
for $(m,n)\not= (0,0)$ and we shall set $\cG(0,0 |\tau)=0$. The integers $m$ and $n$ may be viewed as the components of the world-sheet momentum of the scalar field, which will be denoted $\bp=m\tau + n$, and we shall sometimes use the alternative abbreviated notation $\cG (\bp) = \cG(m,n|\tau )$. The zero mode contribution  $G_0(\tau) = 2 \ln\left(2\pi \left|\eta(\tau) \right|^2\right)$ will drop out of (\ref{2b5}) and any perturbative string calculation  in view of the momentum conservation relation $s+t+u=0$.  This has the immediate consequence that the  term linear in  $G$ will vanish.  Therefore, the world-sheet momentum space Green function $\cG$  contains only non-zero modes.

\sm

Expanding $ {\cal L}(s,t,u; z_i | \tau)^w$ of (\ref{2b5}) in powers of the Green functions  gives  a monomial of weight $w$ in which different powers of $G(z_i-z_j |\tau )$ link the four vertices at $z_1$, $z_2$, $z_3$, $z_4$.  Substituting  the Fourier series (\ref{2c2}) for each Green function and performing the $z_i$ integrations leads to a sum of  world-sheet Feynman diagrams in which each propagator joining vertices $i$ and $j$  denotes a factor of $\cG(\bp_{ij})$ with a sum over all the (discrete) world-sheet momenta $\bp_{ij}$ subject to momentum  conservation at each vertex.

\subsection{Nomenclature for world-sheet diagrams}
\label{sec:diarep}

In the following it will prove convenient to symbolize these Feynman diagrams in different fashions, according to the context in which they are discussed.  The problem is that there is no simple way of labeling the function corresponding to a  general (and complicated) diagram -- the simplest representation is in terms of the diagram itself.  

\sm

A systematic way of labeling an arbitrary diagram, at any order in the low energy expansion, and with any number $N$ of vertices and a total of $w$ propagators is as follows. Each vertex is labelled by an integer $i,j=1,2,\dots,N$, and the number   of propagators that joins the pair of vertices\footnote{If $\ell_{ij}$ were restricted to take only the values  $0$ or $1$ it would be the incidence matrix for the graph - however, in general we have $\ell_{ij} \in \NN$.} $(i,j)$ by $\ell_{ij}$.   
Each propagator joining a given vertex pair $(i,j)$  is assigned a label $(i,j, \lambda_{ij})$ where $1\le \lambda_{ij} \le \ell_{ij}$, and carries world-sheet momentum ${\bf p}_{ij}^{\lambda_{ij}} = (m_{ij}^{\lambda_{ij}} \tau + n^{\lambda_{ij}}_{ij})$.   In that case the modular function associated with the diagram is given by the product of propagators  summed over word-sheet momenta,
\be
\label{precise}
D_{\ell_{12}, \ell_{13}, \dots, \ell_{ij}, \dots , \ell_{N-1\, N} }
=   \sum_{ \{ {\bf p}^{\lambda_{ij}}_{ij}  \} }\, \prod_{1\le i < j \le N} \,\prod_{1\le \lambda _{ij} \le \ell_{ij}}
\cG \left ( {\bf p}^{\lambda_{ij}}_{ij} \right ) \, 
\prod_{1\le i \le N} \delta \left ( \sum_{k \ne i} \sum_{\lambda_{ik}=1}^{\ell_{ik}}{\bf p}^{\lambda_{ik}}_{ik} \right ) 
\ee
where the product of $\delta$'s enforces conservation of momentum at each vertex.

\sm

This notation has redundancies since the modular function associated with the diagram is invariant under an arbitrary relabeling of the vertices.  It is also  needlessly complicated for many purposes, so,  in the following we will label the functions associated with diagrams in a manner that is adapted for the particular problem at hand, bearing in mind the general expression in (\ref{precise}).

\subsection{Diagrammatic representation of modular coefficient functions}
\label{25}
 
 In  this section,  we will  be interested in 
\begin{itemize}
\itemsep = 0.0in
\item those diagrams that contribute to the four-graviton amplitude and,
\item diagrams belonging to a particular infinite subset of all the diagrams that produce higher-point functions and which we can determine for any weight. 
\end{itemize}
 
 \subsubsection{Contributions to the four-graviton amplitude}

In this case a general diagram has $\ell_{ij}$ propagators joining any pair of vertices labelled $i$ and $j$, where $i,j=1,2,3,4$. Therefore there are six $(i,j)$ pairs so a general diagram has $\ell_1, \dots,\, \ell_6$ propagators.  The notation we will adopt is summarized by the following diagram:
 \vskip3pt
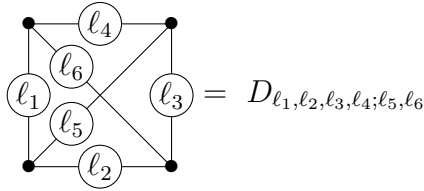
\begin{center}
 \tikzpicture[scale=1.9]
\begin{scope}[xshift=6cm]
\draw (0,0) node{$\bullet$} ;
\draw (1,0) node{$\bullet$} ;
\draw (0,1) node{$\bullet$} ;
\draw (1,1) node{$\bullet$} ;
\draw (0,0) -- (1,0) ;
\draw (0,0) -- (0,1);
\draw (1,1) -- (1,0) ;
\draw (1,1) -- (0,1);
\draw (0,0) -- (1,1) ;
\draw (0,1) -- (1,0) ;
\draw (0,0.5) [fill=white] circle(0.15cm) ;
\draw (0,0.5)  node{$\ell_1$};
\draw (0.5,0) [fill=white] circle(0.15cm) ;
\draw (0.5,0)  node{$\ell_2$};
\draw (1,0.5) [fill=white] circle(0.15cm) ;
\draw (1,0.5)  node{$\ell_3$};
\draw (0.5,1) [fill=white] circle(0.15cm) ;
\draw (0.5,1)  node{$\ell_4$};
\draw (0.3,0.3) [fill=white] circle(0.15cm) ;
\draw (0.3,0.3)  node{$\ell_5$};
\draw (0.3,0.7) [fill=white] circle(0.15cm) ;
\draw (0.3,0.7)  node{$\ell_6$};
\draw(2,0.5) node{$= \ D_{\ell_1,\ell_2,\ell_3,\ell_4;\ell_5,\ell_6}$};
\end{scope}
\endtikzpicture
 \end{center}
 \vskip3pt
where the label $\ell$ on the link  
\begin{tikzpicture} [scale=1.3]
\scope[xshift=0cm,yshift=0cm]
\draw (0,-0.0) node{$\bullet$}   ;
\draw (1,0.0) node{$\bullet$}  ;
\draw (0,0) -- (1,0) ;
\draw (0.5,0) [fill=white] circle(0.15cm) ;
\draw (0.5,0.0) node{$\ell$};
\endscope
\end{tikzpicture}
indicates the number of propagators joining the corresponding pair of vertices. The shape of this diagram is a tetrahedron that is symmetric in all edges and vertices and its weight  is given by the number of propagators, so that 
\be
\label{weightdiag}
w= \sum_{r =1}^6 \ell_r
 \ee
 where $0\le \ell_r $.     When  $\ell_r =0$ for particular values of $r$,  the diagram degenerates to a simpler diagram. In such cases, we shall omit the index that vanishes.\footnote{For example $D_{\ell_1,\ell_2,\ell_3,\ell_4;0,0}= D_{\ell_1,\ell_2,\ell_3,\ell_4}$, while $ D_{\ell,0,0,0,;0,0} = D_{\ell}$, etc.}

Certain classes of  diagrams automatically give vanishing contributions. Thus, any  diagram that has a vertex $z_i$ connected to a single propagator is zero due to the fact that  the integral of  ${\cal G}(z|\tau)$  over $z$ vanishes (so that there is no zero momentum mode in ${\cal G}(m,n|\tau)$).  Likewise, one-particle reducible diagrams vanish since any such diagram  contains a zero-momentum propagator.  Furthermore, diagrams involving the self-contraction of a propagator do not arise since such diagrams have a coefficient proportional to $k_r^2=0$, where $k_r$ is the momentum of an external graviton.
    
 The following is a selection of the non-vanishing diagrams  that contribute to the expansion of the four-graviton amplitude.

 \medskip
{\sl (1) Diagrams involving a pair of vertices.}
 \smallskip
 
When all but one of the $\ell_r=0$ the results is  diagram with $\ell$  propagators joining a pair of points, represented by
\vskip3pt 
\begin{center}
\tikzpicture[scale=1.7]
\scope[xshift=-5cm,yshift=-0.4cm]
\draw (0,0.5) node{$\bullet$}   ;
\draw (1,0.5) node{$\bullet$} ;
\draw (0,0.5) -- (1,0.5) ;
\draw (0.5,0.5) [fill=white] circle(0.15cm) ;
\draw (0.5,0.5) node{$\ell$};
\draw(2,0.5) node{$= \  D_{\ell} $};
\endscope
\endtikzpicture
\end{center}
\vskip3pt

 \medskip
{\sl (2) Diagrams involving three vertices.}
 \smallskip
\vskip3pt
\begin{center}
 \tikzpicture[scale=1.9]
\begin{scope}[xshift=4.5cm]
\draw (0,0) node{$\bullet$} ;
\draw (0,1) node{$\bullet$} ;
\draw (0.7,0.5) node{$\bullet$} ;
\draw (0,0) -- (0,1) ;
\draw (0,0) -- (0.7,0.5);
\draw (0,1) -- (0.7,0.5);
\draw (0.35,0.25) [fill=white] circle(0.15cm) ;
\draw (0.35,0.25)  node{$\ell_1$};
\draw (0.35,0.75) [fill=white] circle(0.15cm) ;
\draw (0.35,0.75)  node{$\ell_2$};
\draw (0,0.5) [fill=white] circle(0.15cm) ;
\draw (0,0.5)  node{$\ell_3$};
\draw(1.5,0.5) node{$=\ D_{\ell_1,\ell_2,\ell_3}$};
\end{scope}

\begin{scope}[xshift=7.5cm]
\draw (0,0) node{$\bullet$} ;
\draw (0,1) node{$\bullet$} ;
\draw (0.7,0.5) node{$\bullet$} ;
\draw (0,0) -- (0.7,0.5);
\draw (0,1) -- (0.7,0.5);
\draw (0.35,0.25) [fill=white] circle(0.15cm) ;
\draw (0.35,0.25)  node{$\ell_1$};
\draw (0.35,0.75) [fill=white] circle(0.15cm) ;
\draw (0.35,0.75)  node{$\ell_2$};
\draw(1.5,0.5) node{$= \ D_{\ell_1}\,D_{\ell_2} $};
\end{scope}
\endtikzpicture
 \end{center}
 \vskip3pt
 
  \medskip
{\sl (3) A few of the diagrams involving four vertices.}
 \smallskip

\vskip3pt
\begin{center}
 \tikzpicture[scale=1.9]
\begin{scope}[xshift=3cm]
\draw (0,0) node{$\bullet$} ;
\draw (1,0) node{$\bullet$} ;
\draw (0,1) node{$\bullet$} ;
\draw (1,1) node{$\bullet$} ;
\draw (0,0) -- (1,0) ;
\draw (0,0) -- (0,1);
\draw (1,1) -- (1,0) ;
\draw (1,1) -- (0,1);
\draw (0,0.5) [fill=white] circle(0.15cm) ;
\draw (0,0.5)  node{$\ell_1$};
\draw (0.5,0) [fill=white] circle(0.15cm) ;
\draw (0.5,0)  node{$\ell_2$};
\draw (1,0.5) [fill=white] circle(0.15cm) ;
\draw (1,0.5)  node{$\ell_3$};
\draw (0.5,1) [fill=white] circle(0.15cm) ;
\draw (0.5,1)  node{$\ell_4$};
\draw(2,0.5) node{$= \ D_{\ell_1,\ell_2,\ell_3,\ell_4}$};
\end{scope}
\begin{scope}[xshift=6.5cm]
\draw (0,0) node{$\bullet$} ;
\draw (1,0) node{$\bullet$} ;
\draw (0,1) node{$\bullet$} ;
\draw (1,1) node{$\bullet$} ;
\draw (0,0) -- (1,0) ;
\draw (1,1) -- (0,1);
\draw (0.5,0) [fill=white] circle(0.15cm) ;
\draw (0.5,0)  node{$\ell_2$};
\draw (0.5,1) [fill=white] circle(0.15cm) ;
\draw (0.5,1)  node{$\ell_1$};
\draw(2,0.5) node{$= \ D_{\ell_1}\,D_{\ell_2}$};
\end{scope}
\endtikzpicture
 \end{center}
 \vskip3pt

\vskip3pt
\begin{center}
 \tikzpicture[scale=1.9]
\begin{scope}[xshift=3cm]
\draw (0,0) node{$\bullet$} ;
\draw (1,0) node{$\bullet$} ;
\draw (0,1) node{$\bullet$} ;
\draw (1,1) node{$\bullet$} ;
\draw (0,0) -- (1,1) ;
\draw (0,0) -- (1,0);
\draw (1,1) -- (1,0) ;
\draw (1,1) -- (0,1);
\draw (0.5,0.5) [fill=white] circle(0.15cm) ;
\draw (0.5,0.5)  node{$\ell_1$};
\draw (0.5,0) [fill=white] circle(0.15cm) ;
\draw (0.5,0)  node{$\ell_2$};
\draw (1,0.5) [fill=white] circle(0.15cm) ;
\draw (1,0.5)  node{$\ell_3$};
\draw (0.5,1) [fill=white] circle(0.15cm) ;
\draw (0.5,1)  node{$\ell_4$};
\draw(2,0.5) node{$= \ D_{\ell_1,\ell_2,\ell_3}\, D_{\ell_4}$};
\end{scope}
\begin{scope}[xshift=6.5cm]
\draw (0,0) node{$\bullet$} ;
\draw (1,0) node{$\bullet$} ;
\draw (0,1) node{$\bullet$} ;
\draw (1,1) node{$\bullet$} ;
\draw (0,0) -- (1,0) ;
\draw (0,0) -- (0,1);
\draw (1,1) -- (1,0) ;
\draw (1,1) -- (0,1);
\draw (0,0) -- (1,1) ;
\draw (0,0.5) [fill=white] circle(0.15cm) ;
\draw (0,0.5)  node{$\ell_1$};
\draw (0.5,0) [fill=white] circle(0.15cm) ;
\draw (0.5,0)  node{$\ell_2$};
\draw (1,0.5) [fill=white] circle(0.15cm) ;
\draw (1,0.5)  node{$\ell_3$};
\draw (0.5,1) [fill=white] circle(0.15cm) ;
\draw (0.5,1)  node{$\ell_4$};
\draw (0.5,0.5) [fill=white] circle(0.15cm) ;
\draw (0.5,0.5)  node{$\ell_5$};
\draw(2,0.5) node{$= \ D_{\ell_1,\ell_2,\ell_3,\ell_4;\ell_5}$};
\end{scope}
\endtikzpicture
 \end{center}
 \vskip3pt

In calculating the contribution of the diagrams of a given weight to a term of given transcendentality in the low energy expansion we need to sum the diagrams of that weight with combinatorial coefficients that were evaluated up to weight 6 in appendix C of \cite{Green:2008uj}, and can easily be evaluated to an  arbitrarily high weight.

\subsubsection{A subclass of diagrams contributing to amplitudes with higher $N$}

The diagrams we have discussed so far are ones that contribute to the four-graviton amplitude,  They are also a sub-class of the diagrams that contribute to $N$-graviton amplitudes with $N>4$.  However, when $N>4$ there are more diagrams for two distinct reasons:  (i) There are more vertices so  there are $N(N-1)/2$ ways of joining a pair of vertices; (ii) There are $2(N-4)$ momentum numerator factors in any diagram.  Many of these can be eliminated by integration by parts, thereby reducing the diagram to one with fewer (or no) momentum factors.  This systematics  was considered in detail in \cite{Green:2013bza}, for the $N=5$ and $N=6$ cases where the diagrams that contribute up to weight 6 were enumerated.  

\sm

The evaluation of the complete set of diagrams for the $N>4$  amplitudes will not be pursued here.  However, in  section~\ref{sec3}  we will consider an infinite subset of diagrams,  in which single propagators are joined to form chains of the form
\vskip3pt 
\begin{center}
\tikzpicture[scale=1.7]
\scope[xshift=-5cm,yshift=-0.4cm]
\draw (0,0.0) node{$\bullet$}   ;
\draw (1,0) node{$\bullet$} ;
\draw (0,0) -- (1,0) ;
\draw (0.35,-0.15) [fill=white] rectangle (0.65,0.15) ;
\draw (0.5,0) node{$a$};
\draw (1.5,0) node{$=$};
\draw (1.9,0) node{$\bullet$} ;
\draw (2.4,0) node{$\bullet$} ;
\draw (2.9,0) node{$\bullet$} ;
\draw (3.4,0) node{$\bullet$} ;
\draw (3.9,0) node{$\bullet$} ;
\draw (1.9,0) -- (2.4,0) ;
\draw (2.4,0) -- (2.9,0) ;
\draw[dashed, thick] (2.9,0) -- (3.4,0) ;
\draw (3.4,0) -- (3.9,0) ;
\draw(5.0,0) node{$= \  D_{1,1, \dots,1} (\tau)$};
\draw [decorate,decoration={brace,mirror,amplitude=8pt},xshift=0pt,yshift=0pt]
(1.9,-0.2) -- (3.9,-0.2)  ;
\draw (2.9,-0.5) node{$a$};
\draw [decorate,decoration={brace,mirror,amplitude=4pt},xshift=0pt,yshift=0pt]
(4.8,-0.15) -- (5.3,-0.15)  ;
\draw (5.05,-0.37) node{$a$};
\endscope
\endtikzpicture
\end{center}
\vskip3pt

A single such chain is zero because of the rule that a diagram with a vertex connected to a single propagator vanishes.  The first nontrivial case is that of two chains of length $a$ and $b$, which are represented by,

\vskip3pt 
\begin{center}
\tikzpicture[scale=1.7]
\scope[xshift=-5cm,yshift=-0.4cm]
\draw (0,0.0) node{$\bullet$}   ;
\draw (1,0) node{$\bullet$} ;
\draw (0.5,0.0) ellipse (0.5 and 0.3);
\draw (0.375,0.175) [fill=white] rectangle (0.625,0.425) ;
\draw (0.5,0.30) node{$a$};
\draw (0.375,-0.425) [fill=white] rectangle (0.625,-0.175) ;
\draw (0.5,-0.30) node{$b$};
\draw (1.5,0) node{$=$};
\draw (1.9,0) node{$\bullet$} ;
\draw (3.9,0) node{$\bullet$} ;
\draw (2.4,0.25) node{$\bullet$} ;
\draw (2.9,0.30) node{$\bullet$} ;
\draw (3.4,0.25) node{$\bullet$} ;
\draw (1.9,0) -- (2.4,0.25) ;
\draw (2.4,0.25) -- (2.9,0.3) ;
\draw[dashed, thick] (2.9,0.3) -- (3.4,0.25) ;
\draw (3.4,0.25) -- (3.9,0) ;
\draw (2.4,-0.25) node{$\bullet$} ;
\draw (2.9,-0.30) node{$\bullet$} ;
\draw (3.4,-0.25) node{$\bullet$} ;
\draw (1.9,0) -- (2.4,-0.25) ;
\draw (2.4,-0.25) -- (2.9,-0.3) ;
\draw[dashed, thick] (2.9,-0.3) -- (3.4,-0.25) ;
\draw (3.4,-0.25) -- (3.9,0) ;
\draw(5.0,0) node{$= \  C_{a,b}(\tau) $};
\endscope
\endtikzpicture
\end{center}
\vskip3pt
This is equivalent to a circular closed chain of length $a+b$.  In  an obvious generalization of the earlier notation this could be denoted $D_{\underbrace{_{1,1,\dots,1}}_{a+b}}(\tau)$, but this is clumsy and does not generalize to cases with more chains.  For that reason we will use a different notation, in which a graph with a number of  chains with lengths $a_1,\dots,a_\rho$ is denoted  $C_{a_1, \cdots, a_\rho}(\tau)$. From the above figure it easily follows that, 
\be 
C_{a,b}(\tau) = \sum_{(m,n)\ne (0,0)} \left(\frac{\tau_2}{\pi\, |m\tau + n|^2}\right)^{a+b}=\
 \tE_{a+b}(\tau)
\label{cabdef}
\ee
which  is the Eisenstein series of weight $a+b$.

\sm

Significant and very interesting challenges arise in determining the functions associated with three or more chains.  The case we will concentrate on in section~\ref{sec3}  will be graphs with three chains, which are denoted $C_{a,b,c}$ with a graphical representation,
\vskip3pt 
\begin{center}
\tikzpicture[scale=1.7]
\scope[xshift=-5cm,yshift=-0.4cm]
\draw (0,0.0) node{$\bullet$}   ;
\draw (1,0) node{$\bullet$} ;
\draw (0,0) -- (1,0) ;
\draw (0.5,0.0) ellipse (0.5 and 0.3);
\draw (0.375,0.175) [fill=white] rectangle (0.625,0.425) ;
\draw (0.5,0.30) node{$a$};
\draw (0.375,-0.125) [fill=white] rectangle (0.625,0.125) ;
\draw (0.5,0) node{$b$};
\draw (0.375,-0.425) [fill=white] rectangle (0.625,-0.175) ;
\draw (0.5,-0.30) node{$c$};
\draw (1.5,0) node{$=$};
\draw (1.9,0) node{$\bullet$} ;
\draw (2.4,0) node{$\bullet$} ;
\draw (2.9,0) node{$\bullet$} ;
\draw (3.4,0) node{$\bullet$} ;
\draw (3.9,0) node{$\bullet$} ;
\draw (1.9,0) -- (2.4,0) ;
\draw (2.4,0) -- (2.9,0) ;
\draw[dashed, thick] (2.9,0) -- (3.4,0) ;
\draw (3.4,0) -- (3.9,0) ;
\draw (2.4,0.25) node{$\bullet$} ;
\draw (2.9,0.30) node{$\bullet$} ;
\draw (3.4,0.25) node{$\bullet$} ;
\draw (1.9,0) -- (2.4,0.25) ;
\draw (2.4,0.25) -- (2.9,0.3) ;
\draw[dashed, thick] (2.9,0.3) -- (3.4,0.25) ;
\draw (3.4,0.25) -- (3.9,0) ;
\draw (2.4,-0.25) node{$\bullet$} ;
\draw (2.9,-0.30) node{$\bullet$} ;
\draw (3.4,-0.25) node{$\bullet$} ;
\draw (1.9,0) -- (2.4,-0.25) ;
\draw (2.4,-0.25) -- (2.9,-0.3) ;
\draw[dashed, thick] (2.9,-0.3) -- (3.4,-0.25) ;
\draw (3.4,-0.25) -- (3.9,0) ;
\draw(5.0,0) node{$= \  C_{a,b,c}(\tau)$};
\endscope
\endtikzpicture 
\end{center}
\vskip3pt
In the next section we will derive a set of coupled inhomogeneous Laplace-eigenvalue equations that these coefficients satisfy, which will lead to solutions for arbitrary $a$, $b$ and $c$.

\sm

There is a variety of obvious generalizations of these classes of diagrams, such as the 
functions $C_{a_1, a_2, a_3, \dots ,a_\rho}$ already defined in the introduction in (\ref{1a1}),  
that has $\rho$ chains of arbitrary length.  Another obvious generalization is to replace 
the single propagator in each link of a chain by a self-energy bubble, so that a single chain has the form,
\vskip3pt 
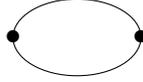
\begin{center}
\tikzpicture[scale=1.7]
\scope[xshift=-5cm,yshift=-0.4cm]
\draw (0,0.0) node{$\bullet$}   ;
\draw (1,0) node{$\bullet$} ;
\draw (0.5,0.0) ellipse (0.5 and 0.3);
\endscope
\endtikzpicture
\end{center}
\vskip3pt
This self-energy bubble diagram will be a key ingredient in the calculation of the diagram $D_{2,2,1}$ by direct summation in appendix \ref{secC}.

\subsection{The Laplacian on modular coefficients via deformation theory} 
\label{sec:laplacegen}

As explained in the Introduction, it is the Laplacian $\Delta$ on the genus-one moduli space $\cM_1$ acting on the partial amplitude $\cB_1$ in (\ref{2b5})  that will be of central interest in this paper. In terms of the standard parametrization of $\cM_1$ by a modulus $\tau=\tau_1 +i \tau_2 $ in the complex upper half plane, the Laplacian takes the form $\Delta = 4 \tau_2^2 \p_\tau \p_{\bar \tau}$, and therefore involves a mixed holomorphic -- anti-holomorphic derivative on moduli. Such derivatives can be evaluated with the help of the theory of deformations of complex structures in terms of associated Beltrami differentials $\mu$ and $\bar \mu$ which, for the torus $\Sigma$, may be chosen to be constant on $\Sigma$. 

\sm

The Beltrami differential parametrizes an infinitesimal deformation of the flat metric $|dz|^2$ on a torus with modulus $\tau$ to a metric $|dz + \mu d \bar z|^2$. Under the quasi-conformal transformation $z'=(z+\mu)/(1+\mu)$ the torus with modulus $\tau$ and metric $|dz+\mu d\bar z|^2$ is mapped to a torus with modulus $\tau' = (\tau + \mu \bar \tau)/(1+\mu)$ and metric $|dz'|^2$ (up to a constant Weyl factor which is immaterial thanks to conformal invariance of the amplitudes). The Laplacian $\Delta$ may then be simply expressed in terms of deformations by $\mu$ via the formula,
\bea
\Delta = \p_\mu \p_{\bar \mu}
\eea
In this formalism, it is the metric that is being deformed, but the  coordinates that are left unchanged.  Thus, the only thing that needs to be varied in the modular function $\cB_1$ of (\ref{2b5}) is the Green function $\cG$. Its deformations to first order in $\mu$ and $\bar \mu$ are given as follows,\footnote{In this section, to shorten the notations, we shall  exhibit neither the dependence of $\cG, \cB_1$ and $ \cL$ on $\tau$, nor the dependence of $\cB_1$ and $ \cL$ on $s,t,u$.}
\bea
\label{2g1}
\p _\mu \cG(z_i-z_j) \Big |_{\mu=0} =  - { 1 \over \pi} \int _\Sigma d^2 z \, \p_z \cG(z-z_i) \p_z \cG(z-z_j)
\eea
along with the complex conjugate equation. The second order mixed variation $\p_\mu \p_{\bar \mu} \cG(z_i-z_j)$ evaluated at $\mu=0$ vanishes in the interior of moduli space. 

\sm

Therefore, it is now straightforward to express the Laplacian of $\cB_1$ in terms of these deformations, and we find, 
\bea
\label{2g2}
\Delta \cB_1 = \sum _{w=2} ^\infty { w(w-1) \over w!} \, { 1 \over  \tau_2^4} \int _{\Sigma ^4} \prod _{i=1}^4 d^2 z_i \, 
\cL ^{w-2} (\p_\mu \cL) ( \p_{\bar \mu} \cL)
\eea
The deformation formula for $\cL$ is given by,
\bea
\label{2g3}
\p_\mu \cL = - { 1 \over \pi} \sum _{1 \leq i < j \leq 4} s_{ij} \int _\Sigma d^2 z \, \p_z \cG(z-z_i) \p_z \cG(z-z_j)
\eea 
At low weight orders, this formula may be simplified and worked out quite explicitly. We consider, for example, the case of the Feynman diagrams $D_\ell$, given by
\bea
\label{2g4}
D_\ell = \int _\Sigma { d^2 z_1 \over \tau_2} \int _\Sigma {d^2z_2\over \tau_2} \, \cG(z_1-z_2)^\ell
\eea
Applying these techniques yields their Laplacian, and after some simplifications, we obtain,
\bea
\label{2g5}
\Delta D_\ell & = & - \ell (\ell-3)D_\ell  + 
 { \ell (\ell-1) \over \pi^2 \tau_2^2} \prod _{i=1}^4 \int _\Sigma d^2 z_i \, 
\cG(z_1-z_3) \cG(z_2-z_4) 
\\ && \hskip 1in \times 
\p_{z_3} \cG(z_2-z_3) \p_{\bar z_4} \cG(z_1-z_4) \p_{z_1} \p_{\bar z_2} \cG(z_1-z_2)^{\ell-2} 
\no
\eea
When $\ell=2$, the integral on the right hand side vanishes in view of the last factor of the integrand, and we obtain $\Delta D_2=2D_2$ in accord with the fact that we already know $D_2$ to be simply equal to the Eisenstein series $ \tE_2$. For $\ell=3$, the first term vanishes, and the integral may be simplified since the last factor of the integrand evaluates to
\bea
\label{2g6}
\p_{z_1} \p_{\bar z_2} \cG(z_1, z_2) = \pi \delta (z_1-z_2) - {\pi \over \tau_2}
\eea
The integration over the $\delta$-function term in (\ref{2g6}) sets $z_1=z_2$ so that the integrand becomes a total derivative in $z_3$ and $z_4$ and the integral of this contribution vanishes.  The integration over the constant term in (\ref{2g6}) gives,
\bea
\label{2g7}
\Delta D_3 =
 { 6 \over \pi \tau_2^3} \prod _{i=1}^4 \int _\Sigma d^2 z_i \, 
\cG(z_1-z_3) \cG(z_2-z_4)  \p_{z_3} \cG(z_2-z_3) \p_{\bar z_4} \cG(z_1-z_4) 
\eea
Transforming the derivative in $\bar z_4$ into a derivative in $\bar z_1$ by translation invariance, 
and then integrating by parts in both $z_1$ and $z_3$ exposes the combination $\p_{\bar z_1} \p_{z_3} \cG(z_1-z_3)$ which is evaluated with the help of (\ref{2g6}). This time the contribution from the constant term cancels out, and its is the $\delta$-function which survives and gives,
\bea
\label{2g8}
\Delta D_3 =
 { 6 \over  \tau_2^3} \prod _{i=2}^4 \int _\Sigma d^2 z_i \, 
\cG(z_2-z_3) \cG(z_3-z_4) \cG(z_4-z_2) 
\eea
The right hand side readily evaluates to an Eisenstein series, and we find $\Delta D_3=6 \tE_3$, which will also be arrived at by a slightly different procedure in the next section.

\sm

The evaluation of higher order contributions by this method has not yet been systematically investigated, as we shall use more algebraic techniques to go to higher order in the subsequent sections.

\section{The modular functions $C_{a,b,c}$}
\setcounter{equation}{0}
\label{sec3}

This section will be devoted entirely to the study of the modular functions $C_{a,b,c}$ which corresponds to the case $\rho=3$ in (\ref{1a1}). The functions $C_{a,b,c}$ evaluate an infinite class of Feynman diagrams with three chains, as explained in section \ref{25}. It is for this case that we will be able to obtain the most general and the most precise results, for which we shall present complete proofs in this section, and the subsequent section \ref{sec9} with the help of a generating function for $C_{a,b,c}$. 

\sm

To start, we consider a generalization of the set of functions $C_{a,b,c}$ defined as follows,
\bea
\label{3a1}
C_{a,b,c}^k (\tau) = \sum _{ (m_r,n_r)\not=0 }
{ \delta _{m,0} \delta _{n,0} (m_1 n_2 - n_1 m_2)^{2k} \,  (\tau_2/ \pi) ^{a+b+c}
\over 
|m_1 \tau+n_1|^{2a} |m_2 \tau+n_2|^{2b} |m_3 \tau+n_3|^{2c} }
\eea
for any non-negative integer $k$. The original function $C_{a,b,c}$ corresponds to the special case $k=0$. The Kronecker $\delta$ factors enforce the vanishing of the total momentum components $m=m_1+m_2+m_3$ and $n=n_1+n_2+n_3$.   Of interest here will be the case where $a,b,c$ are integers such that the sums are absolutely convergent, but the functions $C_{a,b,c}^k$ may be defined for complex values of $a,b,c$ as well. The functions $C_{a,b,c}$ are invariant under permutations of the indices $a,b,c$ for all $k$  in view of the fact that momentum conservation $m=n=0$ implies $(m_1 n_2 - n_1 m_2)^2=(m_2 n_3 - n_2 m_3)^2=(m_3 n_1 - n_3 m_1)^2$.  In the following we will only consider integer values of $a$, $b$ and $c$, and for the most part we will  follow the convention of using this permutation  symmetry to label the indices so that $a\ge b \ge c$.  Finally, the functions $C^k_{a,b,c} (\tau)$ are modular invariant because  the combination $(m_1 n_2 - n_1 m_2)^2$ is a symplectic invariant.

\subsection{Algebraic representation of the Laplacian}

The key tool that we shall use in the analysis of the modular functions $C_{a,b,c}^k$ is the fact that the Laplace-Beltrami operator $\Delta$ on the complex upper half plane, 
\be
\label{3b1}
\Delta = \tau_2^2\, (\partial_{\tau_1}^2 +\partial_{\tau_2}^2)
\ee
acting on the functions $C_{a,b,c}^k$ admits an algebraic representation. Indeed, for a function $C_{a,b,c}^k $ of weight $w=a+b+c$, we find the following result by explicit calculation, 
\bea
\label{3b2}
\Big (\Delta - w(w-1) \Big ) C^k _{a,b,c} 
 =
- 4 ab \, C^{k+1} _{a+1,b+1,c} 
- 4 bc \, C^{k+1} _{a,b+1,c+1}
- 4 ca \, C^{k+1} _{a+1,b,c+1}
\eea
The functions for successive values of $k$ are related by the following algebraic identity,
\bea
\label{3b3}
- 4 C^{k+1} _{a,b,c}  = 
C^k _{a-2,b,c} + C^k _{a,b-2,c} + C^k _{a,b,c-2} - 2 C^k _{a,b-1,c-1} 
- 2 C^k _{a-1,b,c-1}  - 2 C^k _{a-1,b-1,c} 
\eea
which is easily proven by applying the relations for momentum conservation $m=n=0$ to derive the following polynomial identity
valid under the summation in (\ref{3a1}), 
 \bea
 \label{3b4}
 \sum _{r=1}^3 |m_r\tau +n_r|^4 - \sum _{r\not= r'}  |m_r\tau +n_r|^2 |m_{r'}\tau +n_{r'}|^2 = - 4 \tau_2^2 (m_1 n_2-n_1m_2)^2
\eea
The results of (\ref{3b2}) and (\ref{3b3}) may be combined to eliminate the functions with upper index $k+1$, and produce a relations between functions of upper index $k$ only, 
\bea
\label{Lap}
&& \Big (\Delta - a(a-1)-b(b-1)-c(c-1) \Big ) C^k _{a,b,c} 
\no \\ && \hskip .2in 
=
+ ab \Big ( C^k _{a-1,b+1,c} + C^k _{a+1,b-1,c} + C^k _{a+1,b+1,c-2} 
 - 2 C^k _{a,b+1,c-1} - 2 C^k _{a+1,b,c-1}  \Big )
\no \\ && \hskip 0.4in 
+ bc \Big ( C^k _{a-2,b+1,c+1} + C^k _{a,b-1,c+1} + C^k _{a,b+1,c-1} 
- 2 C^k _{a-1,b,c+1} - 2 C^k _{a-1,b+1,c}  \Big )
\no \\ && \hskip 0.4in 
+ ca \Big ( C^k _{a-1,b,c+1} + C^k _{a+1,b-2,c+1} + C^k _{a+1,b,c-1} 
- 2 C^k _{a,b-1,c+1}  - 2 C^k _{a+1,b-1,c}  \Big ) \quad
\eea
This expression is closed on the functions $C^k_{\cdot, \cdot, \cdot}$ and we may set $k=0$ to recover the functions in which we are interested.  The application of the Laplacian is now reduced to an algebraic operation. Note that, although the functions of central interest in string theory have positive integer $a,b,c$, the above formulas are valid for $a,b,c \in \CC$ as long as their values are such that all functions involved correspond to convergent sums.  

\subsection{Evaluating  $C_{a,b,0}$ and $C_{a,b,-1}$}

The right hand side of (\ref{Lap}) may end up involving a lower index which vanishes or which equals -1. Both of these functions will appear in applying the Laplace operator to functions $C_{a,b,c}$ with $a,b,c \geq 1$. We shall now evaluate these terms separately. The corresponding sums then simplify considerably, and reduce to, 
\bea
\label{3c1}
C_{a,b,0} (\tau) & = &  \sum _{(m_r,n_r) \not=(0,0)}  
{ \tau_2 ^{a+b}
\over  \pi^{a+b} |m_1 \tau+n_1|^{2a} |m_2 \tau+n_2|^{2b}  }
\no \\
C_{a,b,-1} (\tau ) & = &  \sum _{(m_r,n_r) \not=(0,0)}   
{ \tau_2  ^{a+b-1}\,  |(m_1+m_2) \tau + n_1 + n_2|^2
\over \pi^{a+b-1} |m_1 \tau+n_1|^{2a} |m_2 \tau+n_2|^{2b}  }
\eea
In $C_{a,b,0}$, the sum excludes the values $(m_2,n_2)=(-m_1, -n_1)$; restoring their contribution allows  evaluating the sums in terms of Eisenstein series. In $C_{a,b, -1}$ the exclusion of the values $(m_2,n_2)=(-m_1, -n_1)$ is automatic thanks to the presence of the numerator, and  decouples the sums into a product of Eisenstein series. The final results are as follows,
\bea
\label{3c2}
C_{a,b,0}  & = &  \tE_a \, \tE_b  -  \tE_{a+b} 
\no \\
C_{a,b,-1} & = &  \tE_{a-1} \,  \tE_b   +  \tE_a \,  \tE_{b-1} 
\eea
Having established these special values we will now proceed to determining  the equations satisfied by the modular functions  $C_{a,b,c}$.  To start with, we will study the functions with weights $w=3,4,5$, before analyzing the case of general weight $w$.

\subsection{Laplace equation for $C_{a,b,c}$ at low weights}
\label{sec:cabclow}

Of central interest in string theory applications will be the functions $C_{a,b,c}$ with  integer $a,b,c\geq 1$. We shall now derive explicitly the Laplace operator on the corresponding functions for $w=a+b+c= 3,4$, and $5$. 

\subsubsection{The case $C_{1,1,1} = D_3$}

In this case we have $a=b=c=1$.  Formally, the  Laplace equation derived from (\ref{Lap}) becomes $\Delta C_{1,1,1} =  3 C_{2,2,-1} - 6 C_{2,1,0}$. Actually, both functions on the right hand side of this equation are divergent, which may be attributed to a logarithmic divergence of the sum over world-sheet momenta, as may be seen by evaluating them with the help of (\ref{3c2}) and the fact that $ \tE_1$ is divergent.
\sm

To carry out the evaluation, we combine the sums of both terms,\footnote{An alternative procedure is to regularize all functions in the equation, for example  with the help of a cutoff $\Lambda$ on the world-sheet momenta $|m|, |n| \leq \Lambda$. The divergent function $ \tE_1$ then gets replaced by its momentum cutoff version $ \tE_1^\Lambda$ which will cancel out of all calculations of the Laplacian. An alternative regularization is by continuing the functions $C_{a,b,c}$ to non-integer values by infinitesimal shifts $a\to a+\ep$ etc.} and observe that the combined sum is convergent, and given by, 
\bea
\label{3q1}
C_{2,2,-1} - 2 C_{2,1,0} =  \sum _{(m_r,n_r) \not=0} { \delta _{m,0} \delta _{n,0} \, \tau_2 ^3 \, 
 ( (m_1 \tau + n_1) (m_2 \bar \tau + n_2)+ {\rm c.c.} )
\over \pi^3 \, |m_1 \tau + n_1|^4 |m_2 \tau + n_2|^4}
\eea
where we still use the notation $m=m_1+m_2+m_3$ and $n=n_1+n_2+n_3$.
The sum excludes the terms with  $(m_1, n_1) = (-m_2, -n_2)$. Restoring this contribution to the  double sum makes each factor in the double sum odd under $(m_r , n_r) \to (-m_r, -n_r)$.  Hence the unrestricted double sum vanishes, leaving only the contribution from $(m_3,n_3)=(0,0)$, which gives $C_{2,2,-1} - 2 C_{2,1,0}  = 2  \tE_3$.   As a result, the Laplace equation becomes, 
\bea
\label{3q2}
\Delta C_{1,1,1} = 6  \tE_3
\eea
 Using (\ref{1a3}), we recast the equation as $\Delta \left ( C_{1,1,1} -  \tE_3 \right ) =0$. Its solution  is given by, 
\be
\label{3d3a}
C_{1,1,1} =  \tE_3 +  \zeta(3)
\ee
where we have used  knowledge of the power-behaved terms in the large-$\tau_2$ expansion of $C_{111}$, given in equation (B.8) in \cite{Green:2008uj},
\be
\label{c111expand}
C_{1,1,1}  = \frac{2  y^3}{945}  +  \zeta(3) + \frac{3   \zeta(5)}{4 y^2} + \cO(e^{-2 y})
\ee 
In this equation, and many of the following, we have defined the quantity,
\be
\label{ydef}
y =\pi\, \tau_2
\ee
which eliminates many factors of $\pi$ from subsequent formulas.
The result (\ref{3d3a}) has  been obtained earlier by Don Zagier \cite{Zagier:2014} by direct calculation of the sums over $m,n$.

\subsubsection{The case  $C_{2,1,1} = D_{2,1,1}$}
\label{sec:C211}

Using the expression for the Laplacian in (\ref{Lap}), we readily find again a formal relation,
\bea
\left ( \Delta -2 \right ) C_{2,1,1} =    - 4 C_{3 ,1,0} + 4 C_{3 ,2 ,-1} - 5 C_{2,2,0}
\eea
where the expressions $C_{3,1,0}$ and $C_{3,2,-1}$ are separately logarithmically divergent through terms linear in $ \tE_1$, which require regularization to $ E^\Lambda_1$. In the combined contribution we may directly apply (\ref{3c2}), resulting in the cancellation of all dependence on $ \tE_1^\Lambda$. The Laplace equation then reduces to, 
\bea
\label{c211lap}
(\Delta -2)\, C_{2,1,1} =   9  \tE_4 -  \tE_2^2\
\eea
This equation has the same structure as the inhomogeneous Laplace-eigenvalue equation that arose in a rather different context in  \cite{Green:2005ba}.

\subsubsection{The cases  $C_{2,2,1} = D_{1,1,1,1;1}$ and $C_{3,1,1} = D_{2,1,1,1}$}

We will now consider the two weight-5 modular  functions  $C_{2,2,1}$ and $C_{3,1,1}$.
Using the expression for the Laplacian  (\ref{Lap}), we readily find the relations,
\bea
\Delta C_{2,2,1} & = &4 C_{3,3,-1} - 8 C_{3,2,0}
\no \\
(\Delta -6) C_{3,1,1} & = &  3 C_{2,2,1} - 6 C_{4,1,0} -10 C_{3,2,0} + 6 C_{4,2 ,-1} 
\eea
Again, the relation on the second line is formal and requires regularization as in the earlier cases.
Substituting the corresponding expressions from (\ref{3c2}) gives,
\bea
\label{c221lap}
\Delta C_{2,2,1} & = & 8  \tE_5
\no \\
\left ( \Delta -6 \right ) C_{3,1,1} & = &  3 C_{2,2,1} + 16  \tE_5 - 4  \tE_2  \tE_3
\label{c311lap}
\eea
In particular, using $\Delta  \tE_5 = 20  \tE_5$, it follows that $\Delta ( C_{2,2,1} - { 2 \over 5}  \tE_5)=0$, and thus,
\bea
\label{c221sol}
C_{2,2,1} = { 2 \over 5}  \tE_5 + \frac{\zeta(5)}{30}
\eea
where the coefficient of the $\zeta(5)$ term is determined from knowledge of the terms that are power behaved in the limit $y\to \infty$, given in \cite{Green:2008uj}. Substituting this in (\ref{c311lap}) leads to,
 \bea
 \label{D2111}
(\Delta -6)\,C_{3,1,1}
=  \frac{86}{5}   \tE_5 - 4   \tE_2  \tE_3 +\frac{\zeta(5)}{10} 
 \eea
which is analogous to (\ref{c211lap}) and will be discussed further in section~\ref{sec5}.

\subsection{Poincar\'e series representations}
\label{sec:poincare}

The inhomogeneous Laplace eigenvalue equations satisfied by the modular functions discussed in this paper have a structure analogous to the equation that arises in the context of S-dual properties of  the coefficient of the $\cD^6\, \cR^4$ interaction  obtained in \cite{Green:2005ba}.   The solution to this equation was obained in the form of a Poincar\'e series in appendices A and B of  \cite{Green:2014yxa}.  Here we will derive a similar Poincar\'e series representation for the specific example of the modular function $C_{2,1,1}$, but the same methods can be applied to obtain solutions for any of the Laplace equations that arise in subsequent sections.

\sm

We begin by recalling the classic Poincar\'e series representation for the Eisenstein series, 
\bea
\label{EisenPoin}
E_s (\tau) = { 2 \zeta (2s) \over \Gamma (s)} \sum _{\gamma \in \Gamma _\infty \backslash SL(2,\ZZ)} 
\Big ( \Im \gamma (\tau ) \Big )^s
\eea
where $\gamma $ acts on $\tau$ by M\"obius transformations, 
\bea
\label{gam}
\gamma (\tau) = { a\tau + b \over c\tau+d}
\hskip 1in 
\gamma = \left ( \matrix{ a & b \cr c & d \cr} \right ) \in SL(2,\ZZ)
\eea
and $\Gamma _\infty$ the subgroup of $SL(2,\ZZ)$ which stabilizes $\tau=i\infty$. 
Note that we can also represent any of the modular functions $C_{a,b,c}$ by a sum over $2\times 2$ matrices,
\bea
C^k_{a,b,c}(\tau) = { 1 \over \pi^{a+b+c}}
\sum _{\gamma \in M(2,\ZZ)} (\det \gamma )^{2k-a-b-c} 
\Big (  \Im \gamma (\tau ) \Big ) ^a 
\Big (  \Im S \gamma (\tau ) \Big ) ^b 
\Big ( \Im ST \gamma (\tau ) \Big ) ^c
\label{cabcseries}
\eea
where $S(\tau)=-1/\tau$, $T(\tau)=\tau+1$ and $M(2,\ZZ)$ is the set of $2 \times 2$ matrices with integer
entries, and non-vanishing determinant.   Although this expression is not a Poincare series it can be recast as one in the following manner.

\sm

We will also use the  Fourier series for a non-holomorphic Eisenstein series, valid for any~$s$,
\bea
\label{fouriereis}
\tE_s(\tau) & = & {2\zeta(2s)\over \pi^s} \, \tau_2^s
+ {2 \Gamma(s-\frac12) \zeta(2s-1) \over \Gamma(s)\pi^{s-\frac12}} \, \tau_2^{1-s}
\no \\ && +
{4\sqrt{\tau_2}\over\Gamma(s)} \sum_{N\neq0} |N|^{s-\frac12} \, \sigma_{1-2s} (|N|)\, K_{s-\frac12}(2\pi|N|\tau_2) e^{2i\pi N\tau_1}
\eea
where $K_\nu (z)$ is the modified Bessel function of order $\nu$, and the divisor sum is given by $\sigma _k (n) = \sum _{d | n} d^k$  for positive divisors $d$.  With our normalisation, the Eisenstein series satisfies the functional equation $\Gamma(s)\tE_s(\tau)=\Gamma(1-s)\tE_{1-s}(\tau)$ for all $\tau$ in the upper half-plane, and $s \in \CC$. For integer values $n$ of $s$, the function $K_{n-\half}$ reduces to a spherical Bessel function, and we have the following simplified form, expressed in terms of $q=e^{2 \pi i \tau}$ and $y = \pi \tau_2$, 
\bea
\label{a1}
\tE_n (\tau) & = &(-1)^{n-1} {B_{2n}\over (2n)!} (4y)^n +  {4 \,(2n-3)!\over (n-2)!(n-1)! } \zeta (2n-1) \, (4y)^{1-n}
\\ &&
\no + {2\over (n-1)!}  \sum _{N=1}^\infty  N^{n-1}
\sigma_{1-2n} (N) \,(q^N+\bar q^N) \sum _{m=0}^{n-1} { (n+m-1)! \over m! \, (n-m-1)! }
(4  N y)^{-m}
\eea
where $B_{2n}$ are Bernoulli numbers.  We see that when $s=n$ is an integer the series in $m$ in~(\ref{a1})  truncates to a finite sum with $n$ terms and $ 0 \leq m< n$ for any value of $N$. 
The purely power behaved terms, together with the leading exponential corrections (the terms  in the sum with $N=1$) will play an important part of our considerations in sections~\ref{sec4} and~\ref{sec5}.

\sm 

Consider now the inhomogeneous Laplace equation for $C_{2,1,1}$ in (\ref{c211lap}), which we will rewrite as 
\bea
\label{c211lapnew}
(\Delta -2)\, \left(C_{2,1,1} -\frac{2}{3}\, \tE_4\right) =   \frac{7}{3}\,  \tE_4 -  \tE_2^2\
\eea
where we have subtracted $20 E_4/3$ from each side (using the fact that $(\Delta -2) E_4 = 10 E_4$) so that the source term on the right hand side now has leading behavior  $O(y)$ in the limit $y\to \infty$, as is seen from the expansion (\ref{a1}) for the cases $n=2$ and $n=4$. 

\sm

We shall now find a solution to  (\ref{c211lapnew}) expressed as a Poincar\'e series of the form
\bea
\label{poincc211}
  C_{2,1,1} (\tau)  -\frac{2}{3}\, \tE_4(\tau) =  \sum_{\gamma\in
    \Gamma_\infty\backslash SL(2,\ZZ)} \Phi(\gamma\tau)  
\eea
To determine the function $\Phi(\tau)$ we recall the Poincar\'e series representation of the Eisenstein series of (\ref{EisenPoin}). 
Substituting this expression for the cases $s=2$ and $s=4$ in the source term on the right-hand side of (\ref{c211lapnew}) and substituting the expression (\ref{poincc211}) on the left-hand side of (\ref{c211lapnew}) leads to the  inhomogeneous Laplace equation for $\Phi$
\bea
\label{a8}
  (\Delta-2) \Phi(\tau) =\frac{y^4}{2025}- {y^2 \over 45 } \,\tE_2(\tau)  
\eea
where  the leading behavior of the right hand side is of order $\cO(y)$ as $y\to \infty$.

Noting that  $q^m$ and $\bar q^m$ are annihilated by the Laplace operator, $ \Delta q^m= \Delta \bar q^m= 0$,   
we seek a solution to (\ref{a8})  of the  form
\bea
\label{poinceq}
  \Phi(\tau)= \Phi_0(y)+ \sum_{n=1}^\infty  \Phi_n(y) \, (q^n + \bar q^n)  
\eea
Given that $\lim_{y\to \infty} \Phi(\tau) =O(y)$, it follows from Lemma~2.9 in~\cite{Green:2014yxa} that $\Phi(\tau)$ must satisfy the boundary conditions   
$\lim_{y\to0} \Phi_n(y)=O(y^0)$ for all $n \geq 0$.  This determines that the solution of (\ref{poinceq}) is uniquely given by
\bea
\Phi(\tau)  =
 {\zeta(3)\over90} y +{y \over90}\,\sum_{n\geq1} {1\over n^3} \left({q^n\over
    1-q^n}+{\bar q^n\over 1-\bar q^n}\right)
\eea 
and therefore  the  Poincar\'e series representation of $C_{2,1,1}$ is determined by substituting $\Phi(\gamma\tau)$ in  (\ref{poincc211}).

\subsection{Laplace equations for $C_{a,b,c}$ at any weight}
\label{sec34}

For general weight $w=a+b+c \geq 5$ with $a,b,c \geq 1$ the Laplace equations of (\ref{Lap}) involve several functions $C_{a,b,c}$.  The corresponding system of coupled  equations can be diagonalized in order to determine the  spectrum of eigenvalues and eigenvectors of the Laplacian acting on the space of $C_{a,b,c}$ for weight $w$. With the help of MAPLE we have carried out this analysis  up to weight $w = 34$ included.   The results may be summarized by the following,

\subsubsection{{\bf \large Theorem 1} : Structure of the Spectrum} 

(a) The eigenvalues of the Laplacian acting on the space of functions $C_{a,b,c}$ with weight $w=a+b+c$ for $a,b,c \geq 1$ are given by $s(s-1)$ where $s=w-2\mm$ and  $\mm$ runs over all integers in  the range, 
\bea
\label{3d1}
1 \leq \mm \leq \left [ { w-1 \over 2} \right ]
\eea

\noindent
(b) The multiplicity $n_s$ of the eigenvalue $s(s-1)$ equals 
\bea
\label{3d2}
n_s= \left [ { s +2 \over 3} \right ] 
\eea
Here, $[x]$ denotes the integral part of $x$, defined to be $n \in \ZZ$ for $x$ in the range $n \leq x < n+1$.
A complete analytical proof  will be given  in section \ref{sec9} with the help of a generating function $\cW$ for the modular functions $C_{a,b,c}$.

\subsubsection{Illustration of the spectrum up to weight $w=12$}

To illustrate the above result on the structure of the spectrum, 
we shall list below the allowed values of $s$ for weights $w$ in the range $3 \leq w \leq 12$, 
along with the eigenvalue $s(s-1)$ and the degeneracy $n_s$ in the notation $s(s-1) ^{(n_s)}$,
\bea
\label{3d3}
w=3 \hskip 0.5in s=1  \hskip 0.75in & \hskip 0.8in & 0^{(1)}
\\
w=4 \hskip 0.5in s=2  \hskip 0.75in &  & 2^{(1)}
\no \\
w=5 \hskip 0.5in s=1,3  \hskip 0.6in &  & 0^{(1)} \oplus 6^{(1)} 
\no \\
w=6 \hskip 0.5in s=2,4  \hskip 0.6in &  & 2^{(1)} \oplus 12^{(2)}
\no \\
w=7 \hskip 0.5in s=1,3,5  \hskip 0.45in &  & 0^{(1)} \oplus 6^{(1)}  \oplus 20^{(2)}
\no \\
w=8 \hskip 0.5in s=2,4,6 \hskip 0.45 in &  & 2^{(1)} \oplus 12^{(2)} \oplus 30 ^{(2)}
\no \\
w=9 \hskip 0.5in s=1,3,5,7 \hskip 0.3 in &  & 0^{(1)} \oplus 6^{(1)}  \oplus 20^{(2)} \oplus 42^{(3)}
\no \\
w=10 \hskip 0.42in s=2,4,6,8 \hskip 0.3 in &  & 2^{(1)} \oplus 12^{(2)} \oplus 30 ^{(2)} \oplus 56^{(3)} 
\no \\
w=11 \hskip 0.42in s=1,3,5,7,9 \hskip 0.15 in &  & 0^{(1)} \oplus 6^{(1)}  \oplus 20^{(2)} \oplus 42^{(3)} \oplus 72^{(3)}
\no \\
w=12 \hskip 0.44in s=2,4,6,8,10 \hskip 0.05in &  & 2^{(1)} \oplus 12^{(2)} \oplus 30 ^{(2)} \oplus 56^{(3)} 
\oplus 90^{(4)}
\no
\eea
MAPLE also provides the full inhomogeneous Laplace-eigenvalue equations. We shall list these in the subsections below up to eigenvalue 12, which is the lowest eigenvalue at which the multiplicity exceeds one.

\subsubsection{The full inhomogeneous eigenvalue equations for eigenvalue 0}

For eigenvalue 0, namely $s=1$, the multiplicity is one for each odd weight (and zero for each even weight), and we list the corresponding inhomogeneous Laplace-eigenvalue equations  below for each odd weight, up to weight $w=9$,
\bea
\label{3d4}
\Delta  C_{1,1,1} & = & 6  \tE_3
\no \\
\Delta C_{2,2,1} & = & 8  \tE_5
\no \\
\Delta \left ( C_{3,3,1} + C_{3,2,2} \right ) & = & 18  \tE_7
\no \\
\Delta \left ( 9 C_{4,4,1} + 18 C_{4,3,2} + 4 C_{3,3,3} \right ) & = & 288   \tE_9
\eea
The first two cases are already familiar from  our analysis at low weights in (\ref{3q2}) and (\ref{c311lap}).
These equations were integrated with the help of the eigenvalue equation $\Delta   \tE_s= s(s-1) \tE_s$ for the Eisenstein series and the integration constant was fixed by matching the asymptotics at the cusp.

\sm

Clearly, the equation at odd weight $w$ associated with eigenvalue 0 exhibits a right hand side proportional to the Eisenstein series $ \tE_w$, and may thus be integrated to obtain a purely algebraic relation between the $C_{a,b,c}$ at weight $w=a+b+c$. The algebraic equations corresponding to the differential equations of (\ref{3d4}) are given as follows, 
\bea
\label{3d5}
C_{1,1,1} & = &  \tE_3 + \zeta (3)
\no \\
30 C_{2,2,1} & = & 12  \tE_5 + \zeta (5) 
\no \\
14 C_{3,3,1} + 14 C_{3,2,2} & = & 3  \tE_7 + g_7
\no \\
9 C_{4,4,1} + 18 C_{4,3,2} + 4 C_{3,3,3} & = & 4  \tE_9 + g_9
\eea
where $g_7, g_9$ are integration constants which are determined by matching 
the asymptotic behavior of both sides of the equation at the cusp. A general formula for these
algebraic relations valid for any odd weight will be derived in section \ref{sec39}.

\subsubsection{The full inhomogeneous eigenvalue equations for eigenvalues 2 and 6}

For eigenvalue 2, which has multiplicity 1 for every even weight starting at weight 4, we have
\bea
\label{3d6}
(\Delta -2) C_{2,1,1} & = & 9  \tE_4 -  \tE_2^2
\no \\
(\Delta -2) \left ( 4 C_{3,2,1} + C_{2,2,2} \right ) & = & 52  \tE_6 - 4 \tE_3^2
\no \\
(\Delta -2) \left ( 6 C_{4,3,1} + 3C_{4,2,2} + 5 C_{3,3,2} \right ) & = & 153  \tE_8 - 9 \tE_4^2
\no \\
(\Delta -2) \left ( 8 C_{5,4,1} + 8C_{5,3,2} + 7 C_{4,4,2} + 8 C_{4,3,3} \right ) & = & 336  \tE_{10} - 16 \tE_5^2
\eea
The first equation is already familiar from (\ref{c211lap}).
For eigenvalue 6, which has multiplicity 1 for every odd weight starting at weight 5, we have, 
\bea
\label{3d6a}
(\Delta -6) \left ( 2 C_{3,1,1} + C_{2,2,1} \right ) & = & 40  \tE_5 - 8 \tE_2  \tE_3
\no \\
(\Delta -6) \left ( 6 C_{4,2,1} + 2 C_{3,3,1} + 3 C_{3,2,2} \right ) & = & 156  \tE_7 - 24 \tE_3  \tE_4
\no \\
(\Delta -6) \left ( 8 C_{5,3,1} + 4C_{5,2,2} + 3 C_{4,4,1} + 10 C_{4,3,2} + 2 C_{3,3,3} \right ) & = & 384  \tE_9 - 48 \tE_4  \tE_5 \qquad
\eea

\subsubsection{The full inhomogeneous eigenvalue equations for eigenvalue 12}

For eigenvalue 12, the multiplicity is 2 for every even weight starting at weight 6.
At weight 6 and 8 we have respectively, 
\bea
\label{3d7}
(\Delta - 12) \left [ \matrix{ 
6 C_{3,2,1} - C_{2,2,2} \cr 6 C_{4,1,1} + C_{2,2,2} \cr} \right ] 
& = & 
 \left [ \matrix{108  \tE_6 - 36  \tE_3^2 \cr 120  \tE_6 +  \tE_3^2 - 36  \tE_2  \tE_4 \cr} \right ]
\no \\ 
(\Delta - 12) \left [ \matrix{ 
2 C_{4,3,1} + C_{4,2,2} \cr 8 C_{5,2,1} + C_{4,2,2} + 2 C_{3,3,2} \cr} \right ] 
& = & 
\left [ \matrix{66  \tE_8 - 18  \tE_4^2 \cr 190  \tE_8 +18  \tE_4^2 - 48  \tE_3  \tE_5 \cr} \right ]
\eea
and at weight 10, 
\bea
&&
(\Delta - 12) \left [ \matrix{ 
12 C_{5,4,1} + 12 C_{5,3,2} + 3 C_{4,4,2}+ 2 C_{4,3,3} \cr
40 C_{6,3,1} + 20 C_{6,2,2} + 16 C_{5,3,2} + 9 C_{4,4,2} + 14 C_{4,3,3} \cr} \right ] 
\no \\ && \hskip 2.2in 
= 
\left [ \matrix{624  \tE_{10} - 144  \tE_5^2 \cr 1656  \tE_{10} +144  \tE_5^2 - 360  \tE_4  \tE_6 \cr} \right ]
\eea

\subsection{Structure of the eigenspaces of the Laplacian}
\label{sec36}

In this section, we shall  infer the structure of the eigenspaces of functions $C_{a,b,c}$ from MAPLE calculations up to weight $w=34$.  Recall from Theorem 1 that the eigenvalues of the Laplacian are given by  $s(s-1)$, where $s=w-2\mm$ and the allowed values for $\mm$ are given by $\mm=1,2,\dots , [(w-1)/2]$, and  corresponding eigenspace $\cV(w,s)$ has dimension,
\bea
\label{3e1}
\dim \cV(w,s) = \left [ { s+2 \over 3} \right ]
\eea
In addition to this general structure of the spectrum, one may provide a more detailed description of the eigenfunctions  leading to statements that will be proven in full using the generating function in section \ref{sec9}.

\subsubsection{Structure of the  space $\cV(w,1)$ for eigenvalue 0}

For zero eigenvalue, we have $s=1$ and therefore the weight is given by $w=2\mu+1$ with $\mm=\mu$ for any positive integer $\mu$.  The corresponding eigenspace of eigenvalue zero is then generated by the following modular functions,
\bea
\label{3e2}
C_{\mu - \mp,\mu - \mp- \mq, 2\mp+\mq+1} & \hskip 0.5in & 0 \leq \mp, \quad 0 \leq \mq, 
\qquad
3\mp+2\mq \leq \mu -1
\eea
The range may be explicitly parametrized by 
\bea
\label{3e3}
\mp=0,1,\dots, \left [ { \mu -1 \over 3} \right ], \qquad \mq=0,1,\dots, \left [ { \mu -1 -3 \mp \over 2} \right ]
\eea
Clearly the structure depends on $\mm$ modulo 6.

\subsubsection{Structure of the space $\cV(w,s)$ for arbitrary $w,s$}

For odd weight $w=2\mu +1 \geq 3$ we have odd values $s=2\ell+1$ with $\ell< \mu $ and $\mm=\mu -\ell$. The eigenspace is restricted by the condition that only the functions  $C_{a,b,c}$ can contribute 
for which the value of $a$ does not exceed $\mu +\ell$.
 This set may be parametrized as follows,
\bea
C_{\mu +\ell-\mp, \mu +\ell-\mp-\mq, 2\mp+\mq+1-2\ell} & \hskip 0.5in & 0 \leq \mp, \quad 0 \leq \mq
\no \\
&& 3\mp+2\mq \leq \mu + 2 \ell -1
\eea
For even weight $w=2\mu \geq 4$ and thus even values $s=2\ell$ with $\ell < \mu$,  the eigenspace is again
restricted by the condition that only the functions  $C_{a,b,c}$ can contribute 
for which the value of $a$ does not exceed $\mu +\ell$. 
This set may be parametrized as follows,
\bea
C_{\mu +\ell -\mp, \mu +\ell -\mp-\mq, 2\mp+\mq-2\ell} & \hskip 0.5in & 0 \leq \mp, \quad 0 \leq \mq
\no \\
&& 3\mp+2\mq \leq \mu + 2 \ell 
\eea

\subsection{Embedding eigenspaces at  successive weights}
\label{sec37}

We shall establish the existence of a map, denoted $\mE$, 
which embeds an eigenspace $\cV(w,s)$ into the eigenspace $\cV(w+2, s)$ of the same 
eigenvalue but weight increased by 2, 
\bea
\label{3f1}
\mE :  \cV(w,s) \to \cV(w+2,s) 
\eea
for all $s=w-2\mm$ with $\mm=1,2,\dots , [(w-1)/2]$. The map on $\cV(w,s)$ may actually be defined on the larger space of all functions $C_{a,b,c}^k$ with $a,b,c \geq 1$ and $k \geq 0$, via the relation,
\bea
\mE \left ( C_{a,b,c}^k (\tau) \right ) = 
ab \, C_{a+1,b+1,c}^k(\tau) + bc \, C_{a, b+1, c+1} ^k (\tau) + ca \, C_{a+1, b, c+1}^k (\tau)
\eea
The key observation in the proof of the above assertion is  that the functions $C_{a,b,c}^k$ 
satisfy the same Laplace-eigenvalue  equation (\ref{Lap}) for all value of $k \geq 0$.  
To see this, let a linear combination $C_\cJ ^k$ with coefficients $\cJ_{a,b,c}$ of 
$C_{a,b,c}^k(\tau)$-functions at fixed weight $w=a+b+c$ satisfy the eigenvalue equation for one value of $k$, 
\bea
\label{2h6}
\Big ( \Delta - s(s-1) \Big ) C_\cJ ^k (\tau) \approx 0
\hskip 1in 
C_\cJ ^k (\tau) = \sum _{[a,b,c]} \cJ _{a,b,c} C_{a,b,c} ^k (\tau) 
\eea
Here, $\approx 0$ stands omitting inhomogeneous terms linear and  bilinear in Eisenstein series, 
and the sum in $[a,b,c]$ is over all $a \geq b \geq c \geq 1$ with $a+b+c=w$. 
If   (\ref{2h6}) is satisfied for one value of $k \geq 0$, then it will be satisfied for all 
values of $k$ {\sl with the same coefficients $\cJ_{a,b,c}$}. 

\sm

Now on the one hand, $C_\cJ^k$ has eigenvalue $s(s-1)$ as written in (\ref{2h6}), 
while on the other hand from (\ref{3b2}), we have, 
\bea
\label{2h7}
\left (\Delta - w(w-1) \right ) C_\cJ ^k (\tau) = - 4 \mE ( C_\cJ ^{k+1} )
\eea 
Applying $(\Delta - w(w-1)) $ to the left hand side of (\ref{2h6}), and using the fact that 
this operator commutes with the operator $( \Delta - s(s-1) )$, we obtain the relation  $
 ( \Delta - s(s-1)  ) \mE  ( C_\cJ^{k+1}  ) \approx 0$.
Since the same Laplace-eigenvalue equation is obeyed irrespective of $k$, it follows that we 
also have the eigenvalue equation,
\bea
\left ( \Delta - s(s-1) \right ) \mE \left ( C_\cJ \right ) \approx 0 
\eea
where it is understood that $k=0$, and the weight of $\mE ( C_\cJ)$ is $w+2$.

\subsection{Corollaries}

A number of results follow right away from the existence of the embedding map $\mE$ exhibited in the previous section. We shall describe those next.

\sm

(1)  In summary, if $C_\cJ$ is an eigenvector of weight $w$ with Laplace-eigenvalue $s(s-1)$
for $s=w-2\mm$ and $\mm=1,2, \dots, [(w-1)/2]$, then for $\mu \geq 1$, the function $\mE^\mu (D_\cJ)$ 
is an eigenvector of weight $w+2\mu$ with the same eigenvalue $s(s-1)$. 
It is manifest by inspection that the map $\mE$ is injective. Thus, we have 
\bea
\label{3g1}
\cV(w+2,s) = \mE (\cV(w,s) )  & \hskip 0.7in & s=w-2\mm 
\no \\
\dim \cV(w+2,s) = \dim \cV(w,s)
&&
\mm = 1, 2, \dots, \left [ {w-1 \over 2} \right ]
\eea

\sm

(2) Given the first emergence of the eigenvalue $s(s-1)$ at weight $s=w-2$ by the eigenvector 
combinations $C_\cJ$, the eigenvectors corresponding to the same eigenvalue $s(s-1)$ 
and higher weight $w=s-2+2\mu$ are obtained by $\mE^\mu (C_\cJ)$. To prove the structure of the 
spectrum, it will thus suffice to prove that,
\bea
\label{3g2}
\dim \cV (w, w-2) = \left [ { w \over 3} \right ]
\eea
MAPLE calculations up to weight $w=34$ confirm this assertion holds. A complete proof of this structure will be given in section \ref{sec9}.

\sm

(3) The map $\mE$ may be expressed in terms of the mutually commuting shift operators $T_a, T_b$, and $T_c$
acting on the variables $a,b,c$ and on the functions $C_{a,b,c}$ by, 
\bea
T_a C_{a,b,c} & = & C_{a+1,b,c} \hskip 1in [T_a, a]=T_a
\eea
with analogous relations for $T_b$ and $T_c$,  so that we have 
\bea
\mE = ab T_a T_b + bc T_b T_c + ca T_c T_a
\eea
The $\mu$-th power of the map $\mE$ is obtained via the trinomial formula and we find,
\bea
\label{3g6}
\mE^\mu (C_{a,b,c} ) (\tau) 
& = &  \sum _{\mu_1, \mu_2=0} ^\mu 
{ \mu !  \Gamma (a+\mu -\mu _2) \Gamma (b+\mu _1 +\mu _2) \Gamma (c+\mu -\mu_1)  
\over 
\mu_1 ! \, \mu_2 !  \, \Gamma (\mu-\mu_1 -\mu_2 +1) \, \Gamma (a) \, \Gamma (b) \, \Gamma (c) }
\no \\ && \hskip 1in \times 
C_{a+\mu-\mu_2, b+\mu_1+\mu_2, c+\mu-\mu_1} (\tau)
\eea
The predictions of this formula have been tested against explicit MAPLE calculations at weights up to $w=34$, and agree. We shall limit ourselves here to providing examples up to weight 11, and make contact with the examples given to this weight order in section \ref{sec34}.

\subsubsection{Illustrations up to eigenvalue $12$}

At eigenvalue 0, with multiplicity 1 for all odd weights, we find, 
\bea
\mE ( C_{1,1,1} ) & = & 3 C_{2,2,1}
\\
\mE( C_{2,2,1}) & = & 4 C_{3,3,1} + 4 C_{3,2,2}
\no \\
\mE(C_{3,3,1} + C_{3,2,2}) & = & 9 C_{4,4,1} + 18 C_{4,3,2} + 4 C_{3,3,3}
\no
\eea
Comparison of the right hand sides with the entries of (\ref{3d4}) shows agreement on the top three lines, while the last line above predicts the correct linear combination for weight $w=13$. Similar comparisons at higher eigenvalue further confirm the validity of the general formula. At eigenvalue 2, we have, 
\bea
\mE ( C_{2,1,1} ) & =  & 4 C_{3,2,1} + C_{2,2,2}
\\
\mE( 4 C_{3,2,1} + C_{2,2,2} ) & = & 4(6 C_{4,3,1} + 3 C_{4,2,2} + 5 C_{3,3,2})
\no \\
\mE \left ( 6 C_{4,3,1} + 3 C_{4,2,2} + 5  C_{3,3,2} \right ) 
& = &  9(8 C_{5,4,1} + 8 C_{5,3,2} + 7 C_{4,2,2} + 8 C_{4,3,3})
\no
\eea
which agrees with (\ref{3d6}). Skipping eigenvalue 6, which is analogous to eigenvalue 2, we have for eigenvalue 12 where the first degeneracy occurs, 

\sm

For eigenvalue 12, we shall just list here the map between weight 6 and weight 8,
\bea
\mE ( 6C_{3,2,1} - C_{2,2,2} ) & = & 18(2 C_{4,3,1} +  C_{4,2,2})
\no \\ 
\mE ( 6 C_{4,1,1} + C_{2,2,2})
& = & 6(8 C_{5,2,1}  +  C_{4,2,2} + 2 C_{3,3,2})
\eea
which again agrees with (\ref{3d7}).

 \subsection{Explicit algebraic relations for $s=1$ and for all odd weights}
 \label{sec39} 
 
It was established earlier that the zero-eigenvalue equation on the functions $C_{a,b,c}$ for any odd weight $w=a+b+a$
leads to an algebraic equation between the $C$-functions and  the Eisenstein series $\tE_w$. The simplest one of these relations was derived in (\ref{3d3a}), and a list of the first cases up to weight 10 was given in (\ref{3d4}). We shall now present an explicit  formula valid for all weights. 

\sm

Parametrizing the weight by $w=2\mu+3$ for $\mu$ a positive integer, we may obtain the linear combination of $C$-functions at weight $w$ which give the eigenfunction of the Laplacian for weight 0 by applying the embedding operator $\mE^\mu$ to the function with zero eigenvalue at weight 3, namely $C_{1,1,1}$ with the help of the general formula (\ref{3g6}). The right hand side of the zero eigenvalue equation at weight $w$ is proportional to $\tE_w$. Therefore, the Laplace- eigenvalue equation for zero eigenvalue can be integrated to an algebraic equation at any odd weight $w$, and this equation is given by,
\bea
\label{3g5}
 \sum _{\mu_1, \mu_2=0} ^\mu \! \! 
{  (\mu -\mu _1)! \,  (\mu -\mu _2)! \, (\mu _1 +\mu_2) !
\over 
\mu _1 ! \, \mu _2 !  \, (\mu -\mu _1 -\mu _2 )!  } \, 
C_{1+\mu -\mu _1, 1+\mu -\mu _2, 1+\mu _1+\mu _2} (\tau) = f_w \tE_w (\tau) + g_w
\quad 
\eea
The summand vanishes for $\mu_1+\mu_2 >\mu $ so that the sum truncates, as expected.
Here, $f_w$ and $g_w$ are constants which depend only on $w$, and may be determined from the asymptotic behavior near the cusp; their values for weight up to 9 are given in (\ref{3d5}).

\section{The generating function for  $C_{a,b,c}$}
\setcounter{equation}{0}
\label{sec9}

It is natural to introduce a generating function for the modular functions $C_{a,b,c}$ with $a,b,c \geq 1$ in terms of three independent real variables $t_1, t_2, t_2$,
\bea
\cW(t_1, t_2, t_3 |\tau) = \sum _{a,b,c=1}^\infty t_1 ^{a-1}  \, t_2 ^{b-1} \, t_3^{c-1} \, C_{a,b,c} (\tau)
\eea
The lowest order term is independent of $t_1, t_2, t_3$ and reduces to $C_{1,1,1}(\tau)$. Since the $C_{a,b,c}(\tau)$ are $SL(2,\ZZ)$-invariant modular functions of $\tau$, so will $\cW$ be provided the variables $t_1, t_2, t_3$ transform
trivially under $SL(2,\ZZ)$, which is what we shall henceforth assume. 
Using the explicit expression for $C_{a,b,c}(\tau)$ given in terms of its defining sum in (\ref{3a1}), we may carry out the corresponding geometric series, and we find the following expression for $\cW$, 
\bea
\cW (t_1, t_2, t_3 |\tau) 
=
\sum _{(m_r, n_r) \not= (0,0)} \delta _{m,0} \, \delta_{n,0} \, \prod _{r=1}^3 { \tau _2 \over \pi |m_r \tau + n_r |^2 - t_r \tau_2}
\eea
Recall that we use the notation $m=m_1+m_2+m_3$ and $n=n_1+n_2+n_3$ for the components of total momentum, and that the Kronecker $\delta$ factors enforce the vanishing of $m$ and $n$.

\sm

Setting $-t_r=M_r^2$ would give the 2-loop sunset diagram on a torus with Euclidean signature and modulus $\tau$ for three free scalar Green functions with masses $M_1^2, M_2^2$, and $M_3^2$. For fixed $\tau=\tau_1+i \tau_2$ with $\tau_2>0$, the above sum is absolutely convergent for small enough $t_1, t_2, t_3$. This may be established as follows.  Contributions to the sum with $m_r \not=0$ are absolutely convergent provided $|t_r| < \pi \tau_2$, while those with $m_r=0$ are absolutely convergent provided $|t_r| < \pi/\tau_2$. 
Thus we have convergence when $|t_i| $ is less than both $\pi \tau_2$ and $\pi/\tau_2$. This convergence is uniform in $\tau_1$ for fixed $\tau_2$, but fails to be uniform in $\tau$ throughout $\cM_1$, as the sum diverges for $\tau _2 \to \infty$.

\subsection{Differential equation for the generating function}

We shall now translate the algebraic representation of the Laplace-Beltrami operator $\Delta$ on the functions $C_{a,b,c}(\tau)$ given in (\ref{Lap})  into a representation on the function $\cW$ by a linear differential operator in $t_1, t_2, t_3$. To do so, we use the algebraic representation of $\Delta$ on $C_{a,b,c}$ to compute its action on $\cW$. This may be done term by term in (\ref{Lap}). A first example is given by the first term on the left side of (\ref{Lap}), for which we have, 
\bea
\sum _{a,b,c=1}^\infty  t_1 ^{a-1}  \, t_2 ^{b-1} \, t_3^{c-1} \, a(a-1) \, C_{a,b,c} 
=  t_1 \p_1 ^2 (t_1 \cW)
\eea
where we use the notation $\p_r = \p / \p t_r$ for $r=1,2,3$. Clearly, the terms given by $b(b-1)$
and $c(c-1)$ are obtained by permutations of $t_1, t_2, t_3$. A second example is provided by the first term on the right side of (\ref{Lap}), for which we have, 
\bea
\sum _{a,b,c=1}^\infty  t_1 ^{a-1}  \, t_2 ^{b-1} \, t_3^{c-1} \, ab \, C_{a-1,b+1,c} 
=
\sum _{a=0}^\infty  \sum _{b,c=1}^\infty  t_1 ^a  \, t_2 ^{b-2} \, t_3^{c-1} \, (a+1)(b-1) \, C_{a,b,c} 
\eea
This expression is compared with the action of the following differential operator on $\cW$,
\bea
\p_1 \p_2 (t_1^2 \cW) 
= 
\sum _{a,b,c=1}^\infty  \, (a+1) (b-1) t_1 ^{a}  \, t_2 ^{b-2} \, t_3^{c-1} \, C_{a,b,c}
\eea
The two expressions match, except for the contribution from $a=0$, which may be expressed purely in terms of Eisenstein series with the help of  the first equation in (\ref{3c2}). Note that it will involve the regularized quantity $E_1 ^\Lambda$, which in the end will consistently cancel out of all results, as we shall show explicitly below.  The second term on the first line of (\ref{Lap}) as well as the first two terms on the second and third lines may all be obtained by permutations of $t_1, t_2, t_3$. The same analysis applies to the remaining terms on the right side of (\ref{Lap}). Putting all together, we have, 
\bea
\Delta \, \cW -  \mL^2 \, \cW = \cR
\eea
where we have introduced the  following operators,
\bea
\mD  & = &  t_1 \p_1 + t_2 \p_2 + t_3 \p_3
\no \\
\mL^2 & = & ( \mD^2 + \mD ) 
+ ( t_1^2 + t_2 ^2 + t_3^2 -2 t_1t_2 -2t_2t_3 -2 t_3t_1 ) \Big (  \p_1 \p_2 +  \p_1 \p_3  + \p_2 \p_3  \Big )   
\eea
The right hand side function $\cR$ is given by, 
\bea
\cR = 
 \sum _{a,b=0}^\infty  
\Big ( t_1 ^a \, t_2 ^b +  t_2 ^a \, t_3 ^b +  t_3 ^a \, t_1 ^b \Big ) \cR^0 _{ab}
+ \sum _{a,b=0}^\infty 
\Big ( t_1 ^a \, t_2 ^b \, t_3 +  t_2 ^a \, t_3 ^b \, t_1 +  t_3 ^a \, t_1 ^b \, t_2 \Big ) \cR^1 _{ab} 
\eea
in terms of the following combinations of Eisenstein series, 
\bea
\cR^0 _{ab} & = & 3 a(b+1) E_{a+1} E_{b+2} + 3 (a+1)b E_{a+2} E_{b+1}  + (2 -a - b - 4ab) E_{a+b+3}
\no \\
\cR^1 _{ab} & = & ab \Big ( E_{a+2} E_{b+2} - E_{a+b+4} \Big ) 
\eea
Note that $\cR^0_{ab}$ and $\cR^1_{ab}$ are symmetric in $a,b$, and that they never involve $E_1^\Lambda$ in view of the fact that the corresponding coefficients that enter $\cR^0_{ab}$ for $a=0$ or $b=0$ cancel.

\subsection{Adapted coordinates and realization of $\mS_3$}

The operators $\mD$ and $\mL^2$ are manifestly invariant under rotations of the variables $t_1, t_2, t_3$ which leave the direction $(1,1,1)$ invariant. Cyclic permutations of these variables then correspond to 120 degree
rotations around this axis, while permutations with odd signature correspond to reflections through any plane that contains this axis. We shall now make a linear change of variables in which these transformations are realized naturally,
\bea
u & = & { 1 \over \sqrt{3}} \left ( t_1 + t_2 + t_3 \right )
\no \\
v & = & \sqrt{{2 \over 3}}  \left ( t_1 + \ep t_2 + \ep^2 t_3 \right ) \hskip 1in \ep = e^{2\pi i/3}
\no \\
\bar v & = & \sqrt{{2 \over 3}} \left ( t_1 + \ep^2 t_2 + \ep t_3 \right )
\eea
In terms of these variables, we have,
\bea
\label{inv}
 t_1^2 + t_2 ^2 + t_3^2 -2 t_1t_2 -2t_2t_3 -2 t_3t_1 & = & - u^2 + 2 v \bar v
 \no \\
 \p_1 \p_2 +  \p_1 \p_3  + \p_2 \p_3 & = & \p_u^2 - 2 \p_v \p_{\bar v}
 \eea
and the operators become,\footnote{We could  stay with real coordinates, and decompose $v$ into real coordinates $x,y$ by $v=(x+iy)/\sqrt{2}$. We would then have $\mD= u \p_u + x \p_x + y \p_y$ as well as $\mL^2= \mL_1^2 + \mL_2^2 - \mL_0^2$ with $\mL_1= u \p_x + x \p_u$, $\mL_2= u \p_y + y \p_u$ and $\mL_0=x \p_y - y \p_x $, so that each one of these operators is manifestly real, thereby justifying reference to the algebra $SO(2,1;\RR)$. The operators $\mL_1$ and $\mL_2$ are related to $\mL_\pm$ by $\mL_\pm = (\mL_1 \pm i \mL_2) /\sqrt{2}$.}
\bea
\mD  & = &  u \p_u + v \p_v + \bar v  \p_{\bar v} 
\no \\
\mL^2 & = & \mL_+ \mL_-  + \mL_- \mL_+  - \mL_0^2 
\eea
where the first order operators $\mL_\pm, \mL_3$ are expressed as follows,
\bea
\mL_+   =  u \p_{\bar v} + v \p_u \hskip 0.1in 
& \hskip 1in & 
[\mL_0, \mL_+ ] = + i \mL_+  
\no \\
\mL_-   =  u \p_v + \bar v \p_u \hskip 0.1 in
& \hskip 1in & 
[\mL_0, \mL_- ] = - i \mL_-  
\no \\
\mL_0  =  i v \p_v - i \bar v \p_{\bar v} 
& \hskip 1in & 
[\mL_+ , \mL_-  ] = - i \mL_0
\eea
The algebra is that of $SO(2,1)=SO(2,1; \RR)$, and $\mL^2$ is its quadratic Casimir operator.
Both combinations in (\ref{inv}) are invariant under $SO(2,1)$.
\sm

Under the permutation group $\mS_3$, the variables $u$, $v$ and $\bar v$ transform as, 
\bea
(123) (u, v, \bar v) = ( u, v, \bar v) \hskip 0.22in & \hskip 0.8in & (132) (u, v, \bar v) = (u, \bar v, v)  
\no \\
(231) (u, v, \bar v) =( u, \ep^2 v,  \ep \bar v ) && (213) (u, v, \bar v) = (u, \ep \bar v, \ep^2 v) 
\no \\
(312) (u,  v, \bar v) = (u, \ep v, \ep^2 \bar v) && (321) (u, v,\bar v) = (u, \ep^2 \bar v, \ep v) 
\eea
The generating function $\cW$ is real; it is a sum of monomials of the form $u^\alpha v^\beta \bar v ^\gamma$; and it is invariant under the group $\mS_3$. We shall now use this information to constrain the form of $\cW$ when expressed in terms of the variables $u, v, \bar v$. To do so, we associate to an arbitrary monomial
$u^\alpha v^\beta \bar v ^\gamma$ of weight $w=\alpha + \beta +\gamma +3$ a unique corresponding $\mS_3$-invariant polynomial obtained by summing over the six images of the monomial under permutations $\mS_3$,
\bea
\mS_3 \, : \, u^\alpha \, v^\beta \, \bar v ^\gamma \quad 
 \longrightarrow \quad u^\alpha \, 
\Big ( v^\beta \bar v ^\gamma + v^\gamma \bar v ^\beta \Big ) 
\Big ( 1 + \ep ^{\beta - \gamma} + \ep ^{2 (\beta -\gamma)} \Big ) 
\eea
Clearly, the last parenthesis vanishes when $\beta - \gamma \not \equiv 0 \, (\rm{mod} \, 3)$,
in which case the monomial produces no $\mS_3$-invariant contribution to $\cW$.
Therefore, all the non-trivial $\mS_3$-invariant contributions of weight $w$ to $\cW$ must take the form,
\bea
 u^{s-1-3\mp}  (v \bar v)^{\mm-1}  \Big ( v^{3 \mp } + \bar v ^{3 \mp } \Big ) 
 \hskip 1in 
 w= s+2 \mm
\eea
where $\mm \geq 1$ and $\mp \geq 0$ with   $s-1 \geq 3\mp$.
Changing basis we can equivalently work with the following set of functions, 
\bea
 u^{s-1-3\mp} (u^2-2v \bar v)^{\mm-1} \Big ( v^{3 \mp } + \bar v ^{3 \mp} \Big ) 
\eea
This basis has the  advantage that the combination $u^2-2v \bar v$ is invariant under $SO(2,1)$, so that $\mL_\pm (u^2-2v \bar v)= \mL_0 (u^2-2v \bar v)=0$. 

\sm

The image of a monomial $u^{s-1-3\mp} (u^2-2v \bar v)^{\mm-1}  ( v^{3 \mp } + \bar v ^{3 \mp} ) $ under $\mL_0$ is not invariant under $\mS_3$, so that $\mL_0$ does not have a good action on the $\mS_3$-invariant  functions $\cW$. However, the action of the operator $\mL_0^2$ is well-defined on this space. Thus, we shall choose to simultaneously diagonalize $\mD, \mL^2, \mL_0^2$.

\subsection{Diagonalizing $\mD, \mL^2, \mL_0^2$}

We choose to simultaneously diagonalize the three commuting operators, $\mD, \mL^2$, and $\mL_0^2$, and denote the corresponding eigenvalues respectively by $(w-3)$, $s(s-1)$, and $-9 p^2$ for $p \geq 0$,
\bea
\label{eigen}
\mD \, \cW_{w;s;\mp} & = & (w-3) \, \cW _{w; s;\mp}
\no \\
\mL^2 \, \cW _{w;s;\mp} & = & s(s-1) \, \cW_{w;s;\mp} 
\no \\
\mL_0^2 \, \cW _{w;s;\mp} & = & - 9 \, \mp^2 \, \cW_{w;s; \mp} 
\eea
The first and last eigenvalue equations in (\ref{eigen}) are manifestly satisfied by construction. 
The representation theory of $SO(2,1)$ will relate these eigenvalues, including $s$, as we shall establish shortly. Weight $w$ signifies that the eigenfunction $\cW_{w;s; \mp}$ is a homogeneous polynomial of degree $w-3$  in the coordinates $(t_1,t_2,t_3)$, or equivalently in $(u,v,\bar v)$. Therefore, the corresponding representation of $SO(2,1)$ will be finite-dimensional.

\sm

The eigenfunctions $\cW_{w;s;\mp}$ take the form, 
\bea
\cW_{w;s;\mp} (u,v,\bar v) =
\sum _{k=\mm_- }^{\mm_+}  \cF_{w;s;\mp} (k) (u^2-2v \bar v)^{k-1} u^{w-2k-1-3\mp}  \Big ( v^{3 \mp } + \bar v ^{3 \mp} \Big ) 
\eea
for as yet undetermined real coefficients $\cF_{w;s;\mp} (k)$.
The integers $\mm_\pm$ are subject to the conditions $1 \leq \mm_-
\leq \mm_+ \leq w-3\mp-1$. We shall assume that $\mm_-$ is the lowest
value for $k$ with non-zero contribution, while $\mm_+$ is the highest
value, so that $\cF_{w,s,\mp} (\mm_\pm ) \not= 0$, but otherwise
$\mm_\pm$ remain to be determined.  It is straightforward to compute
the operator $\mL^2$ on these functions. The eigenvalue equation for
$\mL^2$ in (\ref{eigen}) then reduces to the following equations. For
$\mm_- < k < \mm_+$ we have, 
\bea
&& 
(w-s-2k)(s+w-2k-1) \cF_{w,s,\mp}(k) 
\no \\ && \hskip 0.4in 
= (w-2k-3\mp+1)(w-2k-3\mp) \cF_{w,s,\mp}(k-1) 
\eea
while for the extremities, we have the conditions (using the assumption $\cF_{w,s,\mp} (\mm_\pm ) \not= 0$), 
\bea
(w-s-2\mm_-)(s+w-2\mm_- -1) & = & 0
\no \\
(w-2\mm_+ -3\mp -1) (w-2\mm _+ -3\mp-2)  & = & 0
\eea
The first equation forces two solutions for $s$ which are equivalent as they yield the same eigenvalue $s(s-1)$ for $\mL^2$. We shall make the choice $s=w-2\mm_-$.  The solution of the second equation is easily identified as well, and assembling the full solution, we have, 
\bea
\label{mm}
s= w-2 \mm_-  \hskip 1in \mm_+ = \left [ { w-3\mp-1 \over 2} \right ]
\eea
with $\mm_- \geq 1$, $s \geq 1$,  and $\mp \geq 0$.

\subsection{Proof of Theorem 1}

The structure of the spectrum of the Laplace operator, which had been inferred on the basis of MAPLE calculations in Theorem 1 of section \ref{sec34}, may now be proven with the help of the above results on the structure of the generating function $\cW$. In fact, the structure may be read off from the relations of (\ref{mm}). 

\sm

For given weight $w$, the allowed eigenvalues $s(s-1)$ of the Laplacian are governed by $s=w-2 \mm_-$ with $\mm_- = 1, \cdots , [(w-1)/2]$. Identifying $\mm_-$ with $\mm$ in Theorem 1, we see that we have the same set of eigenvalues for given weight $w$. Their degeneracy may be read off from (\ref{mm}) as well. Combining the condition $\mm_- \leq \mm_+$ with the second equation in (\ref{mm}), and expressing $w$ in terms of $s$ and $\mm_-$ gives, 
\bea
\mm_- \leq \left [ { s+2 \mm_- -3 \mp -1 \over 2} \right ]
\eea
Subtracting $\mm_-$ on both sides then gives a condition which equivalent to $0 \leq 3\mp \leq s-1$, which in turn implies that the degeneracy of the eigenvalue $s(s-1)$ is given by $[(s+2)/3]$, as indeed announced in Theorem 1. 

\sm

In addition, we obtain an explicit form for the eigenfunctions given by their coefficients,
\bea
\label{rec}
\cF_{w;s;\mp}(k) = 
 { \Gamma \left ( k+{ 3\mp-w+1 \over 2} \right ) \Gamma \left ( k+ { 3\mp-w +2 \over 2} \right ) 
\over \Gamma \left ( k+ {s-w +2 \over 2} \right ) \Gamma \left ( k - {s-w -3 \over 2} \right ) } \, 
\cF_{w;s;\mp} ^0
\eea
where $\cF_{w;s;\mp} ^0$ is an arbitrary constant, and $k$ is in the range $\mm_- < k \leq \mm_+$.
 
 \sm
 
Note that the map $\mE$ of section \ref{sec37} corresponds to multiplication by a factor $(u^2 - 2 v \bar v)$, which leaves the $SO(2,1)$ representation unchanged, but increases the weight by 2.

\subsection{Illustration for low weight $w$}

Taking the normalization $\cF_{w;s;\mp}(\mm_-)=1$, we may now write down the basis functions 
$\cW_{w;s;\mp}(u,v,\bar v)$ up to weight $w=6$,
\bea
\cW_{3;1;0} & = & 1
\no \\
\cW_{4;2;0} & = & u
\no \\
\cW_{5;3;0} & = & u^2 - { 1 \over 3}  (u^2 - 2 v \bar v) 
\no \\
\cW_{5;1;0} & = & u^2 - 2 v \bar v
\no \\
\cW_{6;4;1} & = & v^3 + \bar v^3
\no \\
\cW_{6;4;0} & = & u^3 -{ 3 \over 5} u (u^2 - 2 v \bar v) 
\no \\
\cW_{6;2;0} & = & u (u^2 - 2 v \bar v) 
\eea
The expansion coefficients of $\cW$ in this basis are the eigenfunctions $\mC_{w;s;\mp} (\tau)$ referred to in (\ref{1a4}) of the Introduction, so that the generating function may be written as follows,
\bea
\cW(t_1, t_2, t_3|\tau) 
= \sum _{w=3} ^\infty ~ \sum _{\mm=1}^{[{w-1 \over 2} ]} ~ \sum _{\mp=0} ^{[{ w+2 \over 3} ]-\mm }
\cW_{w;w-2\mm;\mp} (u,v,\bar v) \, \mC _{w;w-2\mm;\mp}(\tau)
\eea
where the eigenvalue $s$ is given by the combination $s=w-2\mm$. 
The functions $\mC_{w;s,\mp} (\tau)$ may  be read off by decomposing $\cW$ onto the basis functions $\cW_{w;s;\mp}$, and we find, 
\bea
\mC_{3;1;0} & = & C_{1,1,1}
\no \\
\mC_{4;2;0} & = & \sqrt{3} C_{2,1,1}
\no \\
\mC_{5;3;0} & = & {3 \over 4 } \Big ( 2 C_{3,2,1}+ C_{2,2,1} \Big ) 
\no \\
\mC_{5;1;0} & = & \half C_{2,2,1}
\no \\
\mC_{6;4;1} & = & {\sqrt{3} \over 36} \Big ( 3 C_{4,1,1} - 3 C_{3,2,1} + C_{2,2,2} \Big )
\no \\
\mC_{6;4;0} & = & {\sqrt{3} \over 36} \Big ( 30 C_{4,1,1} + 24 C_{3,2,1} + C_{2,2,2} \Big )
\no \\
\mC_{6;2;0} & = & { \sqrt{3} \over 10} \Big ( 4 C_{3,2,1} + C_{2,2,2} \Big )
\eea
Comparing the entries with $s=1,2, 3$ with the expression for $s=1$ given in (\ref{3d4}), $s=2$ given in (\ref{3d6}), and $s=3$ given in (\ref{3d6a}),  we find perfect agreement. For $s=4$ we find perfect agreement with (\ref{3d7}) upon taking linear combinations which had been left unspecified in (\ref{3d7}), but which are here dictated by the fact that the eigenfunctions also diagonalize $\mL_0^2$.

\section{Modular functions in weight-4 interactions}
\setcounter{equation}{0}
\label{sec4}

In this section we shall consider the weight-4 Feynman diagrams that contribute to $j^{(2,0)}(\tau)$.  Their integrals over $\cM_1$ will determines the coefficient of the one-loop $D^8 \cR^4$ interaction.   

\sm

The weight-4 diagrams that will enter this calculation are $D_2^2$, $D_{1,1,1,1}$, and $D_4=C_{1,1,1,1}$.
The precise rational coefficients with which they contribute to $j^{(2,0)}$ will be listed below in   (\ref{weight4comb}).  The modular function $D_2$ equals the Eisenstein series $ \tE_2$, while  $D_{1,1,1,1}$ equals $ \tE_4$, but the modular function $D_4 = C_{1,1,1,1}$ has a novel form beyond the framework of the functions $C_{a,b,c}$ studied in the preceding section, though it does fit into the framework of the functions defined in (\ref{1a1})  of the Introduction for the case for $\rho=4$. 

\sm

We have not found a useful closed algebraic representation for the Laplace operator acting on the modular functions $C_{a_1, \dots, a_\rho}(\tau)$ for $\rho \geq 4$. Thus, we do not have generalization of the powerful tool  (\ref{Lap}) at our disposal. It would be very valuable, of course, to find such an algebraic representation, if it exists at all.
Instead, our procedure will be based on expanding the multiple sums that define the modular function $D_4$ in the vicinity of the cusp in the large $\tau_2$ limit, and using the resulting expansion to conjecture the form of the Laplace operator acting on $D_4$. In section \ref{sec5} we shall extend this procedure to modular functions of weight 5.

\subsection{Asymptotic expansion near the cusp $\tau \to i \infty$}

We shall begin by producing the general form of the expansion near the cusp $\tau \to i \infty$.
To minimize the occurrence of factors of $\pi$ throughout, it will be  convenient to express the asymptotic expansion  of the modular functions $C$, $D$, and $E$ in terms of the parameter,
\be
y=\pi \tau_2
\eea
The asymptotic expansion near the cusp $y \to \infty$ of a modular function $D$ (and similarly of the functions  $C$ and $E$) takes on the following form, 
\bea
D(\tau) = \sum _{k, \bar k =0} ^\infty \cD^{(k,\bar k )} (y) \, q^k \, \bar q^{\bar k}
\hskip 1in 
\overline{\cD^{(k,\bar k)} (y) } = \cD^{(\bar k, k)} (y)
\eea
where $q=e^{2 \pi i \tau}$ and each coefficient function $\cD ^{(k, \bar k)}(y)$ consists of a finite sum of integer powers of $y$, namely it is a  Laurent polynomial in $y$ of finite degree.  The series is absolutely convergent at any point in the interior of $\cM_1$. Our conjectured relations will be verified by comparing the pure power-behaved terms for $k=\bar k=0$ as well as the leading exponential corrections for $k+\bar k=1$.
For this purpose, it will be convenient to have also the expansion of the Laplacian of $D(\tau)$ available, which is given by, 
\bea
\Delta D(\tau) & = & \sum _{k, \bar k =0} ^\infty (\Delta \cD) ^{(k,\bar k )} (y) \, q^k \bar q^{\bar k}
\no \\
(\Delta \cD) ^{(k,\bar k )} (y) & = & y^2 \p_y^2 \cD^{(k,\bar k )} (y)
- 4  (k+\bar k) y^2 \p_y \cD^{(k,\bar k )} (y) + 16 k \bar k y^2 \cD^{(k,\bar k )}(y)
\eea
The Fourier transform of the function $D(\tau)$ in the variable $\tau_1$  at fixed value of $y$ is then obtained by the following  infinite sums, 
\bea
\int _{\RR/\ZZ} d \tau_1 \, e^{-2\pi i N \tau_1} \, D(\tau)
=
\sum _{ k, \bar k =1}^\infty \delta _{k-\bar k,N}\,  \cD^{(k,\bar k )} (y) \, |q|^{k + \bar k}
\eea
An important special case is the expansion is for the Eisenstein series $E_n$ itself given in (\ref{a1}) where the $N$-th Fourier modes gets a contribution from the single power $q^N$.  More generally an infinite number of coefficient functions $\cD^{(k,\bar k)}$ (with $k- \bar k =N$) contribute to a given mode.  These appear to correspond to the effect of an infinite number of world-sheet ``instanton--anti-instanton'' pairs (an interpretation which deserves further study).

\subsection{The case $C_{2,1,1}=D_{2,1,1}$}

It will prove of interest to first consider a known example and examine the properties of the weight-4 modular function $C_{2,1,1} = D_{2, 1,1}$ with the help of the asymptotic expansion procedure discussed above. Its Feynman diagram is given as follows, 
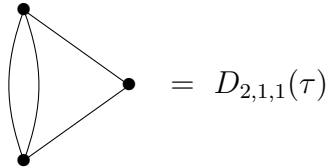
\begin{center}
\tikzpicture[scale=2.0]
\begin{scope}[xshift=4.5cm]
\draw (0,0) node{$\bullet$} ;
\draw (0,1) node{$\bullet$} ;
\draw (0.7,0.5) node{$\bullet$} ;
\draw (0,0) to [bend left=20] (0,1) ;
\draw (0,0) to [bend left=-20] (0,1) ;
\draw (0,0) to (0.7,0.5);
\draw (0,1) to (0.7,0.5);
\draw(1.5,0.5) node{$= \ D_{2,1,1}(\tau)$};
\end{scope}
\endtikzpicture
\end{center}
The function $C_{2,1,1}$ satisfies the Laplace equation  (\ref{c211lap}), but will  not arise directly in the expression for $j^{(2,0)}$ to be given in (\ref{weight4comb}).  We can usefully rewrite the associated Laplace-eigenvalue equation  as,
\be
(\Delta -2) \left ( C_{2,1,1} - \frac{9}{10}   \tE_4 \right ) = -  \tE_2^2
\label{newc211}
\ee
The power-behaved terms $\cC_{2,1,1}^{(0,0)} (y)$ in the  expansion of  $C_{2,1,1}$   near the cusp $y=\infty$ can be obtained from this equation  in a straightforward manner by expanding the source term and matching the power-behaved terms. The  $y^{-1}$ and  $y^2$ terms are in the kernel of $(\Delta-2)$ and are determined by enforcing appropriate boundary conditions, or alternatively may be determined by carrying out explicitly the multiple sum which defines $C_{2,1,1}$. One finds,
\be
\label{powerc211}
\cC_{2,1,1} ^{(0,0)} =  \frac{2   y^4 }{14175}  +  \frac{ \zeta(3) y}{45}  + \frac{5  \zeta(5)}{12 y}  -\frac{ \zeta(3)^2}{4 y^2} +  \frac{9 \zeta(7)}{16  y^3} 
 \ee
 The coefficient $ \cC_{2,1,1} ^{(1,0)}$ of the leading exponential correction terms may be obtained by considering the first Fourier mode of (\ref{newc211}) truncated to its lowest order exponential part, and doing so results in the  expression,
\bea
 \cC_{2,1,1} ^{(1,0)} = {y \over 45} +{1 \over 3}  + {11 \over 12 y }  - {\zeta(3) \over 2 y^2} + { 9 \over 8 y^2} +{9 \over 16 y^3} 
 \label{firstmode}
\eea
Here we have already enforced the condition of power growth near the cusp to set to zero a contribution which would grow  exponentially at the cusp. A  term proportional to $1+1/(2y)$ in the formula above is in the kernel of $(\Delta -2)$ and may be determined by boundary conditions. Alternatively, this expansion may be obtained by direct approximation of the multiple sums that define $C_{2,1,1}$.  The power-behaved terms were analyzed in this manner in \cite{Green:2008uj}, and the two approaches are found to agree. Here we have also determined the leading exponential correction.  

\sm

The above exercise is somewhat redundant in the case of $C_{2,1,1}$ since we have derived the Laplace equation that it satisfies.  However, the direct expansion of other $C$- and $D$-functions will prove significant in the sequel where the Laplace equations are conjectural. We now turn to consider the function $D_4$, which will enter in the expression for $j^{(2,0)}$ in (\ref{weight4comb}).

\subsection{The case $D_{4} = C_{1,1,1,1}$}

The Feynman diagram representation of $D_4$ is as follows,
\begin{center}
\tikzpicture[scale=1.7]
\scope[xshift=-5cm,yshift=-0.4cm]
\draw (0,0.5) node{$\bullet$}   ;
\draw (1,0.5) node{$\bullet$} ;
\draw (0,0.5) to [bend left=60] (1,0.5) ;
\draw (0,0.5) to [bend left=20] (1,0.5) ;
\draw (0,0.5) to [bend left=-20] (1,0.5) ;
\draw (0,0.5) to [bend left=-60] (1,0.5) ;
%
\draw(2,0.5) node{$=  \ D_{4}(\tau) $};
\endscope
\endtikzpicture
\end{center}
Although we have not determined the functional form of $D_4(\tau)$ directly  from its expression as a multiple sum, we will arrive at a convincing conjecture for its form,  based on examining its behavior near the large $y$ cusp.  The power-behaved part of the expansion was given already in equation (B.9) of \cite{Green:2008uj},
\be
\label{powerd4}
\cD_4^{(0,0)} (y) = \frac{  y^4}{945}  + \frac{2  \zeta(3) y}{3}   + \frac{10 \zeta(5)}{ y} - \frac{3  \zeta(3)^2}{y^2} + \frac{9  \zeta(7)}{4 y^3}
\ee
The leading exponential corrections may be evaluated in a similar manner, as will be detailed in appendix \ref{secB}, and we find, 
\bea
\label{expd4}
\cD_{4}^{(1,0)} (y) = { 4 y^2 \over 15 }  +{2  y \over 3}  +2 + { 4 \over y} +{12 \zeta (3) \over y} 
-{ 6 \zeta (3) \over y^2} +{ 9 \over 2 y^2}   +{ 9 \over 4 y^3} 
\eea 
Since this function is manifestly real, we have $\cD_{4}^{(0,1)} (y) =\cD_{4}^{(1,0)} (y)$.
It may be verified by explicit calculation that the above pure power part and leading exponential corrections in the expansion of $D_4$ satisfy the pure power part and leading exponential corrections of the following equation, 
\be
\label{d4laplace}
(\Delta -2) \left ( D_4 - 3   \tE_2^2 - \frac{18}{5}   \tE_4 \right )  = - 24   \tE_2^2
\ee
Based on this evidence, along with its overall modularity and weight, we conjecture that this equation holds true exactly for all $\tau$ in the upper half plane.

\sm

Assuming the conjectured equation (\ref{d4laplace}) holds true exactly leads to a new algebraic equation involving $D_4$. Indeed, eliminating the $ \tE_2^2$ term between the right hand side of  (\ref{d4laplace})
and the right hand side of (\ref{newc211}) leads to the following Laplace-eigenvalue equation, 
\bea
(\Delta -2) \left ( D_4 - 24 C_{2,1,1} +18  \tE_4 - 3  \tE_2^2 \right )=0
\eea
Given that the argument of the above equation has power behavior near the cusp, we know that its solution must be proportional to $ \tE_2$. Matching the pure power behavior suffices, however, to show that the $ \tE_2$ contribution is in fact absent, so that the unique solution is given by the algebraic relation,
\bea
\label{d4def}
D_4 = 24 C_{2,1,1} -18  \tE_4 +3  \tE_2^2
\eea
In the next section we shall produce analogous conjectures for all the weight-5 Feynman diagrams by expressing them in  terms of modular functions $C_{a,b,c}$ of weight $a+b+c=5$ and Eisenstein series.

\section{Modular functions in weight-5 interactions}
\setcounter{equation}{0}
\label{sec5}

We will now consider the weight-5 modular functions $D_5 = C_{1,1,1,1,1}$, $D_{3,1,1} = C_{2,1,1,1}$, $D_{2,2,1}$ and $D_{2,1,1,1} = C_{3,1,1}$ that contribute non-trivially to $j^{(1,1)}$ and will contribute to the interaction of order $D^{10}\, \cR^4$.  As in the last section, we will not determine these functions directly from their definition in terms of integrated powers of Green functions, but we will study their asymptotic behaviour near the cusp.  In this manner we will be led to relationships that will express the functions $D_5$, $D_{3,1,1}$ and $D_{2,2,1}$ in terms of $C_{3,1,1}$, which satisfies the Laplace equation  (\ref{D2111}). 
We will now consider each of these in turn.

\subsection{The case $D_{2,1,1,1} = C_{3,1,1} $}
 \label{d2111}

The Feynman diagram representation for $C_{3,1,1}$ is as follows,
 \begin{center}
\tikzpicture[scale=1.7]
\scope[xshift=-5cm,yshift=-0.4cm]
\draw (0,0.5) node{$\bullet$};
\draw (1,0.5) node{$\bullet$};
\draw (.33,0.72) node{$\bullet$};
\draw (0.66,0.72) node{$\bullet$};
\draw (0,0.5) to [bend left=60] (1,0.5) ;
\draw (0,0.5) to  (1,0.5) ;
\draw (0,0.5) to [bend left=-60] (1,0.5) ;
%
\draw(2,0.5) node{$=\  C_{3,1,1}(\tau) $};
\endscope
\endtikzpicture
\end{center} 
 This modular function satisfies the Laplace-eigenvalue equation  (\ref{D2111}), which may be re-written in the following form,
 \bea
 \label{D2111summ}
(\Delta -6)\,\left(C_{3,1,1}- \frac{43}{35}   \tE_5 + \frac{ \zeta(5)}{60} \right) =  - 4   \tE_2   \tE_3
 \eea  
For later reference  we give the pure power part (which was also given in (B.53) of \cite{Green:2008uj}),   
\bea
 \label{D2111expan}
 \cC_{3,1,1} ^{(0,0)} (y) 
 =
 \frac{2  y^5}{155925} 
  +\frac{2   \zeta (3) y^2 }{945} 
   -\frac{\zeta (5)}{180}
    +\frac{7 \zeta (7)}{16 y^2}
 -\frac{\zeta (3)  \zeta (5)}{2 y^3}
 +\frac{43  \zeta (9)}{64 y^4}
 \eea
Furthermore, the leading exponential correction term can be extracted directly from the definition of $C_{3,1,1}$ as a multiple sum, or by considering  the first mode of the equation (\ref{D2111summ}).  The results agree and take the form,
\bea
\label{D2111mode1} 
\cC_{3,1,1} ^{(1,0)}  (y) 
&=&
{ 2 y^2 \over 945} +{ y \over 45} +{2 \zeta (3) \over 3} -{ 64 \over 45} 
+ { \zeta (3) \over  y} -{ 43 \over 24  y} 
- {  \zeta (3) \over 2y^3} - { \zeta (5) \over 2y^3} 
\no \\ && 
+ { 43  \over 32 y^3} + { 43 \over 64 y^4}
+ c_5 \left ( 4 +{6 \over y} + { 3 \over y^2} \right )
\eea
The constant $c_5$ is not determined by the mode expansion of (\ref{D2111summ}), but may be evaluated directly from the sum representation of $C_{3,1,1}$  and we find, 
\bea
\label{c5}
c_5 = - { 1 \over 6} \zeta (3) + { 19  \over 48}
\eea
Since $\cC_{3,1,1}^{(1,0)} $ is real, we have $\cC_{3,1,1} ^{(0,1)}  (y) =\cC_{3,1,1} ^{(1,0)}  (y) $.

 \subsection{The case $D_5 = C_{1,1,1,1,1} $}
 \label{d5} 
 
 The Feynman diagram representation of $D_5$ is given as follows,
\begin{center}
\tikzpicture[scale=1.7]
\scope[xshift=-5cm,yshift=-0.4cm]
\draw (0,0.5) node{$\bullet$}   ;
\draw (1,0.5) node{$\bullet$} ;
\draw (0,0.5) to [bend left=60] (1,0.5) ;
\draw (0,0.5) to [bend left=20] (1,0.5) ;
\draw (0,0.5) to  (1,0.5) ;
\draw (0,0.5) to [bend left=-20] (1,0.5) ;
\draw (0,0.5) to [bend left=-60] (1,0.5) ;
%
\draw(2,0.5) node{$= \  D_{5}(\tau) $};
\endscope
\endtikzpicture
\end{center}
Here again we have not obtained a Laplace equation for $D_5$ directly from its definition as a multiple sum.  However, as with $D_4$,  we conjecture a Laplace equation for $D_5$  based on its asymptotic expansion near the cusp.   The power part of the expansion in this limit was given in (B.10) of \cite{Green:2008uj}, and takes the form,
 \bea
 \label{powerd5}
 \cD_5^{(0,0)} (y) & = & 
 \frac{4 y^5}{18711} 
 +\frac{10 y^2 \zeta (3)}{27} 
 +\frac{95  \zeta (5)}{6} 
 +\frac{10 \zeta (3)^2}{ y}  
 \no \\ && 
 +\frac{105 \zeta  (7)}{4 y^2} 
 -\frac{45  \zeta (3)  \zeta (5)}{2 y^3} 
 + \frac{225 \zeta (9)}{16 y^4} 
 \eea
while  the leading exponential correction takes the form, 
\bea
\label{expd5}
 \cD_5^{(1,0)} (y) & = & 
 {8 y^3 \over 189} 
 + {10 y^2 \over 27} 
 +{5 y \over 3} 
 + 60  \zeta (3) 
 - {535 \over 6} 
+{ 80  \zeta (3) \over y} 
- { 255  \over 2 y} 
\no \\ &&
+{15  \zeta (3) \over y^2} 
+{15  \zeta (5) \over y^2}
-{45 \over  y^2}
 -{ 45 \zeta (3) \over 2 y^3} 
 - { 45 \zeta (5) \over 2 y^3} 
 +{ 225 \over 8  y^3} +{ 225 \over 16 y^4}
\no \\ && 
+ 60 \, c_5 \left ( 4  +{6  \over y} 
+ { 3 \over y^2} \right )
\eea
where the constant $c_5$ was determined from the sum representation of $D_5$ in appendix \ref{secB}, and is found to be given by (\ref{c5}). Since $\cD_5^{(1,0)}(y)$ is manifestly real, we have $\cD_{5}^{(0,1)} (y) =\cD_{5}^{(1,0)} (y)$.

\sm

If we now assume that $D_5$ is related to a superposition of the other weight-5 functions that arise in the four-graviton amplitude, then one is led to conjecture the following Laplace-eigenvalue relation,
\bea
(\Delta - 6) \left ( D_5-10\tE_2C_{1,1,1} \right ) = 360 \tE_5 - 240 \tE_2\tE_3-90\zeta (5)
\eea
Eliminating the product $\tE_2\tE_3$ between this equation and the Laplace-eigenvalue equation for $C_{3,1,1}$
established in (\ref{D2111summ}), and moving $\tE_5$ under the operator $\Delta -6$ using the eigenvalue equation 
$(\Delta - 20)\tE_5=0$, we obtain the following relation, 
\be
\label{laplaceD5}
(\Delta-6) \Big (D_5 -60 C_{3,1,1} -10  \tE_3\, \tE_2 + 48  \tE_5 -16 \zeta(5) \Big ) =0 
\ee
One verifies that within the pure power and leading exponential correction approximation, the above equation is indeed satisfied. With this conjecture (and noting that the large $y$ expansion  fixes the coefficient of the solution of the homogeneous equation to be zero)   we conclude that $D_5$ is given in terms of $D_{2,1,1,1}= C_{3,1,1}$ by the following algebraic relation, 
 \bea
 \label{D5}
 D_5 =60 C_{3,1,1} + 10  \tE_2 C_{1,1,1} -48  \tE_5 +16 \zeta(5) 
\eea
which bears a striking similarity with the corresponding equation for $D_4$ in (\ref{d4def}).

 \subsection{The case $D_{3,1,1} = C_{2,1,1,1}$}
 \label{sec:D311}


\begin{center}
\tikzpicture[scale=2]
\begin{scope}[xshift=4.5cm]
\draw (0,0) node{$\bullet$} ;
\draw (0,1) node{$\bullet$} ;
\draw (0.7,0.5) node{$\bullet$} ;
\draw (0,0) to [bend left=40] (0,1) ;
\draw (0,0) to (0,1) ;
\draw (0,0) to [bend left=-40] (0,1) ;
\draw (0,0) to (0.7,0.5);
\draw (0,1) to (0.7,0.5);
\draw(1.5,0.5) node{$= \ D_{3,1,1}(\tau)$};
\end{scope}
\endtikzpicture
\end{center}

Once again, we have evaluated the complete pure power part for this function in the large-$y$ limit (which was partially given in  (B.34) of \cite{Green:2008uj}).  This has the form, 
\be
\label{D311}
\cD_{3,1,1}^{(0,0)} (y) = 
\frac{2y^5}{22275}
+ \frac{y^2 \zeta(3)}{45}
+ \frac{11  \zeta(5)}{60}  
+ \frac{105  \zeta(7)}{32 y^2}
- \frac{3  \zeta(3) \zeta(5)}{2 y^3} 
+ \frac{81 \zeta(9)}{64 y^4} 
  \ee  
The asymptotic expansion leads to  strong constraints on the possible form of the modular function, and  we are led to conjecture the following algebraic relation involving $D_{3,1,1}$, 
\be
\label{D113twosumm}
40 D_{3,1,1}= 300 C_{3,1,1} + 120  \tE_2 \tE_3 -276  \tE_5    + 7  \zeta(5) 
\ee
We have  verified this equation only to pure power  order.

\subsection{The case $D_{2,2,1}$}

The Feynman diagram representation od $D_{2,2,1}$ is given as follows,
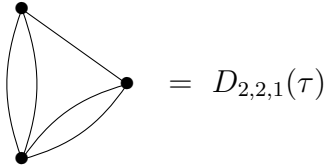
\begin{center}
 \tikzpicture[scale=2.]
\begin{scope}[xshift=4.5cm]
\draw (0,0) node{$\bullet$} ;
\draw (0,1) node{$\bullet$} ;
\draw (0.7,0.5) node{$\bullet$} ;
\draw (0,0) to [bend left=20] (0,1) ;
\draw (0,0) to [bend left=-20] (0,1) ;
\draw (0,0) to [bend left=20] (0.7,0.5);
\draw (0,0) to [bend left=-20] (0.7,0.5);
\draw (0,1) to (0.7,0.5);
\draw(1.5,0.5) node{$= \ D_{2,2,1}(\tau)$};
\end{scope}
\endtikzpicture
 \end{center}
This is an example of a modular function that is not of the form $C_{a_1, \dots, a_\rho}$ for any value of $\rho$.  However, we will see that its asymptotic behavior leads us to conjecture a relation again to the modular function $D_{2,1,1,1} = C_{3,1,1}$ of weight 5, along with Eisenstein series.  In this case the expansion in the large $y$ limit gives (noting some errors in the terms in  (B.35) of \cite{Green:2008uj}), 
 \bea
 \label{powerd221}
\cD_{2,2,1} ^{(0,0)} (y) 
= 
\frac{8 y^5}{467775}
+\frac{4   \zeta (3) y^2}{945}
+\frac{13 \zeta (5)  }{45} 
+ \frac{7 \zeta(7) }{8  y^2} 
- \frac{\zeta(3)\zeta(5) }{y^3}  
+ \frac{9  \zeta(9) }{8 y^4}
 \eea
The expansion leads to conjecture the following Laplace-eigenvalue equation, 
\be
\label{possD221}
(\Delta-6) \left(D_{2,2,1} - \frac{72}{35}  \tE_5 - \frac{4}{15} \zeta(5) \right)= - 
 8   \tE_3   \tE_2
\ee
Combining  (\ref{D2111summ}) with (\ref{possD221})   leads to the following conjectured algebraic relation
 \be
10 D_{2,2,1} = 20C_{3,1,1} - 4   \tE_5 +3  \zeta(5)
 \label{D221sol}
 \ee
This equation may be used to predict the leading exponential corrections  $\cD_{2,2,1} ^{(1,0)} (y) =\cD_{2,2,1} ^{(0,1)} (y) $,
and we find, 
\bea
\label{expD221}
\cD_{2,2,1} ^{(1,0)} (y) =
{ 4 y^2 \over 945} +{2 y \over 45} +{13 \over 45} +{1 \over y} -{ \zeta (3) \over y^2} +{ 2 \over y^2}
- {\zeta (3) \over y^3} - { \zeta (5) \over y^3} +{9 \over 4 y^3} +{9 \over 8 y^4}
\eea
By directly evaluating this sum, we have checked in appendix \ref{secC} that all terms indeed match. 

\sm

This concludes our study of the weight-5 modular functions that contribute to the higher derivative interaction $D^{10} \cR^4$, and we shall now proceed in the next section to providing further numerical evidence for the conjectured relations, and to evaluating the integrals over $\cM_1$ of all these contributions in section \ref{sec7}.

\section{Low order coefficients in the genus-one expansion}
\setcounter{equation}{0}
\label{sec7}

We will now obtain the integrated coefficients of some low order terms in the low-energy expansion of the amplitude (\ref{2a4}).  The coefficient of a term of order $\sigma_2^p\, \sigma_3^q$  is obtained by integrating $j^{(p,q)}(\tau)$, defined in (\ref{2b6}), over the fundamental domain, ${\cal M}_1$.  Since  $j^{(p,q)}(\tau)$ is   $\cO(\tau_2^a)$ as $\tau_2\to \infty$, with $a\ge 1$, such integrals diverge.  Our discussion will begin with a brief review  of treatment of these divergences, which are related to the occurrence of non-analytic threshold terms in the expansion.    We will also review the explicit determination of the coefficients of terms with weights $w\le 3$ before determining the  coefficients of the weight 4 and weight 5  interactions\footnote{The weight-4 and weight-5  results correct the discussion in the earlier literature \cite{Green:1999pv,Green:2008uj} where only the asymptotic behavior at large $y$ of the relevant modular functions was evaluated and  the values of the coefficients that follow from integration over $\tau$  were consequently not correctly determined.}.

 \subsection{Separation of analytic and non-analytic terms}
 \label{sec:structuref}
 
An important new issue appears in considering the low energy expansion of the amplitude beyond the tree-level approximation.  This concerns the appearance of physical threshold singularities that are branch cuts in the Mandelstam variables, and correspond to non-local contributions to the effective action originating from integrating out massless modes.    In order to make sense of the power series expansion (or local terms in the effective action)  it is necessary to separate the analytic and non-analytic parts of the amplitude, which is straightforward up  to the orders we are considering here.  This is achieved   by splitting the fundamental integration region in  (\ref{pertexpand})  into two sub-domains \cite{Green:2008uj},
 \be
 \cM_1 = \cM_{1L} \cup \cM_{1R}
 \label{splitM}
 \ee 
 where the  first component is the cutoff fundamental domain, 
 \be
 \cM_{1L}  = \{\tau : ~ |\tau_1 |  \le \half, ~ |\tau|\ge 1, ~ \tau_2 \le L \}
 \label{cm1ldef}
 \ee
 and  $L$ is a positive real parameter which may be taken to be large $L \gg 1$.   The remainder 
$\cM_{1R}$ of the fundamental domain $\cM_1$ is the rectangular region that bounds the cusp, 
namely $ \cM_{1R}  = \{\tau : ~  |\tau_1|  \le \half, ~ \tau_2 \ge L\}$.
Accordingly, the one-loop amplitude $\cA_1 ^{(4)} (\e_i, k_i)$ is given by the sum of two parts, 
  \be
 \label{nonanamp}
 \cA^{(4)} _1 (\epsilon_i,k_i ) = \cA^{(4)\, an }_1 (\epsilon_i,k_i ) + \cA^{(4)\, non-an} _1 (\epsilon_i,k_i )
 \ee
The analytic part of the amplitude, $\cA^{(4)\, an}_1 (\epsilon_i,k_i)$, given by, 
\bea
\label{anexpand}
\cA_1 ^{(4) \, an } (\e_i, k_i)
=2\pi\, \kappa _{10}^2 \, \cR^4 
\int _{\M_{1L}} d\mu_1\, 
\cB_1 (s,t,u| \tau)
\eea
admits an expansion in powers of $\sigma_2$ and $\sigma_3$, in view of the fact that
the integrand $\cB_1 (s,t,u |\tau)$ admits a series expansion in $\sigma _2$ and $\sigma_3$,
 \be
\cB_1 (s,t,u| \tau)= \sum_{p,q =0}^\infty j^{(p,q)}(\tau) \, \sigma_2^p \, \sigma_3^q
 \label{b1nexpand}
 \ee
which is {\sl uniform in $\tau$} throughout the  region $\cM_{1L}$. In particular, the integrals 
of the coefficient functions $j^{(p,q)} (\tau)$ over $\cM_{1L}$ are convergent, but do depend upon 
the regulator $L$.

\sm

The non-analytic part of the amplitude $\cA^{(4)\, non-an}_1 (\epsilon_i,k_i)$, is given by, 
\bea
\label{nonanexpand}
\cA_1 ^{(4)non-an } (\e_i, k_i)
=2\pi \, \kappa _{10}^2 \, \cR^4 
\int _{\M_{1R}} d\mu_1\, 
\cB_1 (s,t,u| \tau)
\eea
Although the expansion of (\ref{b1nexpand}) continues to hold pointwise for $\tau$ in the interior of 
$\cM_1$, it does not hold {\sl uniformly} in $\cM_{1R}$, so that the integral of $\cB_1$ over $\cM_{1R}$ 
will fail to be analytic in $s,t,u$ near the origin. In particular, the integrals of the coefficient 
functions $j^{(p,q)}(\tau)$ over $\cM_{1R}$ will diverge for large enough $p,q$, further signaling the breakdown of  analyticity of  $\cA^{(4)\, non-an}_1 (\epsilon_i,k_i)$. Thus, in the region $\cM_{1R}$, 
the expansion (\ref{b1nexpand}) is invalid; the integral needs to be treated non-perturbatively
in $s,t,u$. The  resulting value of $\cA^{(4)\, non-an}_1 (\epsilon_i,k_i)$ will depend upon $L$,
but the dependence on $L$ in the full amplitude will, of course, cancel.

\sm

Concretely, the low-energy behavior of  $\cA^{(4)\, non-an}_1 (\epsilon_i,k_i)$ is determined by expanding the integrand around $\tau_2 = \infty$, thereby generating threshold singularities in $s,t,u$ associated with the propagation of massless intermediate states.  The first of these is the supergravity threshold of the symbolic form $s_{ij} \log (s_{ij})$, which can also be determined directly from dimensional regularization of maximal supergravity in $10-\epsilon$ dimensions.   However, in the string theory context there are further ``stringy" thresholds, which were discussed in detail in \cite{Green:2008uj}.  The first of these arises at order $s_{ij}^4 \log (s_{ij})$.  
   
\sm
   
 The analytic part of the low energy expansion of the one-loop four-graviton amplitude is obtained by integrating (\ref{b1nexpand}) term by term over the cutoff fundamental domain,
\be
J^{(p,q)} (L)= \int_{\cM_{1L}} d \mu_1\,  j^{(p,q)}(\tau)
\label{bigjint}
\ee
In the limit $L\gg 1$ this gives an expression of the form,
\be
\label{Jres}
J^{(p,q)} (L)= \sum_{i} a_i L^{r_i} + c\, \log (L\,\lambda) +\Xi^{(p,q)}+  \cO (1/L)
\ee
where  $a_i$, $c$, $\lambda$ and $r_i$ are constants, with $r_i>0$.  This is a sum of positive powers of $L$, together with a $\log L$ and a constant term.  

\sm

The $L$-dependent terms are cancelled by corresponding terms from the non-analytic part, leaving the sum of $c\, \log \lambda$ and  $\Xi^{(p,q)}$.   The presence of $c\, \log \lambda$ signals the presence of a threshold contribution to the non-analytic part of the  amplitude of the symbolic form $s^a_{ij}\, \log (s_{ij}/\mu)+\dots$, while $\Xi^{(p,q)}$ is the  coefficient of the interaction of order $\sigma_2^p\, \sigma_3^q\,\, \cR^4$.   A redefinition of the scale, $\mu \to e^{-\rho}\, \mu $ of the logarithm leads to a redefinition of the analytic term by the addition of $\rho\, \sigma_2^p\, \sigma_3^q\, \cR^4$.  This generally implies an ambiguity in the definition of certain analytic terms, as we will see in the weight-4 example later in this section.   A more detailed discussion with explicit examples is given in \cite{Green:2008uj}. 
 The fact that integrals over a fundamental domain of functions of moderate growth, such as $j^{(p,q)}(\tau)$, can be interpreted as sensible finite quantities has a respectable mathematical  pedigree, as discussed in \cite{Zagier:1982}.
An example of immediate relevance is that with this interpretation the integral of a non-holomorphic  Eisenstein series, which vanishes, $\int_{{\cal M}_1} d\mu_1  \tE_s=0$, if $s>1$, since $\int_{{\cal M}_{1L}} d\mu_1  \tE_s = 2 s\,\zeta(2s)\, L^{s-1}$, which has no $L$-independent piece.
 The resulting expansion of the amplitude takes the form given in~(\ref{fourgravcons}).

\sm

The modular invariant coefficient functions $j^{(p,q)}$ defined by (\ref{2b6}) are given by sums of Feynman diagrams with coefficients that are simple combinatorial factors that arise from expanding the right-hand side of (\ref{2b5}).  
 The resultant sums of diagrams that arise at low orders in the
 expansion were  presented in appendix C of \cite{Green:2008uj}, and
 are summarized here  for reference in the following
 sections:
\bea
\label{xidefs}
j^{(0,0)} &=& 1 \\
\label{weight2comb}
j^{(1,0)}&=&   \,D_2  \\
\label{weight3comb}
j^{(0,1)} &=& \frac{1}{ 3!}  (8 D_{1,1,1} + 2D_3)  \\
\label{weight4comb}
j^{(2,0)} &=& \frac{1}{ 4!}  (D_4 + 9 D_2^2 + 6 D_{1,1,1,1})  \\
\label{weight5comb}
j^{(1,1)} &=& \frac{5}{ 6!} (2 D_5 +32 D_{3,1,1}-24 D_{2,2,1} + 24 D_{2,1,1,1}
\no \\ && \qquad
- 48 D_{1,1,1,1;1}+28 D_3 D_2 + 96 D_{1,1,1} D_2) 
\eea
All the $D$ functions that enter into these expressions up to weight 5 have been  determined in the previous two sections, so we will now evaluate the integral of $j^{(p,q)}(\tau)$ for all $w= 2p+3q \le 5$. 

\sm

\subsection{Interaction coefficients with weights $\le 3$}
\label{sec:weight02}

The integration of the modular invariant coefficient functions  up to weight 3 over a fundamental domain were determined in \cite{Green:1999pv}.  The results are summarized as follows.

\sm

The weight-0 diagram has no propagators and $D_0=1$, so the
coefficient of $\mathcal R^4$ is given by the volume of the fundamental $\tau$ domain.  In other words,
\be\label{xizero}
\Xi^{(0,0)} = \frac{\pi}{3}
\ee

The only weight-1 diagram is the zero mode of a single propagator, which corresponds to the vanishing of  $D^2 \cR^4$, which follows from the on-shell condition. However, to this order in the expansion there is  a threshold  of the form $\sim s\, \log s $ that corresponds to the non-analytic term that arises in supergravity in dimensional regularization   around $D=10$ dimensions. 

\sm

The weight-2 contribution also comes from a single Feynman diagram, in which the two propagators join the same pair of vertices.  This gives  $D_2 =  \tE_2$, so  
\be\label{xitwo}
\Xi^{(1,0)} = 0
\ee
It is notable that the vanishing of the coefficient of the $1/4$-BPS
interaction,  $D^4 \, \mathcal R^4$, is a special feature of $D=10$ dimensions, where there are no $1/4$-BPS states.   The corresponding coefficient is non-zero for toroidal compactifications to $D=10-d$ dimensions, where $1/4$-BPS states do exist and circulate in the loop 
\cite{Green:2008uj,Green:2010wi}.
 
 \sm
 
The weight-3 contribution arises from the sum of the two contributions
in (\ref{weight3comb}).  The  expressions $D_{1,1,1}
=  \tE_3$ and $D_{3} = C_{1,1,1} =  \tE_3 +\zeta(3)$
(from~(\ref{3d5})) lead to \cite{Green:1999pv}
\be
\label{xithree}
\Xi^{(0,1)} ={\pi\over9}\, \zeta(3)
\ee

\subsection{The sum of all weight-4 terms and the $D^8 \cR^4$ interaction}
\label{sec:fourcomplete}

The expression for $D_4(\tau)$ given in (\ref{d4def}) enters in the
sum of weight-4 terms in $j^{(2,0)}(\tau)$  in (\ref{weight4comb}),
imply that 
 \begin{equation}
   \label{e:j20new}
   j^{(2,0)}(\tau)=C_{2,1,1}(\tau)-\frac12\,\tE_4(\tau)+\frac12\,\tE_2(\tau)^2
 \end{equation}
Using the Laplace equation in~(\ref{1a5}) satisfied by
$C_{2,1,1}(\tau)$ and the one satisfied by the Eisenstein series
in~(\ref{1a3}), we can re-express this quantity as 
\begin{equation}
\label{newj20}
j^{(2,0)}(\tau)=\frac12\, \Delta C_{2,1,1}(\tau)- 5\, \tE_4(\tau)+\tE_2(\tau)^2
\end{equation}
Now we can evaluate the integral over the fundamental domain with
cutoff $L$, by integrating by part the Laplacian 
\begin{equation}
\label{totint40}
\int_{\cM_{1L}}d\mu_1 \, \left(\frac12\, \Delta C_{2,1,1}(\tau)- {5\over12}\, \Delta\tE_4(\tau)\right)
=\left.\int_0^1 \partial_{\tau_2}\left(\frac12
    C_{2,1,1}(\tau)-{5\over12} \tE_4(\tau)\right)\,d\tau_1\right|_{\tau_2=L}
\end{equation}
Using the constant term expansion for the Eisenstein series in~(\ref{fouriereis}) and for $C_{2,1,1}(\tau)$ in~(\ref{powerc211}), we obtain in the large-$L$ limit
\begin{equation}
\label{totint4a}
\int_{\cM_{1L}}d\mu_1 \, \left(\frac12\, \Delta C_{2,1,1}(\tau)- {5\over 12}\, \Delta\tE_4(\tau)\right)
=\frac{\pi}{90}\zeta(3) +{\rm positive\ powers\ of}\  L +\cO(L^{-1})
\end{equation}
The square of the Eisenstein series is integrated using Green's
theorem in the cutoff fundamental domain leading
to~\cite{Zagier:1982,Green:1999pv}
\bea
  \label{e:intEsEsp}
  \int_{\cM_{1L}} \!\!\! {d\tau_1d\tau_2\over\tau_2^2}
  \tE_s(\tau)\tE_{s'}(\tau)
  & = &
  {4\zeta(2s)\zeta(2s')\over \pi^{s+s'}} {L^{s+s'-1}\over s+s'-1}
\\ &&
+{4\zeta(2s-1)\zeta(2s'-1)\Gamma(s-\frac12)\Gamma(s'-\frac12)\over
    \Gamma(s)\Gamma(s')\pi^{s+s'-1}} {L^{1-s-s'}\over s+s'-1} 
\no \\ &&
+  {4\,\zeta(2s)\,\zeta(2s'-1)\,\Gamma(s'-\frac12)\over
    \Gamma(s')\pi^{s+s'-\frac12}}\,  {L^{s-s'}\over s-s'}\cr
&-&{4\,\zeta(2s')\,\zeta(2s-1)\,\Gamma(s-\frac12)\over
  \Gamma(s)\pi^{s+s'-\frac12}} \, {L^{s'-s}\over s-s'}
+{\mathcal C}_L
\no
\eea
The correction term $\mathcal C_L$ is given by the integral
\begin{eqnarray}
\mathcal C_L &=& \int_{\cM_{1R}} \,d\mu_1\,
  \,F_s(\tau)\, F_{s'}(\tau)\\
\no &=& {32\over
    \Gamma(s)\Gamma(s')}\sum_{n>0}  n^{s+s'-1}
    \sigma_{1-2s}(n)\sigma_{1-2s'}(n)\int_{\tau_2>L} K_{s-\frac12}(2\pi n \tau_2)\,
    K_{s'-\frac12}(2\pi n \tau_2) {d\tau_2\over\tau_2}
\end{eqnarray}
where $F_s(\tau)$ is the non-zero Fourier mode part of the Eisenstein
series in the second line of~(\ref{fouriereis}). But bounding the Bessel function
$|K_s(2\pi x)|\leq e^{-2\pi x}/\sqrt x$ for $x\geq1$ one gets
\begin{equation}
  0\leq \mathcal C_L\leq   {32\over
    \Gamma(s)\Gamma(s')\, L}\sum_{n>0}  n^{s+s'-1}
    \sigma_{1-2s}(n)\sigma_{1-2s'}(n) e^{-4\pi n L}\to 0 \qquad L\to\infty
\end{equation}
Specializing this formula to the case $s=s'=2$ we find that
\begin{equation}
\label{methode2e2}
\int_{\cM_{1L}}d\mu_1\,  \tE_2(\tau)^2 = \frac{2\pi}{45} \zeta(3)\, \log(L/  \mu)
+\frac{\pi^4}{6075}\, L^3 +\cO(L^{-1})
\end{equation}
where the scale $\mu$ is given by
\be
\label{logmu}
\log\mu = \half - \log 2 +\frac{\zeta'(3)}{\zeta(3)} - \frac{\zeta'(4)}{\zeta(4)}
\ee
Collecting all the contributions,  the integral of $j^{(2,0)}(\tau)$ over
the cutoff fundamental domain is given by 
\begin{equation}
\label{totint4}
\int_{\cM_{1L}}d\mu_1 \, j^{(2,0)}(\tau)  
= \frac{\pi}{90}\zeta(3) + \frac{2\pi}{45} \zeta(3)\, \log(L/  \mu)
+{\rm positive\ powers\ of}\  L +\cO(L^{-1})
\end{equation}
where we will drop terms that are positive powers of $L$ (since these cancel against other terms arising from the low energy expansion of the non-analytic threshold terms, as described earlier).  The logarithmic term,  $\log  (L/\mu)$  arises from the integral of the term linear in $\tau_2$ in the  large-$\tau_2$ expansion of  $ \tE_2(\tau)^2$.    The $\log L$ cancels upon addition of the  non-analytic  threshold term, which was  given in~\cite[eq.~(4.41)]{Green:2008uj} by,
\bea
 - {4\pi \zeta(3)\over 45} \,
s^4\,\log(-\alpha'sL\pi/\hat \mu_4) + (s\to t) + (s\to u)\, ,
 \label{totupper}
\eea
where $\hat \mu_4$ is another constant.  The  $\zeta(3)/90$ in~(\ref{totint4})  arises as a  boundary term from the integral of the term involving the Laplacian.  It can be absorbed into a redefinition of the scale, $ \mu$, in $\log (L/\mu)$, after recalling that $s^4+t^4+u^4 = \sigma_2^2/2$ (and remembering that the Mandelstam invariants have been rescaled by $\alpha'/2$ in this paper relative to~\cite{Green:2008uj}).

\sm

In summary, the term in the  low energy expansion of order $\sigma_2^2 = \cO(s^4)$ has a logarithmic normal threshold.  An additional constant proportional to $\zeta(3)$ arising from the integral over finite values of $y$ was incorrectly omitted in ~\cite{Green:2008uj}.  However, the separation of the analytic and non-analytic contributions is ambiguous and the additive  constant can be absorbed into the scale of the logarithm.  With this definition, the analytic term has zero coefficient, i.e. $\Xi^{(2,0)} =0$.

\subsection{The sum of all weight-5 terms and the $D^{10}\, \cR^4$  interaction}
\label{sec:fivecomplete}

In section~\ref{sec5} we motivated strong conjectures that express the modular functions that enter into $j^{(1,1)}(\tau)$ in~(\ref{weight5comb}) as the sum of $D_{2,1,1,1}(\tau)= C_{3,1,1}(\tau)$ and sums of products of Eisenstein series,  where $C_{3,1,1}(\tau)$ satisfies the  inhomogeneous Laplace equation~(\ref{D2111}).
  
\sm
  
To be precise,  the two equations for weight 5 coefficients that correspond to $C_{a,b,c}(\tau)$ functions (namely, $C_{2,2,1}(\tau)=D_{1,1,1,1;1}(\tau)$ and $C_{3,1,1}(\tau)= D_{2,1,1,1}(\tau)$) have been rigorously derived.  The other equations  are based on the assumption that knowledge of all the terms that are powers of $1/y$, as well as all the power-behaved corrections to terms of order $e^{-2y}$, in  the large-$y$ limit, are sufficient to determine the full modular function.  
Using this information it follows that the total contribution of the
weight-5 terms to $j^{(1,1)}(\tau)$  is  given by, 
\begin{equation}
\label{totalnew}
j^{(1,1)}(\tau)={5\over 6!}\,\left(
336\, C_{3,1,1}(\tau)+240  \tE_2(\tau)\, \tE_3(\tau)+48  \tE_2(\tau)\,\zeta (3)-\frac{1632  }{5} \tE_5(\tau)+\frac{144 }{5}\zeta (5)\right)
\end{equation}
From~(\ref{D2111summ}) we have,
    \bea
 \label{D2111def}
C_{3,1,1} (\tau)= \frac{1}{6}\Delta C_{3,1,1}(\tau) -\frac{43}{15}   \tE_5(\tau)+ \frac{2}{3}  \tE_2\,   \tE_3(\tau) -\frac{\zeta(5)}{60} 
 \ee 
so that we find, 
\begin{equation}
  \label{e:j11new}
  j^{(1,1)}(\tau)={5\over 6!}\,\left(56\Delta C_{3,1,1}(\tau) -\frac{6448}{5}
                     \tE_5(\tau)+ 464  \tE_2(\tau)\,   \tE_3(\tau)+48 \tE_2(\tau)\zeta(3) +\frac{116\zeta(5)}{5}\right)
\end{equation}
We will now integrate this expression over  the cutoff fundamental domain $\cM_{1L}$.
The integral of $\tE_2(\tau)\,   \tE_3(\tau)$ is obtained by setting
$s=2$ and $s'=3$ in the~(\ref{e:intEsEsp}) resulting into
\begin{equation}
\label{e2e3int}
\int_{\cM_{1L}}d\mu_1\,  \tE_2(\tau)\,  \tE_3(\tau) = 
\frac{2 \pi^{10}}{42525}\, L^{4}
+\frac{2\pi^7}{945}\, \zeta(3) L +\cO(L^{-1})
\end{equation}
As described earlier, terms that are powers of $L$ cancel with corresponding terms arising from the low energy expansion of the non-analytic terms that come from the integral over $\cM_{1R}$.  The absence of an 
$L$-independent term  in~(\ref{e2e3int}) implies it does not contribute to the weight-5 coefficient.  Likewise, the integrals of the terms proportional to $ \tE_5(\tau)$ and $ \tE_2(\tau)$ in~(\ref{totalnew}) do not contribute $L$-independent terms.

  \sm
  
The integral of $\Delta C_{3,1,1}(\tau)$ gives a boundary term proportional to $ \partial_{\tau_2}  \,\int d\tau_1 \, C_{3,1,1}|_{\tau_2=L}$, which only contains terms that behave as $L^a$, with $a\ne 0$.  This follows from the fact that there is no term proportional to $\tau_2$ in the  large-$\tau_2$ expansion in~(\ref{D2111expan}).  Thus, we find,
\be
\label{D2111int}
\int_{\cM_{1L}}d\mu_1\, C_{3,1,1}(\tau)  = - \frac{\pi}{3}\, \frac{\zeta(5)}{60} + {\rm powers\ of}\ L+\cO(L^{-1})
\ee
where the powers of $L$ cancel with the integral over ${\cal M}_R$.
Substituting in~(\ref{e:j11new}) we find that,
\be
\label{finaltot}
\Xi^{(1,1)} = \int_{\cM_{1L}}\,d\mu_1\,
j^{(1,1)}(\tau) =\frac{\pi}{3}\,{29\over 180}
\, \zeta(5)\,
\ee
(which differs from the incorrect expression given in~\cite{Green:2008uj}). It will be interesting to compare this value with result from space-time supersymmetry and duality, when they will become available.

\section{Discussion and Comments}
\setcounter{equation}{0}
\label{sec8}

This paper has been concerned with properties of non-holomorphic modular invariants that arise in the   low energy expansion of the four-graviton amplitude in Type II superstring theory. These are defined by Feynman diagrams for a free massless  bosonic field theory on a torus with complex structure modulus $\tau$.   Any diagram of weight $w$ (i.e. of order $(\alpha')^w$ in the low energy expansion)  consists of  $w$ powers of  the world-sheet Green function joining pairs of vertices. 
 If the diagram has $\ell$ loops it reduces to a multiple sum over the  $ 2\ell$ components of the integer-valued  world-sheet loop momenta. The $\ell=1$ diagrams  simply reduce to standard non-holomorphic Eisenstein series. We determined the exact functional form of the infinite class of $\ell=2$ invariants denoted $C_{a,b,c}$  that are defined by setting $k=0$ in (\ref{3a1}) and have weight $w=a+b+c$.  
We have furthermore provided evidence for conjectured expressions for the $w=4$ and $w=5$ functions $D_4$,  $D_{2,2,1}$, and $D_5$, which have  $\ell=3$, $3$ and $4$ loops, respectively.    The conjectured expressions of these functions, which enter into the expansion of the four-graviton amplitude, are 
given in (\ref{d4def}), (\ref{D5}) and (\ref{D221sol}).
Making use of these expressions, we determined the integral over $\tau$ of the sum of all the diagrams that contribute to the low energy expansion of the four-graviton amplitude at weights 4 and 5,  thereby determining the coefficients of the one-loop effective $D^8 \cR^4$ and $D^{10} \cR^4$  interactions. 
  
  \sm
An obvious challenge is to provide a proof of the differential and algebraic 
relations  for the modular functions   $D_4, D_5$ and $D_{2,2,1}$. Referring to (\ref{d4def}), (\ref{D5}) and (\ref{D221sol}) we see that any of these functions of weight $w$ and  $\ell$ loops is expressed as a weight-$w$ polynomial of $D$-functions with $\le \ell$ loops with rational coefficients together with powers of Riemann zeta values\footnote{It is notable that this relationship between weight-$w$ combinations of $D$ functions  has a structure analogous to that of the relationship between multi-zeta functions, with the number of loops, $\ell$, playing the 
r\^ole of  ``depth''.}. 
  These relations were motivated in sections ~\ref{sec4} and \ref{sec5} by analysis of the asymptotic behaviour of the various functions  in the limit $\tau_2\to \infty$, including the first exponential corrections of order $O(e^{-2\pi \tau_2})$. This is well illustrated for the $w=5,$ $\ell=4$ function $D_5$ that is conjectured to satisfy the relation (\ref{D5}).   The right-hand side of this relation is  the unique combination of $D$-functions of weight $w=5$ and depth $2\ell\le 8$ that matches our asymptotic analysis of $D_5$.  Our conjecture is that $F=0$, where $F= D_5 -60 C_{3,1,1} - 10  \tE_2 C_{1,1,1} +48  \tE_5 -16 \zeta(5)$.    
If it turns out that  $F \ne 0$,  then $F$ is a cuspidal function --  a non-holomorphic modular function that has the behavior $O(e^{-4\pi  \tau_2})$ as $\tau_2\to \infty$.   This is not expressible in terms of any combination of the $D$-functions, Eisenstein series and zeta values.
 If, however,  these conjectures are correct they suggest a possible extension to polynomial relations between $D$-functions of any weight.
  
\sm

 In addition, in order to understand the complete basis of such functions at any weight it is necessary to extend the considerations of this paper to consider the one-loop amplitude with $N$ scattering particles, massless and/or massive.  When $N>4$ the Feynman rules for the coefficient modular functions given in section~\ref{sec2} are modified to include $2(N-4)$ factors of world-sheet momenta in the numerators.  Many of the resulting Feynman diagrams can be reduced to diagrams with fewer powers of momenta by integration by parts inside the Feynman diagram,  as  was discussed in \cite{Green:2013bza} in the $N=5$ and $N=6$ cases. There are no further  weight $4$ diagrams beyond those that arise in the expansion of the $N=4$ case discussed in this paper.  At weight $5$, however,  the complete set includes two extra diagrams that first arise when $N= 5$,  At weight $6$ the complete set includes seven new diagrams that first arise when $N=6$.  Understanding the relationships between the diagrams arising at higher weight and higher $N$ is  important for understanding the complete basis at arbitrary weight.

  \sm
  
The results of this paper generalize straightforwardly to the genus-one Type II string theory amplitude compactified on a $d$-dimensional torus, $T^d$.   In this case the amplitude (\ref{2a4}) is modified by multiplying $\cB_1$ in the integrand by the standard lattice factor $\Gamma_{d,d}(\tau,\rho_d)$,  which ensures that Kaluza--Klein charges and string windings are accounted for appropriately.   The quantities  $\rho_d\in SO(d,d,\ZZ)\backslash SO(d,d, \RR)/(SO(d,\RR) \otimes SO(d,d,\RR))$ parameterize the coset space appropriate to the  T-duality group $SO(d,d)$. It is notable that the  part of the genus-one amplitude that is non-analytic in $s$, $t$ and $u$  is independent of the compactification moduli. Consequently, the non-analytic dependence can be eliminated by consideration of compactification on tori of different moduli, $\rho_d$ and $\rho'_d$. This provides a method of uniquely specifying the local part of the genus-one  effective interactions $D^{2w} \cR^4$.  At higher genus the interplay between the analytic and non-analytic parts of the amplitude is more intricate, as can be argued from considerations based on  unitarity.

  \sm
  
There are several directions in which it would be interesting to extend these results, both because of their mathematical interest and because of their connections to string theory.  One direction that would obviously be of interest is to develop a complete understanding of the structure of all the modular invariants at weight $w$ for all values of $w$ that arise in the expansion of the four-graviton amplitude.  One elegant  way of establishing this might be by considering the action of the Laplacian directly on $\cB_1(s,t,u|\tau)$, which is the generating function for the modular invariants, rather than on the individual terms in its expansion, generalizing the discussion in section~\ref{sec:laplacegen}.  

\sm

Another possible approach to determining the low energy expansion of
closed-string one-loop amplitudes may be to exploit the relationship
to the structure  of the open-string low energy expansion discussed in
\cite{Broedel:2014vla}  in a manner analogous to the tree-level KLT
relations.  In the tree-level case the open-string expansion is a
series of terms with  coefficients that are rational multiples of
multi-zeta values, which are special values of multiple
poly-logarithms~\cite{Brown:2013gia}.   In the case of the Type II  string  these generalize
to special values of single-valued multiple poly-logarithms that give
coefficients that are (odd weight) single-valued multi-zeta values~\cite{Schlotterer:2012ny, Stieberger:2013wea}   The one-loop amplitude involves an elliptic generalization of these considerations.  

\sm

A different direction of interest would be to extend the analysis of this paper to the low energy expansion of higher-genus loop amplitudes.  For genus $h\geq 2$, the world-sheets  are really  super Riemann surfaces, and the integrations are to be performed over their associated supermoduli spaces (for overviews see \cite{D'Hoker:1988ta,Witten:2012bh}). For genus $h \geq 3$, it may not be possible to project this integration onto an integration over the corresponding bosonic moduli space \cite{Donagi:2013dua}. Although the modular group $Sp(2h,\ZZ)$ for higher genus  $h\geq 3$ still exists and must play a leading role,  it is unclear at this time which role, if any, will be played by modular forms under this group. 

\sm

However, for genus $h=2$, higher derivative interactions involving only NS-NS fields may  be reduced to integrals of an $Sp(4,\ZZ)$ modular invariant over the moduli space of compact genus 2 Riemann surfaces \cite{D'Hoker:2001zp,D'Hoker:2001qp}.   In this case the leading term in the low energy limit is  $D^4\, \cR^4$, which has a coefficient  that is simply a constant (which was determined in  \cite{D'Hoker:2005ht}) integrated over genus-two moduli space.  It is therefore proportional to the  volume of the Siegel fundamental domain (which coincides with moduli space for $h=1,2,3$). The  next term  in the  low energy expansion is the $D^6 \cR^4$ interaction, which has a coefficient with a  $Sp(4,\ZZ)$-invariant integrand  that  is linear in the genus-two Green function. This was shown\cite{D'Hoker:2013eea} to be proportional to the first power of the genus-two invariant $\f$  introduced by Zhang \cite{Zhang} and Kawazumi \cite{Kawazumi}. The evaluation of the integral  over genus-two moduli space in \cite{D'Hoker:2014gfa} was made possible by the  realization that the integrand $\f$ satisfies an eigenvalue equation $\Delta \f =5 \f $ in the interior of moduli space, where $\Delta$ is the $Sp(4,\ZZ)$-invariant Laplace-Beltrami operator on the genus-two moduli space.  Our hope is that the pattern of Laplace equations satisfied by the genus-one coefficients obtained in this paper, hints at an analogous pattern at at genus two and at  higher genus.

\vskip 0.3in

\noindent
{\bf \large Acknowledgments}

\bigskip

We are grateful to Don Zagier for discussions at various stages of this work and for 
showing us his evaluation of the modular function $C_{1,1,1}$. The methods used there
inspired  some of the calculations in appendix \ref{secC}.   We
would like to thank Axel Kleinschmidt for discussions on the Poincar\'e
representation in section~\ref{sec:poincare}, and  \"Omer G\"urdo\u gan for discussion on the numerical evaluation of the integrals.
We also thank Anders Kolvra, Jorge Russo and Stephen Miller for useful discussions and email interchanges, and Stefan Stieberger for enlightening comments.
The research leading to these results has received funding from
 National Science Foundation grant PHY-13-13986 and from the
European Research Council under the European Community's Seventh
Framework Programme (FP7/2007-2013) / ERC grant agreement
no. [247252], the ANR grant reference QST 12 BS05 003 01, and the CNRS grants PICS number 6430.

\appendix

\section{Asymptotic expansion of $D_\ell$ by integration}
\setcounter{equation}{0}
\label{secB}

The asymptotic expansion near the cusp as $y = \pi \tau_2 $ tends to $\infty$ of the modular functions which occur in this paper may be evaluated with the help of a number of different approaches. The first method relies on carrying out the integrations over the world-sheet of combinations of the propagator $\cG$, and subsequent evaluation of the remaining sums. This method has been used already extensively in \cite{Green:2008uj} to evaluate the pure power part of the asymptotic expansion. In this appendix,  we shall extend the method to evaluating also the leading exponential correction terms in the asymptotic expansion of $D_4$ and $D_5$. In each case, only a single integration over the world-sheet is required.

\sm

For the function $D_{2,2,1}$ two integrations over the world-sheet would be required,
and this makes the integration method less suitable for the calculation of the asymptotics of $D_{2,2,1}$. Therefore, in the next appendix, the asymptotics of $D_{2,2,1}$ will be calculated by direct evaluation of the sums over the world-sheet momenta, using the  decomposition into partial fractions of the denominators entering the sums. 

\subsection{Asymptotic expansion by integration}

In a first approach, we shall integrate combinations of the Green function $\cG$ of (\ref{2c3}) over the world-sheet torus $\Sigma$ with complex structure modulus $\tau$. This method is suitable for the calculation of the asymptotics of the functions $D_\ell$, which is given by,
\bea
D_\ell (\tau) =   \int _\Sigma { i \over 2} dz \wedge d \bar z \, \cG(z | \tau )^\ell
\eea
and we shall describe the method specifically for this case. We shall parametrize the torus $\Sigma$ of modulus $\tau=\tau_1 +i \tau_2 $ by real coordinates $\alpha, \beta \in \RR/\ZZ$ with $z = \alpha + \beta \tau$, and use a representation of the Green function in which $\cG$ is expanded in Fourier modes with respect to $\alpha$. It will be convenient to make a specific choice for the range of $\beta$ given by $|\beta | \leq \half$. We then have the following representation for $G$,
\bea
\label{Gs}
\cG(z |\tau ) = \cG_1(z  |\tau) + \cG_2(z  |\tau )+ \cG_3(z  |\tau)
\eea
where the functions in this decomposition are given as follows,  
\bea
\cG_1(z  |\tau ) & = & 2 \pi \tau_2 \left ( \beta ^2 - |\beta| + { 1 \over 6} \right )
\no \\
\cG_2 (z  |\tau ) & = & \sum _{m \not= 0} { 1 \over |m|}  \exp \Big \{ 2\pi i m (\alpha + \beta \tau_1)
- 2 \pi \tau_2  |m \beta | \Big \}
\no \\
\cG_3(z  |\tau) & = & \sum _{m \not= 0} { 1 \over |m|} \sum _{k \not= 0} 
\exp \Big \{ 2\pi i m (\alpha + \beta \tau_1 + k \tau_1) - 2 \pi \tau_2 |m(k+\beta)| \Big \}
\eea
The above decomposition holds for $|\beta|\leq \half$ and may be extended beyond this domain by periodicity in $\beta$ with period 1. For the evaluation of $D_\ell$, such extension will not be needed, but it will enter for other types of diagrams. The integral over $\Sigma$ of $\cG_1$ vanishes. The function $\cG_3$ is bounded uniformly in $z$ by an exponential in $y$, 
\bea
\label{b4}
|\cG_3(z | \tau )| \leq C(y) \, e^{-  y}
\eea
where $C(y)$ is power-behaved and independent of $z$. Therefore, $\cG_3$ will not contribute to the pure power part 
$\cD^{(0,0)} _\ell$ of the asymptotic expansion, while the leading exponential correction $\cD_\ell ^{(1,0)}$ and its complex conjugate will involve $\cG_3$ to first and second order.

\subsection{Evaluating the pure power part}

We begin by reviewing the calculation of the pure power part of the expansion, following \cite{Green:2008uj}.
The pure power terms  are obtained by retaining the contributions from $\cG_1$ and $\cG_2$ only, 
and extracting the pure power part of the corresponding integral, 
\bea
\cD^{(0,0)} _\ell (y) =  \int _0 ^1 d \alpha \int _{-\half } ^{ \half} d \beta \, 
\Big (\cG_1(z|\tau ) + \cG_2(z|\tau ) \Big )^\ell + \cO(e^{- y})
\eea
while omitting any exponential terms that still arise in this expression. It will be useful to expand
the integrand as follows,
\bea
\cD^{(0,0)} _\ell (y) = \sum _{\ell_1, \ell_2=0} ^\ell \, \delta _{\ell_1+\ell_2, \ell} \, 
{ \ell \, !  \over \ell _1! \, \ell _2!} \, d^{(0,0)} (\ell_1, \ell_2; y )
\eea
where $\delta _{\ell_1+\ell_2, \ell}$ is the Kronecker $\delta$, while the expansion coefficients are,
\bea
d^ {(0,0)} (\ell_1, \ell_2;y ) =  \int _0 ^1 d \alpha \int _{-\half } ^{ \half} d \beta \, 
\cG_1(z |\tau )^{\ell_1} \, \cG_2(z |\tau )^{\ell_2}  + \cO(e^{-y})
\eea
Carrying out first the integration over $\alpha$ gives $d^{(0,0)} (\ell_1, 1;y)=0$ for all $\ell_1$, and also readily produces the contribution with $\ell_2=0$,
\bea
d^{(0,0)} (\ell_1, 0 ;y ) = 2 
\int _0 ^\half \! d \beta \, \cG_1(z|\tau))^{\ell_1}
\! = \left ( {  y \over 3} \right )^{\ell_1} \! {}_2 F_1 \left (
1, -\ell_1; { 3 \over 2}; { 3 \over 2} \right )
\eea
The hypergeometric function ${}_2F_1$ evaluates to a rational number. For $\ell _2 \geq 2$, we have,
\bea
\int _0 ^1 d\alpha  \, \cG_2(z |\tau ) ^{\ell_2} 
= \sum _{m_1, \dots, m_{\ell_2} \not=0}
{ \delta _{m,0} \over |m_1 \dots m_{\ell_2}|} \, e^{ - 2  y M |\beta|}
\eea
where we have used the following notation,
\bea
\label{mM}
m & = & m_1 + m_2 + \dots + m_{\ell_2}
\no \\
M & = & |m_1 | + |m_2| + \dots + |m_{\ell_2}|
\eea
In the remaining integration over $\beta$, the integrand is even in $\beta$, so we shall restriction the range to $0 \leq \beta \leq \half$ and include a factor of 2. Since $M \geq 2$, the integral over $\beta$ from $\half$ to $\infty$ is uniformly bounded by a multiple of $e^{-\pi y}$, so that the pure power part of the expansion precisely corresponds to taking the integration domain to be $0 \leq \beta < \infty $. The corresponding integral over $\beta$ is then given by a polynomial in $(My)^{-1}$ of degree $2 \ell_1+1$, whose coefficients will be denoted by $P(\ell_1, n)$ and are given as follows,
\bea
\label{b11}
\int _0 ^ \infty d \beta \, \left ( \beta ^2 - \beta +{1 \over 6} \right )  ^{\ell_1} \, e^{- 2 y M \beta }
=  \sum _{n =1}^{2\ell_1+1} { P(\ell_1, n) \over M ^n} \, (2 y )^{-n}
\eea
Collecting all contributions, we find, 
\bea
d^{(0,0)} (\ell_1,\ell_2 ;y ) = 2 
\sum _{n =1}^{2\ell_1+1}   P(\ell_1, n)  \,  S(\ell_2, n) \,  (2  y )^{ \ell_1 -n} 
\eea
where the multiple sum $S$ is defined by
\bea
S(\ell_2, n) = \sum _{m_1, \dots, m_{\ell_2} \not=0}
{ \delta _{m,0} \over |m_1 \dots m_{\ell_2}|} { 1 \over M^n}
\eea
These multiple sums may be expressed in terms of multi-zeta functions (see \cite{Green:2008uj} and \cite{Zagier:2014}), which are  functions of complex variables $s_1, \dots, s_\rho$  defined by,
 \bea
 \zeta(s_1, s_2, \dots, s_\rho) 
 = \sum _{n_1 >  \dots > n_\rho \geq 1} { 1 \over n_1 ^{s_1} \dots n_\rho^{s_\rho}}
 \eea
and we have the following relation, 
 \bea
 S(\ell_2,n) = { \ell_2 ! \over 2^n} 
 \sum _{{ a_1, \dots, a_\rho \in \{ 1,2\} \atop a_1+\dots + a_\rho=\ell_2-2}} 2^{2\rho+2-\ell_2}
 \zeta (n+2, a_1, \dots, a_\rho)
 \eea 
For special values of $\rho$ and of the argument, the multi-zeta functions may be  expressed in terms of
polynomials in the ordinary $\zeta$-functions.

\subsection{Evaluating the leading exponential correction}

The contribution to $D_\ell (\tau)$ arising from terms in $\cG_1$ and $\cG_2$ only (while omitting $\cG_3$)
do produce exponential corrections which are independent of $\tau_1$, as may be seen explicitly from the fact that the $\alpha$-integration over $\cG_2^{\ell_2}$ is independent of $\tau_1$. As a result, those parts contribute exponential corrections  $\cD_\ell ^{(k, \bar k)}(y) $  only for $\bar k  =k$, and will not contribute to the leading exponential corrections which have $k+\bar k=1$ with $k_1, k_2 \geq 0$. Therefore, the leading exponential corrections will arise from expanding in powers of $\cG_3$, up to order 2 included, in view of the bound (\ref{b4}), and we have, 
\bea
\cD_\ell ^{(1,0)} (y)  = \sum _{\ell_1, \ell_2=0} ^\ell \, \delta _{\ell_1+\ell_2, \ell} \, 
{ \ell \, !  \over \ell _1! \, \ell _2!} \, d^{(1,0)} (\ell_1, \ell_2;y)
\eea
where $d^{(1,0)} (\ell_1, \ell_2;y)$ is the contribution proportional to $q$ (indicated by $|_q$ below)  in,
\bea
\label{4k3}
d^{(1,0)} (\ell_1, \ell_2;y ) =   \int _0 ^1 d \alpha \int _{-\half } ^{ \half} d \beta \, 
\cG_1(z|\tau )^{\ell_1}  \, \Big ( \cG_2(z |\tau)+ \cG_3(z|\tau) \Big ) ^{\ell_2} \bigg |_q
\eea
By inspection of the integration over $\alpha$, it is clear that we have, 
\bea
d^{(1,0)} (\ell_1, 0;y) = d^{(1,0)} (\ell_1, 1;y) = 0
\eea
for all $\ell_1$. For all $\ell_2 \geq 3$, the term of order $\cG_3^2 $ is multiplied by at least one power of $G_2$, and this suppresses the exponential by at least the order $e^{-3\pi y}$, so that such terms do not contribute to the leading exponential corrections. Therefore, the contribution for $\ell _2 \geq 3$ takes the form,
\bea
\label{4k3a}
d^{(1,0)} (\ell_1, \ell_2;y ) =  \ell_2  \int _0 ^1 d \alpha \int _{-\half } ^{ \half} d \beta \, 
\cG_1(z |\tau )^{\ell_1}  \,  \cG_2(z |\tau)^{\ell_2-1} \cG_3(z|\tau)  \bigg |_q
\eea
while for $\ell_2=2$, we have,
\bea
d^{(1,0)} (\ell_1, 2 ;y ) & = &  2  \int _0 ^1 d \alpha \int _{-\half } ^{ \half} d \beta \, 
\cG_1(z|\tau )^{\ell_1}  \,  \cG_2(z |\tau) \,  \cG_3(z|\tau)  \bigg |_q
\no \\ && + 
 \int _0 ^1 d \alpha \int _{-\half } ^{ \half} d \beta \, 
\cG_1(z|\tau )^{\ell_1}   \,  \cG_3(z|\tau)^2  \bigg |_q
\eea
The evaluation of these integrals proceeds along the same lines as for the pure power part, and makes use again of the formula (\ref{b11}), and we have for $\ell_2\geq 3$, 
\bea
\cD_\ell ^{(1,0)}  (\ell_1, \ell_2; y) = \ell_2  \sum _{n =1}^{2\ell_1+1}  
 P(\ell_1, n) \, S^{(1)} (\ell_2, n) \, (2 y )^{ \ell_1-n }  
\eea
where the sums are gives as follows for $\ell_2 \geq 3$,
\bea
S^{(1)} (\ell_2, n)
= \sum _{m_1, \dots, m_{\ell_2} \not= 0} 
{ \delta _{m,0} \, \delta _{|m_{\ell_2}|,1} \over | m_1 \dots m_{\ell_2} |}
\left ( { 1 \over M^n} + { 1 \over (M-2)^n} \right )
\eea
with $m$ and $M$ defined by (\ref{mM}). Finally, for the special case $\ell_2=2$, we have, 
\bea
\label{b23}
\cD_\ell ^{(1,0)}  (\ell_1, 2; y) = 2  d^{(0,0)} (\ell_1, 0;y) 
+ 2   \sum _{n=1}^ { 2 \ell_1+1}  P(\ell_1, n) \, { 2 \over 2^n} (2 y)^{\ell _1 - n}
\eea
Note that the sums $S^{(1)} (\ell_2, n)$ diverge for $\ell_2=2$ because the combination $(M-2)^n$ is then forced to vanish. In the correct formula for the case $\ell_2=2$ given above, the part of $S^{(1)}(\ell_2, n)$ that involves only the combination $M^n$ survives and gives rise to the factor ${ 2 \over 2^n}$ in (\ref{b23}).

\sm

We shall need to evaluate these sums up to $\ell_2=5$ included to calculate the expansions up to $D_5$.
It is straightforward to establish the following recursion relations, 
\bea
S^{(1)} (3,n) & = & { 1 \over 4} S^{(1)} (3,n-2) + { 8 \over 2^n}  \left ( 1 - \zeta (n) \right )
\no \\
S^{(1)} (4,n) & = & { 1 \over 4} S^{(1)} (4,n-2) + {12 \over 2^n}  \left ( 4 \zeta (n,1) - \zeta (n+2) 
-2n -1 + 2 \sum _{k=2}^{n+1} \zeta (k) \right ) \quad
\eea
along with the initial conditions,
\bea
S^{(1)}(3,0)=8 & \hskip 1in & S^{(1)}(4,0)= 24 + 4 \pi^2 
\no \\
S^{(1)}(3,1) =2 && S^{(1)}(4,1)=12 -\pi^2 + 6 \zeta (3)
\eea
For $\ell_2=5$, we shall only need a single sum, given by,
\bea
S^{(1)}(5,1)= 72  + { 2 \pi^4 \over 15} - 48 \zeta (3)
\eea
From these data, the leading exponential corrections to $D_4$ and $D_5$ may now be readily obtained,
and one finds the results given respectively in (\ref{expd4}) and (\ref{expd5}).

\section{Asymptotic expansion of $D_{2,2,1}$ by summation}
\setcounter{equation}{0}
\label{secC}

One key ingredient in $D_{2,2,1}$ is the appearance of a subgraph which quantum field theorists refer to as the one-loop vacuum polarization diagram, and which in terms of discrete world-sheet momenta is given by the following sum over internal loop momenta $(m,n)$, 
\bea
\cT (M,N |\tau ) = \sum _{{ (m,n) \not= (0,0), \atop \not=  (-M,-N)}} 
{ \tau_2 ^2 \over \pi^2 |m \tau + n|^2 \, |(m+M)\tau +n+N|^2}
\eea
From its definition, we have $\cT(-M,-N |\tau ) = \cT(M,N |\tau )$ and  $\cT(M,N |\tau ) >0$ for all $M,N$.
The summation over the integer $n$ may be carried out explicitly by decomposing the summand in partial fractions with respect to the variable $n$, and using the following elementary summation formula,
\bea
\label{sumq}
\sum _{n \in \ZZ} { 1 \over m \tau +n } = - i \pi \, {1+q^m \over 1-q^m}
\hskip 1in q = e^{2 \pi i \tau}
\eea
This sum is conditionally convergent, and may be naturally defined by taking the limit as $\Lambda \to \infty$ of the finite sum obtained by restricting the range of $n$ to $-\Lambda \leq n \leq \Lambda$, following Eisenstein. When $M=0$ the sum reduces to, 
\bea
\cT(0,N) =   \sum _{n \not= 0, -N}  { \tau_2^2 \over  \pi^2n^2 (N+n)^2}
+ \sum _{m\not= 0} \sum _{n} { \tau_2^2 \over \pi^2 |m \tau + n|^2 \, |m\tau +n+N|^2}
\eea
In particular, we have $\cT(0,0)= \tE_2$. When $N \not= 0$, we decompose 
into partial fractions in $n$ and carry out the sums over $n$, and we find, 
\bea
\cT(0,N) = { 2  \tau_2^2 \over 3 N^2} - { 6 \tau_2^2 \over \pi^2 N^4}
+ \left ( \sum _{m=1} ^\infty { 2  \tau_2 \over \pi m} { 1 \over  (N^2+4m^2\tau_2 ^2) } \, { 1+q^m \over 1- q^m} 
+{\rm c.c.} \right )
\eea
When  $M\not= 0$ the variable $N$ is unrestricted, and the summation over $n$ gives,  
\bea
\label{5b3}
\cT(M,N) 
& = & 
 { 2  \tau_2^2 \over 3 |M\tau +N|^2} - \left ( { i \tau_2 \over \pi^2 M (M\tau+N)^3} 
 - { i \tau_2 \over \pi^2 M (M \bar \tau+N)^3} \right )
\no \\ && 
+ \sum _{m \not = 0} \left [ 
 { \tau_2 \over \pi m} \, { 1 \over (M\tau+N) (M\bar \tau +N - 2 i m\tau_2)} \, { 1+q^m \over 1- q^m} +{\rm c.c.} \right ]
\eea
Note that both series in $m$ are absolutely convergent for $y>0$.
We have for example, 
\bea
D_3 &= & \sum _{(M,N)\not= (0,0) } { \tau_2 \over \pi |M\tau +N|^2} \, \cT(M,N)
\no \\
D_4 & = & \sum _{(M,N)} \cT(M,N)^2
\no \\
D_{2,2,1} & = & \sum _{(M,N)\not= (0,0)} { \tau_2 \over \pi |M\tau +N|^2} \, \cT(M,N)^2
\eea
In terms of $\cT(M,N)$, the summation needed for $D_5$ will involve two internal momenta, which renders this approach less effective for calculating the asymptotic expansion of $D_5$.

\subsection{The power and leading exponential approximations}

As we shall be mostly interested for the moment in computing the pure power and leading order
exponential corrections, we may right away approximate the two-point function to this order. Subsequent summations over the ``external momenta'' $M,N$ will not interfere with the approximation. 
We write the following asymptotic expansion, 
\bea
\cT(M,N) = \cT_0(M,N) + q \, \cT_1(M,N) + \bar q \, \overline{ \cT_1(M,N)} + \cO( |q|^2)
\eea
Note that $\cT_1(M,N)$ need not be real.  These leading components may be read off from the formulas obtained above. For the special cases where $M=0$, we  find, 
\bea
\cT_0(0,0) & = & {  y^2 \over 45} + { \zeta (3) \over  y } 
\no \\ 
\cT_1(0,0) & = & 2  +{1 \over y }
\eea
as well as when $N\not= 0$, 
\bea
\cT_0(0,N) & = & { 2  \tau_2^2 \over 3 N^2 } - { 6 \tau_2^2 \over \pi^2 N^4} + \sum _{m=1}^\infty 
{4 \tau_2 \over \pi m} \, { 1 \over (N^2+4m^2\tau_2 ^2)} 
\no \\
\cT_1(0,N) & = & { 4  \tau_2 \over \pi (N^2+4\tau_2^2)}
\eea
For the  case $M\not= 0$, we obtain, 
\bea
\cT_0(M,N) & = &  { 2  \tau_2^2 \over 3 |M\tau +N|^2} - \left ( { i \tau_2 \over \pi^2 M (M\tau+N)^3} 
- { i \tau_2 \over \pi^2 M (M \bar \tau+N)^3} \right )
\no \\ &&
+ \sum _{m\not= 0}^\infty  \left [ {  \tau_2 \over \pi |m| (M\tau+N) (M\bar \tau +N - 2 i m\tau_2)}    + {\rm c.c.} \right ]
\no \\
\cT_1(M,N) & = & { 2  \tau_2 \over \pi (M\tau +N)} \left ( { 1 \over  M\bar \tau +N - 2i\tau_2} + { 1 \over M \bar \tau +N+2 i \tau_2 } \right )
\eea

\subsubsection{Useful summation  formulas}

When calculating the sums needed to evaluate the $C$- and $D$-functions, it is very convenient to organize the calculation in terms of the poly-logarithms $\Psi (n,z)$, defined by,
\bea
\Psi (n,z) = { d^n \over dz^n} \ln \Gamma (z)
\eea
where $\Psi (z)$ is often used for $\Psi (0,z)$. 
Finite sums are conveniently expressed in terms of these functions, since we have,
\bea
\Psi (n,m) - \Psi (n,1) = (-)^n \Gamma (n+1) \sum _{\ell=1}^{m-1} { 1 \over \ell ^{n+1}}
\eea
with the following explicit evaluations of their values at $z=1$,
\bea
\Psi (0,1) & = & - \gamma 
\no \\
\Psi (n,1) & = & (-)^{n+1} \Gamma (n+1) \zeta (n+1)
\eea
The following sums may be recast in terms of multi-zeta functions,
\bea
\sum _{m=1}^\infty { \Psi (n,m) -\Psi (n,1) \over m^k} = (-)^n \Gamma (n+1) \zeta (k,n+1)
\eea
where convergence requires $k \geq 2$. A further generalization of this result will be needed
for sums quadratic in $\Psi$, and we have, 
\bea
&&
\sum_{m=1} ^\infty \, { 1 \over m^k} \, \prod _{a=1,2} \Big ( \Psi (k_a-1, m) - \Psi (k_a-1,1) \Big )
\no \\ && \quad =
(-)^{k_1+k_2} \Gamma (k_1) \Gamma (k_2) \Big ( \zeta (k, k_1+k_2) + \zeta (k, k_1, k_2) + \zeta (k, k_2, k_1) \Big )
\eea
Finally, the following two sums will also enter,
\bea
\label{a42}
\sum _{m=1}^\infty { \Psi (2m) + \gamma  \over m^6} 
& = &
{ 133 \over 4} \zeta (7) - 16 \zeta (2) \zeta (5) - 4 \zeta (3) \zeta (4)
\no \\
\sum _{m=1}^\infty { \Psi (1,2m) - \Psi (1,1)  \over m^5} 
& = &
98 \zeta (7) - {111\over 2}  \zeta (2) \zeta (5) - 4 \zeta (3) \zeta (4)
\eea
Although evaluations of these sums have been claimed in \cite{Chen-Eie}, the expressions given for them in 
formulas (1.10) and (1.13) of that paper do not stand the test of numerical verification. The validity of the above formulas (\ref{a42}) has then been verified numerically up to a relative precision of $10^{-16}$ using MAPLE.

\subsection{Evaluating the pure power part $D_{2,2,1}$}

We decompose the calculation of $D_{2,2,1}$ into the contributions coming from $M=0$
and those coming from $M \not=0$, 
\bea
\label{7a1}
D_{2,2,1} = \mD^{(0)} + \mD^{(1)}
\eea
where each piece is given by,
\bea
\label{7a2}
\mD^{(0)} & = & \sum _{N \not= 0} { \tau_2 \over \pi N^2} \, \cT(0,N)^2
\no \\
\mD^{(1)} & = & \sum _{M \not= 0} \sum _{N} { \tau_2 \over \pi |M\tau +N|^2} \, \cT(M,N)^2
\eea
We shall now evaluate each contribution in turn.

\subsubsection{Evaluating the power terms in $\mD^{(0)}$}

Retaining only the power terms, the contribution $\cD^{(0)}$ itself splits into three parts, according to whether it involves one, two, or three sums,
\bea
\label{7b1}
\mD^{(0)} = \mD^{(0)}_1 + \mD^{(0)}_2 + \mD^{(0)}_3
\eea
where these terms are the power parts of the following sums, 
\bea
\label{7b2}
\mD^{(0)} _1 & = & \sum _{N\not= 0} { \tau_2 \over \pi N^2} \left ( { 2  \tau_2^2 \over 3 N^2} - { 6 \tau_2^2 \over \pi^2 N^4} \right )^2
\no \\
\mD^{(0)} _2 & = & 2 \sum _{N\not= 0} { \tau_2 \over \pi N^2} \left ( { 2  \tau_2^2 \over 3 N^2} - { 6 \tau_2^2 \over \pi^2 N^4} \right )\sum _{m=1}^\infty { 2  \tau_2 \over \pi m} \, { 1 \over (N^2+4m^2\tau_2^2)}
\no \\
\mD^{(0)} _3 & = & \sum _{N\not= 0} { \tau_2 \over \pi N^2} \sum _{k=1}^\infty \sum _{m=1}^\infty 
{ (2  \tau_2)^2 \over \pi^2 km } \, { 4 \over (N^2+4k^2\tau_2^2) (N^2+4m^2\tau_2^2)}
\eea
The summations over $N$ are carried out with the help of (\ref{sumq}). 
The remaining summations over $k$ and $m$ are then transformed into sums that give rise to $\zeta$-functions and their products. The results are as follows,
\bea
\label{7b3}
\mD^{(0)} _1 & = & { 8 y^5 \over 467775}  
\no \\
\mD^{(0)} _2 & = & 
{ 4 y^2 \over 945} \zeta (3) -{ 2 \zeta (5) \over 45}  + {1 \over 5670 y} 
- {\zeta(7) \over 3 y^2} + { 1 \over  25200 \, y^3}   -{ 3 \zeta(9)  \over 16 y^4} 
\no \\
\mD^{(0)} _3 & = & 
{\zeta(3)^2 \over 3 y}  +{ 1 \over y^2} \Big ( 2 \zeta (7) - \zeta (3) \zeta(4) - \zeta(2) \zeta(5) \Big )  
+ {\zeta (3) \zeta (5) \over 2 y^3} 
\eea

\subsubsection{Evaluating the power terms in $\mD^{(1)}$}

Retaining only the power terms, the contribution $\mD^{(1)}$ itself splits into three parts, according to whether it involves one, two, or three sums in addition to the sum over $M \not= 0$,
\bea
\label{7c1}
\mD^{(1)} = \mD^{(1)}_1 + \mD^{(1)}_2 + \mD^{(1)}_3
\eea
where 
\bea
\label{7c2}
\mD^{(1)} _1 & = & \sum _{M \not= 0} \sum _{N} { \tau_2 \over \pi |M\tau +N|^2} 
\left ( { 2 \tau_2^2 \over 3 |M\tau +N|^2} -\left \{ { i \tau_2 \over \pi^2 M (M\tau +N)^3} +{\rm c.c.} \right \} \right )^2
\no \\
\mD^{(1)} _2 & = & 2 \sum _{M \not= 0} \sum _{N} { \tau_2 \over \pi |M\tau +N|^2} 
\left ( { 2  \tau_2^2 \over 3 |M\tau +N|^2} -\left \{ { i \tau_2 \over \pi^2 M (M\tau +N)^3} +{\rm c.c.} \right \} \right )
\no \\ && \hskip 0.5in 
\times \sum _{m \not= 0} \left ( { \tau_2 \over \pi |m| (M\tau +N) (M\bar \tau +N - 2 i m\tau_2)} + {\rm c.c.} \right )
\no \\
\mD^{(1)} _3 & = & \sum _{M \not= 0} \sum _{N} {  \tau_2 \over \pi |M\tau +N|^2} 
\sum _{k \not= 0} 
\left ( {  \tau_2 \over  \pi |k| (M\tau +N) (M\bar \tau +N - 2 i k\tau_2)} + {\rm c.c.} \right )
\no \\ && \hskip 1in \times
\sum _{m \not= 0}   \left ( { \tau _2 \over \pi |m| (M\tau +N) (M\bar \tau +N - 2 i m\tau_2)} + {\rm c.c.} \right )
\eea
In each case, the sum over $N$ is computed first with the help of (\ref{sumq}).
For $\mD^{(1)}_1$  the power contributions to the remaining sum over $M\not=0$ are computed in 
terms of $\zeta$-functions. For $\mD^{(1)}_2$ one needs to appeal to a few simple  multi-zeta functions.  For $\mD^{(1)}_3$,  the sum over $N$ reveals that its entire contribution is proportional to $y^{-2}$. 

\sm

In summary, the results are as follows,
\bea
\label{7c3}
\mD^{(1)} _1 & =  & { \zeta (5) \over 3}  +{5 \zeta (7) \over 6 y^2}  +{ 21 \zeta (9) \over 16 y^4} 
\no \\
\mD^{(1)} _2 & =  & - { 1 \over 6y } \zeta (6) - {1 \over 3 y} \zeta(3)^2 
-{ 3  \over 2 y^3} \zeta (3) \zeta (5) -{ 3  \over 8 y^3} \zeta (8)
\no \\
\mD^{(1)} _3 & =  & { \cB \over 16 y^2} 
\eea
where $\cB$ is a number which is independent of $\tau$.  Collecting all the contributions to the value of $D_{2,2,1}$ up to the knowledge of $\cB$, which affects only a single power of $y$, we find, 
\bea
\label{7c4}
\cD_{2,2,1} ^{(0,0)} (y) & = & { 8  y^5 \over 467775} +{ 4 y^2 \over 945} \zeta (3) 
+ {13  \zeta (5) \over 45 } -{ \zeta(3) \zeta(5) \over y^3}  +{9 \zeta (9)  \over 8 y^4} 
\no \\ && 
+ {1 \over 16 y^2} \Big ( \cB + 40 \zeta (7) - 16 \zeta(3) \zeta(4) -16 \zeta(2) \zeta(5) \Big ) 
\eea
The contribution $\cB$ is the most challenging to compute. It is given by the power part of the following sum,
\bea
\label{7d1}
\cB ={ 16 \tau_2^5 \over \pi} \sum _{M \not= 0} \sum _{N} {  1 \over |M\tau +N|^2} 
\left \{ \sum _{k \not= 0} 
\left ( { 1 \over |k| (M\tau +N) (M\bar \tau +N - 2 i k\tau_2)} + {\rm c.c.} \right ) \right \}^2
\eea
Within the power part approximation, $\cB$ is a constant, as becomes evident upon carrying out the summation over $N$ using (\ref{sumq}). To carry out the summation over $N$, we decompose into partial fractions. The order of the poles jumps when $M=-k$, and these contributions need to be isolated, and handled separately. We shall omit the details of the calculation here, and simply quote the result, 
\bea
\cB= -26 \zeta (7) + 16 \zeta (3) \zeta (4) + 16 \zeta (2) \zeta (5)
\eea
which yields the result for $\cD_{2,2,1}^{(0,0)}(y)$ quoted in (\ref{powerd221}).

\subsection{Evaluating the leading exponential terms}

The leading exponential contribution to $\cD^{(0)}$ splits into three parts, according to whether it involves zero, one or two sums, and we shall denote these contributions by $\cE$,
\bea
\label{7q1}
\cD^{(0)} \bigg | _q = 
2q \sum _{N \not= 0} { y \over \pi N^2} \cT_0(0,N) \cT_1(0,N) 
= 
q \left ( \cE^{(0)}_1 + \cE^{(0)}_2 + \cE^{(0)}_3 \right )
\eea
where 
\bea
\label{7q2}
\cE^{(0)} _1 & = & 0
\no \\
\cE^{(0)} _2 & = & 2 \sum _{N\not= 0} { y \over \pi N^2} \left ( { 2  y^2 \over 3 N^2} - { 6 y^2 \over \pi^2 N^4} \right ) { 4 y \over \pi (N^2+4y^2)}
\no \\
\cE^{(0)} _3 & = & 2 \sum _{N\not= 0} { y \over \pi N^2} \sum _{m=1}^\infty  
{ 1 \over m } \, { (4 y)^2 \over \pi^2 (N^2+4y^2) (N^2+4m^2y^2)}
\eea
The summations over $N$ are carried out with the help of the formulas of appendix \ref{secB}. 
The remaining summation over  $m$ are then transformed into sums that give rise to $\zeta$-functions. The results are as follows,
\bea
\label{7q3}
\cE^{(0)} _1 & = & 0 
\no \\
\cE^{(0)} _2 & = & 
{ 4  y^2 \over 945}
 -{ 2  \over 45} 
  + {1 \over 6 y} 
- {1 \over 3 y^2}
+ { 3  \over 8y^3}  
-{ 3  \over 16 y^4} 
\no \\
\cE^{(0)} _3 & = & 
{2  \over 3 y} \zeta(3) 
+ { 1 \over y^2} - {\pi^2 \over 6 y^2} - {\pi^4 \over 90 y^2} +{1 \over 2 y^3} \zeta (3) + { 1 \over 2 y^3} \zeta (5)
\eea

\subsubsection{Evaluating  the leading exponential terms in $\cD^{(1)}$}

The leading exponential contribution to $\cD^{(1)}$ splits into three parts, according to whether it involves zero,  one, or two, sums in addition to the sum over $M \not= 0$,
\bea
\label{7s1}
\cD^{(1)} \bigg | _q = q  \left ( \cE^{(1)} _0 + \cE^{(1)}_1  + \cE^{(1)}_2  + \cE^{(1)}_3 \right )
\eea
where 
\bea
\cE^{(1)}_0 = 2 \sum _{M \not= 0} \sum _N { y \over \pi |M \tau +N|^2} \cT_0(M,N) \cT_1(M,N)
\eea
The remaining sums are the parts linear in $q$ of the sums  defined earlier, 
\bea
\label{7s2}
q \cE^{(1)} _1  & = & \sum _{M \not= 0} \sum _{N} { y \over \pi |M\tau +N|^2} 
\left ( { 2 y^2 \over 3 |M\tau +N|^2} -\left \{ { i y \over \pi^2 M (M\tau +N)^3} +{\rm c.c.} \right \} \right )^2
\bigg |_q
\no \\
q \cE^{(1)} _2  & = & 2 \sum _{M \not= 0} \sum _{N} { y \over \pi |M\tau +N|^2} 
\left ( { 2  y^2 \over 3 |M\tau +N|^2} -\left \{ { i y \over \pi^2 M (M\tau +N)^3} +{\rm c.c.} \right \} \right )
\no \\ && \hskip 0.5in 
\times \sum _{m \not= 0} \left ( { y \over \pi |m| (M\tau +N) (M\bar \tau +N - 2 i my)} + {\rm c.c.} \right )
\bigg |_q
\no \\
q \cE^{(1)} _3  & = & \sum _{M \not= 0} \sum _{N} {  y \over \pi  |M\tau +N|^2} 
\sum _{k \not= 0} 
\left ( {  y \over \pi |k| (M\tau +N) (M\bar \tau +N - 2 i ky)} + {\rm c.c.} \right )
\no \\ && \hskip 1in \times
\sum _{m \not= 0}   \left ( { y \over \pi |m| (M\tau +N) (M\bar \tau +N - 2 i my)} + {\rm c.c.} \right )
\bigg |_q
\eea
In each case, the sums over $N$ are computed first, with the help of the formulas of appendix \ref{secB}.
The contributions involving the sums over $k$ and $m$ are further partitioned according to whether 
either $k$ or $m$ or both are equal to $-M$, or not, 
\bea
\cE^{(1)}_0 & = & \cE^{(1)}_{0a} + \cE^{(1)}_{0b} + \cE^{(1)}_{0c}
\no \\
\cE_2^{(1)} & = & \cE^{(1)} _{2a} + \cE^{(1)} _{2b}
\no \\
\cE_3^{(1)} & = & \cE^{(1)} _{3a} + \cE^{(1)} _{3b} + \cE^{(1)} _{3c}
\eea
Here $\cE_{0a}^{(1)}$ is the contribution from the part of $\cT_0(M,N)$ that does not involve a sum over $m$, while $\cE_{0b}^{(1)}$ is the contribution that arises from the sum part of $\cT_0(M,N)$ of the single term $m=-M$, and finally $\cE_{0c}^{(1)}$ is the contribution from the sum part in $\cT_0(M,N)$ with $M\not= -m$. 
Furthermore,  $\cE^{(1)} _{2a}$ is the contribution from $m=-M$ while $\cE^{(1)} _{2b}$ is the contribution from  $m \not= -M$. Similarly, $\cE^{(1)} _{3a}$ is the contribution from $k=m=-M$, while $\cE^{(1)} _{3b}$ is the combined contribution from $k=-M, m\not= -M$ and $k \not= -M, m=-M$, while $\cE^{(1)} _{3c}$ is the contribution from $k, m \not = -M$. These contributions are found to be given by, 
\bea
\label{7s3}
\cE^{(1)}_{0a} & = & -{1 \over 3 y} -{ 3  \over 4 y^3} - { \zeta (5) \over y^3}
\no \\
\cE^{(1)}_{0b} & = & {17  \over 4y^2} -{\pi^2 \over 3 y^2} +{\pi^4 \over 90 y^2}
\no \\
\cE^{(1)}_{0c} & = & -{21  \over 4 y^2} +{\pi^2 \over 2 y^2} +{ \zeta (3) \over y^2}
\no \\
\cE^{(1)} _1 & =  & { 68  y^2 \over 45} +{122  y \over 45} +{8  \over 3} +{8 \over 3y} 
+{37 \over 12 y^2} +{21 \over 8 y^3} + {21 \over 16 y^4}
\no \\
\cE^{(1)} _{2a} & =  & - {128  y^2 \over 45} - 4 y -{ 16 \over 3} -{19 \over 3 y} 
-{ 11 \over 2 y^2} -{11 \over 4 y^3}
\no \\
\cE^{(1)} _{2b} & =  & 4 +{22 \over 3y} -{ 14 \zeta (3) \over 3y} +{6 \over y^2} 
-{ 4 \zeta (3) \over y^2} +{11 \over 4 y^3} -{3 \zeta (3) \over 2 y^3} - { \zeta (5) \over 2 y^3}
\no \\
\cE^{(1)} _{3a} & =  & { 4 y^2 \over 3} + {4 y \over 3} + 3  
+ { 7 \over 2y} + {7 \over 4 y^2}
\no \\
\cE^{(1)} _{3b} & =  & - 4 -{ 6 \over y} +{ 4  \zeta (3) \over y} -{7 \over 2 y^2} +{ 2 \zeta (3) \over y^2}
\no \\
\cE^{(1)} _{3c} & =  & {1 \over 2 y^2} 
\eea
The sum of these contributions yields (\ref{expD221}).

\section{Preliminary numerical study of modular functions}
\setcounter{equation}{0}
\label{secD}

The analysis of the properties of the $C_{a,b,c}$ functions in section~\ref{sec3} and the asymptotic properties of certain $D$ and $C$ functions in the $y\to \infty$ limit in sections \ref{sec4} and \ref{sec5} has led us to a number of conjectured relations between these functions.  Although we have a complete understanding of the relationships between the $C_{a,b,c}$ functions at any weight $w$,  our analysis of other weight-$w$ functions has been limited to conjectured relationships for functions with $w\le 5$.  We do not have a rigorous mathematical procedure for establishing the validity of these conjectures, beyond the striking agreement of many terms in the asymptotic expansions.   In the absence of an analytic method we have approached the problem of establishing such relationships
by numerical methods.   

\sm

Even though at present our attempts are very primitive we believe they demonstrate the possibility of using numerical approximations to the modular functions to sufficient accuracy to test the validity of our conjectures.  The conjectured relations that we want to test are summarized by:

\sm

$ \bullet$ One relation at weight 4, which we recall here from (\ref{1a9}), 
\bea
\label{6a1}
C_{1,1,1,1} = 24 C_{2,1,1} + 3\tE_2^2 + 18 \tE_4
\eea
We also recall the  alternative notation,  $C_{1,1,1,1}=D_4$ and $C_{2,1,1}=D_{2,1,1}$.

\sm

$\bullet$ Three relations at weight 5, which we recall here from (\ref{1a10}), 
\bea
\label{6a2}
40 C_{2,1,1,1} & = &  300 C_{3,1,1} +120  \tE_2 \tE_3 - 276  \tE_5 + 7  \zeta (5) 
\no \\
C_{1,1,1,1,1} & = &   60 C_{3,1,1} +10  \tE_2 \tE_3 - 48  \tE_5+ 10\zeta(3) E_2 + 16 \zeta (5) 
\no \\
10 D_{2,2,1} & = &   20 C_{3,1,1} - 4  \tE_5 + 3  \zeta (5) 
\eea
In the alternative notation we have $C_{2,1,1,1}=D_{3,1,1}$, 
$C_{3,1,1}=D_{2,1,1,1}$, $C_{1,1,1,1,1}=D_5$.

\sm

Our numerical procedure is based on  evaluating
the expressions for the $D$ functions by  using the integral representation in terms of the Green function,
\bea
  \label{e:Ddefnum}
D_{\ell_{12},\dots,\ell_{34}}(\tau) = \left ( \prod_{i=1}^4 \int_\Sigma { d^2z_i \over \tau_2} \right )
     \prod_{1\leq i<j\leq 4} \cG(z_i-z_j|\tau)^{\ell_{ij}} 
\eea
where $\cG(z|\tau)$ was defined in (\ref{2c3}).

\sm

We have obtained the best compromise between precision and running time by
using the {\tt Divonne} implementation for {\tt Mathematica} described
in~\cite{Hahn:2004fe}.\footnote{We would like to thank the Niels Bohr
  Institute and the IH\'ES for allowing us to use their computer resources.}

\sm

 The asymptotic expansions for large $\tau_2$ described earlier include power behaved corrections to the first exponentially suppressed term proportional to $e^{-2\pi \tau_2}$ and ignore corrections of order $e^{-4 \pi \tau_2}$ which are highly suppressed for relatively small values of $\tau_2$.  Therefore, in order to provide a numerical  test that goes beyond the asymptotic expansions  which motivated the conjectured relationships, we need to evaluate the functions with considerable accuracy, which we have found difficult to achieve.
 
 \sm

In order to keep the computation time to at most the order of a few days, in the following we have attempted a precision of
$\epsilon=10^{-1},10^{-2}, 10^{-3}$ or $10^{-4}$ with about  $500\times10^6$ sampling
points, and using four values of the complex structure parameter
$\tau \in \{i,\frac12+i{\sqrt3\over2},i\sqrt2,\frac12+7i\}$.   

\begin{table}[h]
\begin{center}
\begin{tabular}[h]{||c|c|c|c|c|c||}
\hline
Function & $\eps$&$\tau= i$&$\tau=i\sqrt2$&$\tau=\frac12+7i$&$\tau=\frac12+{\sqrt3\over2}i$ \cr         
\hline       \hline                  
$D_3$ &$10^{-4}$&1.0003&0.9995&0.9999& 1.1851\cr
\hline
$D_4$&$10^{-3}$&0.996& 0.998& 1.009& 1.024\cr 
\hline
$D_5$& $10^{-2}$&0.95 & 0.97 & 1.12 & 0.93 \cr
\hline
$D_{2,2,1}$& $10^{-2}$&0.96& 0.92 & 1.07& 0.84 \cr
\hline
$D_{3,1,1}$ & $10^{-2}$&1.10& 0.87& 1.05& 1.37\cr
\hline 
\end{tabular}
\caption{The ratio of left-hand and right-hand side of the identity}
\end{center}
\end{table}

\begin{table}[h]
\begin{center}
\begin{tabular}[h]{||c|c|c|c|c|c||}
\hline
 Function&\  & $\tau= i$&$\tau=i\sqrt2$&$\tau=\frac12+7i$&$\tau=\frac12+{\sqrt3\over2}i$\cr
  \hline \hline
$D_3$&numerical & 1.35189 & 1.42809 & 23.7143 & 1.33559 \cr 
          &asymptotic & 1.35227 & 1.42744 & 23.7119 & 1.33542 \cr 
\hline                                        
$D_4$&numerical &5.59442 & 6.118 & 265.665 & 5.47065 \cr
           &asymptotic & 5.60153 & 6.11683 & 265.577 & 5.46992 \cr
\hline
$D_5$&numerical & 27.5931 & 29.8883 & 1330.99 & 27.0271 \cr
           &asymptotic & 27.6018 & 29.9017 & 1331.95 & 27.0225 \cr
\hline
$D_{2,2,1}$& numerical & 0.417141 & 0.475233 & 90.6985 & 0.398955 \cr
                     & asymptotic & 0.419068 & 0.463207 & 90.7233 & 0.409520 \cr
\hline
$D_{3,1,1}$ & numerical  &0.701572 & 1.05760 & 474.804  & 0.733569 \cr
& asymptotic$^\star$ & 0.769175 & 1.02236 & 474.913 & 0.778507 \cr 
\hline
\end{tabular}
\caption{Comparison of numerical and asymptotic values}
\end{center}
\end{table}

The results that we will now describe  are summarized in Table 1 and Table 2.  
As a benchmark,  we first numerically tested the   weight-3 relation, 
\bea
\label{e:rw23}
  D_3(\tau) = C_{1,1,1}(\tau)=\tE_3(\tau)+ \zeta(3)
\eea
which is known to be correct since we  proved it analytically in sub-section~(\ref{sec:cabclow}).  The results are shown in the first row of each table.   In this case we were able to achieve $\epsilon = 10^{-4}$, as noted in the second column of Table 1.  Columns 3-6 of  Table 1 show the ratio of the left and right-hand sides of equation  (\ref{e:rw23})  obtained by numerical evaluation  of the functions $D_3$,  $E_3$ with the help of the integral  representation in~(\ref{e:Ddefnum}).  This ratio is equal to 1 to the stated accuracy.  In the first row of Table 2 we list the numerical estimate of the  value of the function $D_3(\tau)$ at each value of $\tau$, while the second row gives the value of the function as determined from the asymptotic estimate, which keeps the pure power terms and the first exponential correction in the exact expression in (\ref{e:rw23}).
The results for the four conjectured relations in (\ref{6a1}) and (\ref{6a2})  are presented in the rows of  Tables 1 and 2 underneath the $D_3$ results.   
 
 \sm
 
 Although all the results are consistent with the conjectured relations,  they are not yet more accurate than the asymptotic expansions that include the first exponentially suppressed terms that we studied analytically in the body of the text. 
For this reason we are not able to state with conviction that the numerical data confirms the conjectured relationships beyond the asymptotic analysis, but we have included this appendix since it does suggest that numerical methods are well within reach given a little more sophistication.
 
 \sm

The ${}^\star$ on the last row in Table 2 indicates the fact that in this row only the pure power part of the asymptotic expansion has been included, since the leading exponential correction has not, in fact, been calculated yet.

 

\end{document}